\documentclass[review]{elsarticle}

\usepackage{amsmath}

\usepackage{amsfonts}
\usepackage{amssymb}

\usepackage{comment}

\usepackage{setspace}
\usepackage{graphicx}
\graphicspath{{./}{}}
\usepackage{subcaption}
\usepackage{gensymb}

\usepackage[a4paper,width=150mm,top=25mm,bottom=25mm,bindingoffset=0mm]{geometry}
\usepackage{graphics}
\usepackage{multirow}
\usepackage{adjustbox}
\usepackage [english]{babel}
\usepackage [autostyle, english = american]{csquotes}
\MakeOuterQuote{"}
\usepackage[normalem]{ulem}
\useunder{\uline}{\ul}{}
\usepackage[table,xcdraw]{xcolor}\label{key}
\definecolor{siteint}{RGB}{31,119,180}
\definecolor{sitesub}{RGB}{255,127,14}
\definecolor{sitemix}{RGB}{44,160,44}
\biboptions{numbers,sort&compress}
\usepackage{xr-hyper}
\usepackage[colorlinks]{hyperref}
\hypersetup{
	colorlinks,
	linkcolor={red},
	filecolor={red},
	urlcolor={red},
	citecolor={blue}
}
\usepackage[table]{xcolor}
\usepackage{booktabs}

\begin{document}
	
	\begin{frontmatter}
		\title{Grain boundary segregation of light elements and their effects on cohesion in ferritic steels}
		\author[MPSusMataffiliation]{Han Lin Mai}
		\author[USYDaffiliation]{Xiang-Yuan Cui}
		\author[BAMaffiliation]{Tilmann Hickel}
		\author[USYDaffiliation]{Simon P. Ringer \corref{mycorrespondingauthor}}
		\author[MPSusMataffiliation]{J\"{o}rg Neugebauer \corref{mycorrespondingauthor}}
		\cortext[mycorrespondingauthor]{Corresponding authors: neugebauer@mpie.de, simon.ringer@sydney.edu.au}
		\address[MPSusMataffiliation]{Computational Materials Design Department, Max Planck Institute for Sustainable Materials, Max-Planck-Straße 1, 40237 Dusseldorf, Germany}	
		\address[USYDaffiliation]{School of Aerospace, Mechanical and Mechatronic Engineering \& Australian Centre for Microscopy and Microanalysis, Faculty of Engineering, The University of Sydney, 2006 New South Wales, Australia}
		\address[BAMaffiliation]{BAM Federal Institute for Materials Research and Testing, 12489 Berlin, Germany}
		\begin{abstract}
		Light elements play an important role in influencing the macroscale properties of engineering alloys through grain boundary (GB) segregation phenomena. However, the scarcity and scattered nature of \textit{ab initio} datasets for light elements in steels makes reproduction and extraction of general trends from the literature difficult. Here, we present a comprehensive \textit{ab initio} evaluation of the segregation energies and cohesive effects for H, He, B, C, N, O, P, S, extensively sampling both substitutional and interstitial sites in six model coincident site lattice (CSL) ferritic iron GBs using density functional theory (DFT). Cohesive effects are evaluated in both a quantum-chemistry bond-order and rigid Rice-Wang interfacial cohesive strength framework. Our calculations indicate that, compared at the same concentration, B and C enhance GB cohesion, N, P, H are mildly detrimental, and He, O, S as powerful decohesive agents/embrittlers. Sampling both interstitial and substitutional starting positions is necessary to accurately capture segregation spectra. Commonly utilised sampling criteria such as site volumes prove insufficient for identifying deepest GB binding sites. Solutes placed in either kind of site can induce large relaxations to the same final configuration, resulting in site classification ambiguity. The nearest neighbour distance of a solute to its neighbours after relaxation is shown to be a controlling factor for the lower threshold of segregation energies at sites. The freely available DFT dataset and analysis repositories are expected to advance understanding of GB segregation behaviours of light elements in steels and serve as a resource for developing machine learning interatomic potentials.
		\end{abstract}
	\end{frontmatter}

\section{Introduction}
Grain boundaries (GBs) are critical defect structures that exist in almost all engineering alloys. The mechanical  behaviours of a material can be dominated by GBs, specifically by an elevated concentration of impurities or alloying elements that occur near these defects, known as GB segregation. Many of the most important impurities or alloying additions are the elements that lie at the beginning of the periodic table, often known as the light elements. Some of the most famous examples include C, which is a critical alloying element in steels, and H, which can attack alloys. H causes embrittlement of metallic alloys, a phenomena known as hydrogen embrittlement, and is one of the oldest active research topics in metallurgy \cite{johnsonIIRemarkableChanges1875, johnsonRemarkableChangesProduced1875}. Further examples include that of temper embrittlement of steels, which many have attributed to the presence of segregated S and P at GBs \cite{kalderonSteamTurbineFailure1972}, as well as radiation damage phenomena in nuclear reactors, where much more significant exposure to both H and He can be expected than in other applications. 
\\\\
Due to their importance, many simulation studies have investigated the segregation of these light elements at GBs \cite{yamaguchiDecohesionIronGrain2007, yamaguchiFirstprinciplesStudyGrain2011,  yamaguchiFirstprinciplesStudyGrain2011a, yamaguchiMobileEffectHydrogen2012, yamaguchiIntergranularDecohesionInduced2014, tahirHydrogenEmbrittlementCarbon2014, azocarguzmanEffectsMechanicalStress2024, lejcekInterfacialSegregationGrain2017, sakicInterplayAlloyingTramp2024, xuEffectAlloyingSolutes2024}. Such studies exist for H \cite{duFirstprinciplesStudyInteraction2011a, yamaguchiFirstprinciplesStudyGrain2011a,yamaguchiMobileEffectHydrogen2012},  He \cite{zhangFirstprinciplesStudyHelium2009, zhangFirstprinciplesStudyHe2010, suzudoAtomisticModelingHe2013}, B \cite{yamaguchiFirstprinciplesStudyGrain2011, wachowiczEffectImpuritiesStructural2011}, C \cite{tahirHydrogenEmbrittlementCarbon2014, wangFirstprinciplesStudyCarbon2016a, yamaguchiFirstprinciplesStudyGrain2011a, wachowiczEffectImpuritiesStructural2011, hatcherParameterizedElectronicDescription2014}, N\cite{wachowiczEffectImpuritiesStructural2011}, O \cite{wachowiczEffectImpuritiesStructural2011}, P \cite{yamaguchiFirstprinciplesStudyGrain2011}, S \cite{yamaguchiFirstprinciplesStudyGrain2011}. However, prior studies of these elements have often focused on a single/few sites, usually selected on an ad-hoc basis, or at a single or few model GBs, with the most work performed on the $\Sigma3[110](1\bar{1}1)$ model GB \cite{yamaguchiDecohesionIronGrain2007, wachowiczEffectImpuritiesStructural2011, suzudoAtomisticModelingHe2013, yamaguchiIntergranularDecohesionInduced2014, itoElectronicOriginGrain2020, kholtobinaEffectAlloyingElements2021, xuEffectAlloyingSolutes2024}. This can be attributed to the computational cost of density functional theory (DFT) calculations, which often limited the structural configurations and model GBs that could be studied. However, to comprehensively understand segregation phenomena, it has been shown that single/few segregation energies at a single model GB cannot be generally representative of how an element may be expected to behave overall at GBs \cite{maiSegregationTransitionMetals2022, maiPhosphorusTransitionMetal2023, wagihViewpointCanSymmetric2023}. Importantly, the scarcity of datasets with extensive site sampling has so far limited our ability to extract and compare trends on the behaviour of these light elements in segregation phenomena at GBs.
\\\\
To address the time and scale limitations of ab-initio methods, empirical interatomic potentials are often used to study GB segregation-related phenomena \cite{tehranchiAtomisticStudyHydrogen2017, reiners-sakicInterstitialsKeyIngredient2025}, but they often suffer from inaccuracy \cite{wagihGrainBoundarySegregation2024}. The recent advent of machine learning interatomic potentials (MLIPs) has promised accuracy approaching that of ab-initio calculations at a fraction of their cost \cite{tuchindaGrainBoundaryEmbrittlement2025, shuangUniversalMachineLearning2025, itoMachineLearningInteratomic2024}, but questions remain on their accuracy \cite{dengSystematicSofteningUniversal2025}, particularly on out-of-domain evaluations \cite{liProbingOutofdistributionGeneralization2025}, as well as in steels \cite{ echeverrirestrepoApplicabilityUniversalMachine2025}. The current lack of open access to challenging, out-of-distribution ab-initio structure-containing datasets to benchmark solute-GB interactions despite the wealth of prior published work is problematic. This data scarcity hampers the community's ability to benchmark the accuracy of MLIPs with respect to defect-GB interactions \cite{liProbingOutofdistributionGeneralization2025}.  This scarcity combined with inconsistent calculation parameter selections also makes reproduction of the results and extraction of general trends from the literature time-consuming and difficult.
\\\\
For the light elements specifically, much attention has been given to their site preferencing behaviour, e.g. substitutional vs. interstitial \cite{cernySegregationPhosphorusSilicon2024, wachowiczEffectImpuritiesStructural2011}, or the different types of interstitial sites \cite{wagihSpectrumInterstitialSolute2023, reiners-sakicInterstitialsKeyIngredient2025, mengHighlyTransferableEfficient2024a, mengNeuralNetworkInteratomic2025, lejcekInterstitialSubstitutionalSolute2016}. However, the cost of extensive sampling in ab-initio methods has so far limited the amount of data on the site preferences of these elements at GBs, with studies utilising empirical potentials or only single/few model GBs with DFT \cite{cernySegregationPhosphorusSilicon2024}. To provide a consistent and comprehensive dataset that may be used for further understanding of site preferencing behaviour, one must extensively study the sites available across a range of GBs.
\\\\ 
In this study, we investigate the segregation behaviours and induced cohesive effects of technologically relevant light elements (X = H, He, B, C, N, O, P, S) across a carefully chosen set of ferritic iron GBs. For accuracy, we utilise ab-initio DFT to perform the calculations in this study. By analysing the generated data, we derive trends that govern segregation, and comment on commonly purported relationships proposed in the literature. The open-source dataset generated by this study, in combination with our prior work \cite{maiHighthroughputInitioStudy2025a}, serves as a comprehensive and chemically complete ab-initio dataset for solute-GB interactions in ferritic iron. This data will not only serve as a fundamental resource for the understanding and engineering of elemental segregation in GBs in ferritic steels, but also as a valuable reference dataset for the future development of MLIPs for steels.
\section{Methodology}
\subsection{DFT calculation details}
We performed first principles calculations based on DFT using the projector augmented wave (PAW) method \cite{blochlProjectorAugmentedwaveMethod1994} as implemented in the Vienna \textit{Ab initio} Simulation Package (VASP) \cite{kresseEfficiencyAbinitioTotal1996, kresseEfficientIterativeSchemes1996}. Spin polarization was accounted for in all calculations performed in this study. We utilized the generalized gradient approximation (GGA) via the Perdew-Burke-Ernzerhof (PBE) functional \cite{perdewGeneralizedGradientApproximation1996}. The Brillouin-zone integrations for all GBs employed $\Gamma$-centred $\textbf{k}$-point meshes, with an energy cut-off of 400 eV for the plane wave basis set. Relaxations were performed with a $\textbf{k}$-point mesh with minimum allowed \textbf{k}-point spacing of 0.5 \AA$^{-1}$, before a final static calculation with the denser mesh indicated in Table \ref{tab:grainboundary_pureprops}. For the final static calculation, we compared the final segregation energies computed with the two different \textbf{k}-point mesh configurations, and found that deviations in the total energy were on average less than 0.03 eV. A more detailed technical discussion on the effects of k-point density on the segregation energies is presented in the Supplementary Information. A first order Methfessel-Paxton scheme with a smearing width of 0.2~eV was adopted for all calculations. The electronic minimisation convergence criterion was set to 1x$10^{-5}$~eV, and the relaxation calculations were deemed converged when atomic forces were below 0.01~eV/\AA. 
\\\\
Justifications for our selections of the exchange-correlation functional, $\textbf{k}$-point meshes, plane wave energy cut-off and grain lengths were presented in our prior published work \cite{maiSegregationTransitionMetals2022}. Note that in some cases for segregation and surface calculations used in the cohesion analysis, the supplied \textbf{k}-point meshes do not allow for calculation convergence. In these cases, we fall back to a minimum allowed \textbf{k}-point spacing of 0.5 \AA$^{-1}$. Calculation post-processing was done with a mixture of the pymatgen \cite{ongPythonMaterialsGenomics2013} and pyiron \cite{janssenPyironIntegratedDevelopment2019} VASP scrapers. The post-processed data is available in a FAIR format \cite{wilkinsonFAIRGuidingPrinciples2016} in a GitHub repository linked in the data availability section. This data includes the raw energies, forces, stresses and structures used to produce the analysis in this paper. Analysis scripts and notebooks for reproducing the figures are additionally included. The pseudopotential files used, i.e. their specific VASP POTCAR filenames, are tabulated in Table \ref{tab:potcar} in the Appendices.
\subsection{Grain boundary models}
We consider six coincident site lattice (CSL) type GBs in this study. These were the $\Sigma3[110](1\bar{1}1)$, $\Sigma3[110](1\bar{1}2)$, $\Sigma5[001](210)$, $\Sigma5[001](310)$, $\Sigma9[110](2\bar{2}1)$ and $\Sigma11[110](3\bar{3}2)$ GBs. In this study, we utilise cells which only contain 1 GB interface, with the ends of the grains not forming the GB interface terminated by a vacuum slab, of varying lengths as indicated in Table \ref{tab:grainboundary_pureprops}. Other details on the model cells and properties for the pure GBs are detailed in Table \ref{tab:grainboundary_pureprops}. We have verified that these GBs are stable against vacancies, by removing single Fe atoms at the GB systematically in the range of segregation, with positive formation energies observed overall (see S.I.). We find that the vacancy formation energies at GBs tend to be lower than that of the bulk bcc-Fe phase. More detailed discussion on the vacancy formation energies at GBs is presented in the Supplementary Information.
\begin{figure}[h!]
	\centering
	\includegraphics[height=8cm]{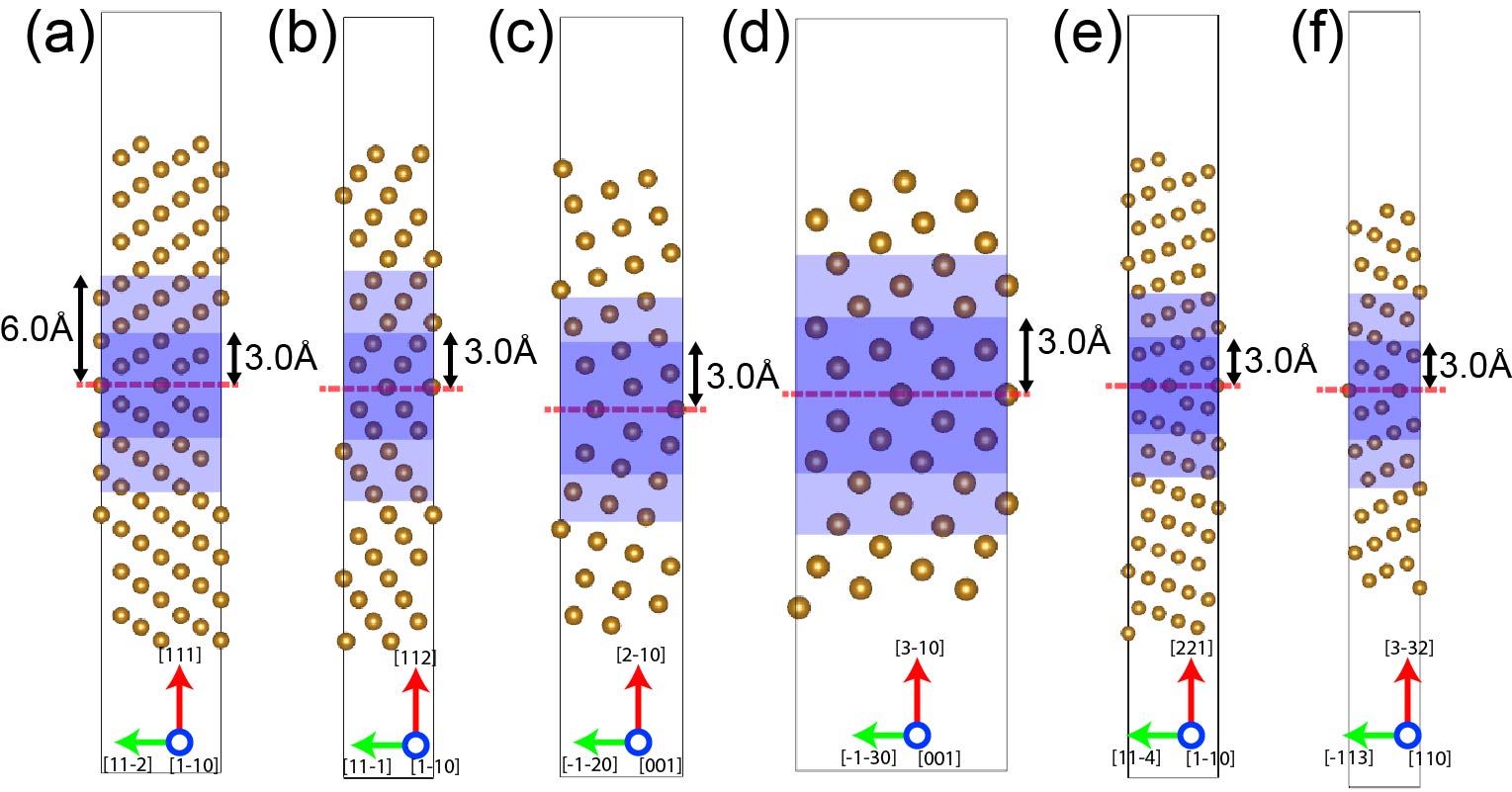}
	\caption{The atomic structures of the six coincident-site-lattice model GBs investigated in this study, the (a) $\Sigma3[110](1\bar{1}1)$,  (b) $\Sigma3[110](1\bar{1}2)$, (c) $\Sigma5[001](210)$, (d) $\Sigma5[001](310)$, (e) $\Sigma9[110](2\bar{2}1)$ and the (f) $\Sigma11[110](3\bar{3}2)$ GBs, respectively. The GB interface planes are highlighted by the dashed red lines. The shaded ranges indicate the range of studied sites for segregation, darker blue for interstitial sites, and the extended range in light blue for substitutional. The b-c projection (see Table \ref{tab:grainboundary_pureprops}) is presented in this Figure. The dimensions of these cells are given in Table \ref{tab:grainboundary_pureprops}. Structures were visualised using VESTA  \cite{mommaVESTAThreedimensionalVisualization2008}}
	\label{fig:GBstructures}
\end{figure}
\begin{table}[h!]
	\centering
	\renewcommand{\arraystretch}{1.5}
	\makebox[\linewidth]{\begin{tabular}{ccccccccccc}
			\hline 
			System & 
			n$_\text{GB}$ & 
			\begin{tabular}[c]{@{}c@{}}a\\ (\AA)\end{tabular} & 
			\begin{tabular}[c]{@{}c@{}}b\\ (\AA)\end{tabular} & 
			\begin{tabular}[c]{@{}c@{}}c$_\text{GB}$\\ (\AA)\end{tabular} &
			\begin{tabular}[c]{@{}c@{}}Vacuum \\ (\AA)\end{tabular} & 
			\begin{tabular}[c]{@{}c@{}}Area \\ (\AA$^2$)\end{tabular} &
			\begin{tabular}[c]{@{}c@{}}$\gamma_\text{GB}$ \\ (J/m$^2$)\end{tabular} &
			\begin{tabular}[c]{@{}c@{}}W$_\text{sep}^\text{RGS}$ \\ (J/m$^2$)\end{tabular} &
			\begin{tabular}[c]{@{}c@{}}\textbf{k}-points \end{tabular} \\ \hline
			$\Sigma3[110](1\bar{1}1)$ & 72 & 4.005 & 6.937 & 28.740 & 15.40 & 27.78 & 1.58 & 4.19 & $6\times3\times1$\\
			$\Sigma3[110](1\bar{1}2)$ & 48 & 4.005 & 4.905 & 27.995 & 14.96 & 19.64 & 0.45 & 4.88 & $6\times6\times1$\\
			$\Sigma5[001](210)$ & 76 & 5.664 & 6.332 & 23.64 & 15.69 & 35.86 & 1.62 & 3.96 & $3\times3\times1$\\ 
			$\Sigma5[001](310)$ & 80 & 5.664 & 8.955 & 18.06 & 13.85 & 50.72 & 1.69 & 3.73 & $3\times3\times1$\\ 
			$\Sigma9[110](2\bar{2}1)$ & 68 &  4.005 & 6.332 & 23.640 & 19.32 & 24.06 & 1.75 & 4.22 & $6\times4\times1$  \\
			$\Sigma11[110](3\bar{3}2)$ & 42 & 4.005 & 4.696 & 24.770 & 15.13 & 18.81 & 1.45 & 4.27 & $6\times6\times1$ \\ \hline
	\end{tabular}}
	\caption{The  number of atoms, sizes of the cells used in this study, vacuum in the GB cells, the cross-sectional areas (for calculating the cohesive effects), their corresponding GB energy ($\gamma_\text{GB}$), rigid work of separation, and corresponding k-points are listed for all considered GBs.} 
	\label{tab:grainboundary_pureprops}
\end{table}
\subsection{Site selection}
\begin{figure}[h!]
	\centering
	\includegraphics[height=6cm]{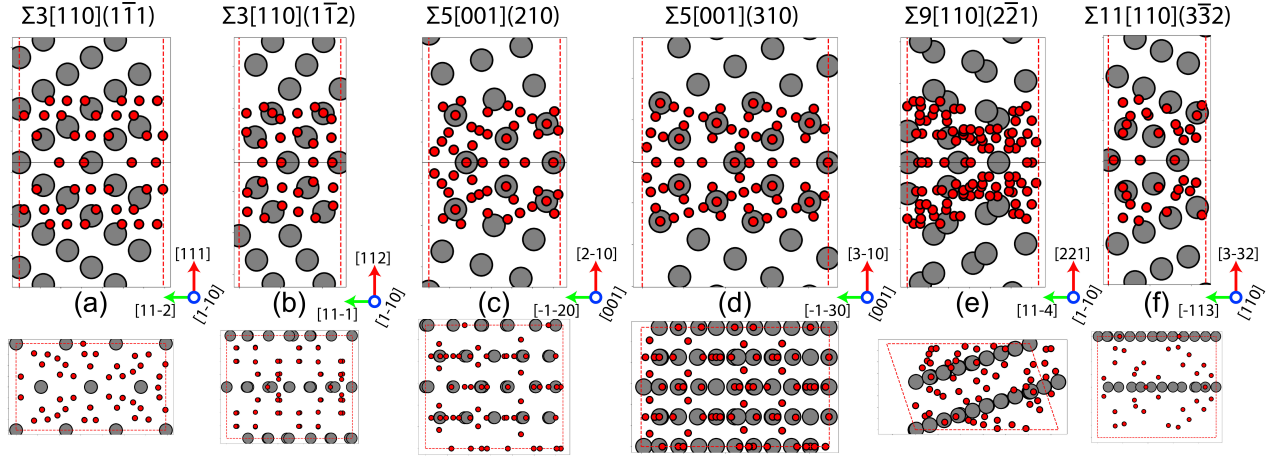}
	\caption{The starting positions considered for interstitial sites in this study for the (a) $\Sigma3[110](1\bar{1}1)$,  (b) $\Sigma3[110](1\bar{1}2)$, (c) $\Sigma5[001](210)$, (d) $\Sigma5[001](310)$, (e) $\Sigma9[110](2\bar{2}1)$ and (f) $\Sigma11[110](3\bar{3}2)$ GBs, respectively.}
	\label{fig:starting_interstitial_sites}
\end{figure}

We studied two types of sites - conventionally known as "substitutional" and "interstitial" classes of sites at our GBs. "Substitutional" sites refer to the structures that are attained through an on-lattice swap with an Fe atom at the optimised pure Fe GB. The data presented in this study for the "substitutional" datapoints, (i.e. on-lattice swaps with Fe atoms on the relaxed pure Fe GBs), were taken from our previous study \cite{maiHighthroughputInitioStudy2025a}. "Interstitial" sites are those that are attained through an insertion of an additional atom into the optimised pure Fe GB. We computed segregation energies of the solutes across both kinds of sites at each GB for all elements (H, He, B, C, N, O, P, S). "Substitutional" sites were studied up to 6 \AA\ away from the interface. "Interstitial" sites were considered up to 3\ \AA\ away from the interface. These distances are indicated in the Figure \ref{fig:GBstructures}.
\\\\
The "interstitial" sites were selected according to a procedure based on Voronoi tessellation of the sites at a GB. The vertices of the tessellated Voronoi polyhedra were considered as starting positions for the interstitial segregation energies. This was achieved through the use of a program written by Guzman \cite{azocarguzmanEffectsMechanicalStress2024}, which was developed on top of the software packages pyscal \cite{menonPyscalPythonModule2019} and Voro++ \cite{rycroftVoroThreedimensionalVoronoi2009}. After initial site generation, we merged all sites that were within 0.35\ \AA\ center to center distance of each other. The results of this generation method are listed in Table \ref{tab:GB_SegEnergy_Results}. A side view of the selected starting positions of the interstitial sites are shown in Fig. \ref{fig:starting_interstitial_sites}. The starting structures are attached in the S.I. The structures with interstitially placed solutes are then relaxed until the force convergence criterion is met.

\subsection{Segregation energies}
The segregation energy of an atom in a GB site refers to the difference in total energy that may be attained by placing it at the GB with respect to its energy in the bulk. The segregation energy of solute atom X (X = H, He, B, C, N, O, P and S), denoted $\text{E}_\text{seg}(\text{X})$ herein, can be calculated by taking the difference in energy in the segregated GB with a pure GB with that of the solute in the bulk, with:
\begin{equation}
	\text{E}_\text{seg}(\text{X}) = \text{E}_\text{GB}\big[\text{Fe},\text{X}\big] - \text{E}_\text{GB} - \Big[ \text{E}_\text{Bulk}\big[\text{Fe},\text{X}\big] - \text{E}_\text{Bulk} \Big] - n\times\mu_{\text{Fe}}\quad ,
\end{equation}
where $E_{\rm GB}[{\rm Fe},X] - E_{\rm GB}$ is the energy change when a single solute atom $X$ is introduced into the grain‐boundary (GB) supercell, relative to the pure‐Fe GB supercell. In the case where the site preference is different at a GB compared to in the bulk, e.g. substitutional in the GB but interstitial at the bulk, or vice-versa, the difference in the number of Fe atoms needs to be adjusted such that the reference system does not contain more or less Fe atoms. So in this case, the integer \textit{n} accounts for any net change in the number of Fe atoms between the GB and bulk cells, so that the total Fe count remains balanced, evaluated as:
\begin{equation}
	n \;=\;\bigl[n_{\rm GB}({\rm Fe},X)-n_{\rm GB}\bigr]
	\;-\;\bigl[n_{\rm Bulk}({\rm Fe},X)-n_{\rm Bulk}\bigr].
\end{equation}
\\\\
For the reference energy of the solute in the bulk, there are two cases, one in which the solute energetically prefers an on-lattice substitutional site, and the other case in which it prefers an interstitial position in the bulk. $E_{\rm Bulk}[{\rm Fe},X] - E_{\rm Bulk}$ is the analogous defect‐formation energy in a bulk supercell, with $X$ placed at its lowest‐energy site. We explicitly calculated both interstitial (octahedral/tetrahedral) and substitutional configurations in the bulk for all the elements in this study, and found that the most energetically favourable bulk position is interstitial for H, C, N, O and substitutional for all the other solutes. $\mu_{\rm Fe}$ is the chemical potential of a single Fe atom in the bcc lattice.
\\\\
In this study, negative segregation energies indicate that a solute gains enthalpy moving from its preferred site in the dilute bulk bcc-Fe phase to the GB, whereas positive values indicate a loss of enthalpy.

\subsection{Duplicate removal}
We first treat our data to remove sites which are highly similar or effectively duplicates after relaxation. To achieve this, we first featurise the atomic environment of the solutes present in the structures using the Smooth Overlap of Atomic Positions (SOAP) formalism, using the dscribe library \cite{himanenDScribeLibraryDescriptors2020} The SOAP vectors were calculated with 5\ \AA\ radial cutoff, 10 radial basis functions and the maximum degree of spherical harmonics set to 10. No averaging was performed as we are interested in the local site descriptors. We then perform a principal component analysis (PCA), retaining enough components to account for 99\% of the variation in the data, and then scale each component to unit variance (i.e. whiten) the generated PCA vectors (i.e.\ scale each principal component to unit variance). These vectors are then compared using a cosine similarity metric. Sites are marked as possible duplicates when their cosine similarity exceeded 95\%. These duplicate candidates were then compared through their segregation energies, and if the energies were within 0.05 eV of each other, we subsequently discard the site. This methodology was mostly adapted from Wagih and Schuh's prior study \cite{wagihLearningGrainBoundary2020}. After duplicate removal, we additionally exclude structures from the cohesion analysis when the calculated segregation energy is greater than \(-0.1~\mathrm{eV}\). Such values correspond to only shallow trapping and therefore are unlikely to produce substantial grain-boundary occupancy at common operating temperatures in ferritic alloys. From the McLean isotherm \cite{mcleanGrainBoundariesMetals1958},
\[
\frac{X_{GB}}{1-X_{GB}}
=
\frac{X_b}{1-X_b}
\exp\!\left(-\frac{\Delta E_{\mathrm{seg}}}{k_B T}\right),
\]
a segregation energy of \(-0.1~\mathrm{eV}\) gives an enrichment factor of only \(\sim 48\) at \(300~\mathrm{K}\), \(\sim 10\) at \(500~\mathrm{K}\), \(\sim 5.3\) at \(700~\mathrm{K}\), and \(\sim 3.2\) at \(1000~\mathrm{K}\). For a representative dilute bulk concentration of \(X_b = 10^{-3}\) (\(0.1~\mathrm{at.}\%\)), this corresponds to site occupation probabilities at the GB of only \(\sim 4.6\%\), \(1.0\%\), \(0.52\%\), and \(0.32\%\), respectively. This concentration is higher than one can expect for most of these solutes in real engineering steels, and thus indicates an upper bound on the expected occupation at the interface. We therefore treat such sites as too weakly trapping to be relevant candidates for the present cohesion analysis.
\subsection{Cohesion}
\subsubsection{Bond order evaluation}
We calculated the DDEC6 bond orders in each GB. We have previously introduced the area-normalised summed bond orders as a method of quantifying the effects of solutes on the strength of interfacial cohesion \cite{maiSegregationTransitionMetals2022}. The ANSBO sums all bond orders crossing a candidate cleavage plane, normalised by the cross-sectional area of the cell, and therefore captures both the strength of individual bonds and their density across the plane. The fact that it is an integral over the bond-breaking events that occur during the cleavage process makes it a natural analogue of the Rice-Wang work of separation quantity.
\\\\
The bonding-based area-normalised summed bond orders (ANSBO) quantity is defined as:
\begin{align}
	\sum_{\substack{\text{frac}\\\text{path}}}\text{BO} = \sum_{i,j}^{i \neq j} \text{BO}[\text{X}_i,\text{X}&_j] + \frac{1}{2}\sum_{k,l}^{k \neq l} \text{BO}[\text{X}_k,\text{X}_l] \nonumber \\
	\text{where}\ \text{X}_{1}(z) <&\ z_\text{CP} < \text{X}_{2}(z) \nonumber
\end{align}
\begin{equation}
	\text{ANSBO} = \sum_{\substack{\text{frac}\\\text{path}}}\text{BO} / \text{A}\quad.
\end{equation}
Here $\sum_{\substack{\text{frac}\\\text{path}}}\text{BO}$ is the chargemol \cite{manzIntroducingDDEC6Atomic2017} calculated summed DDEC6 bond orders of the electronic bonds participating in the cohesion of an arbitrary fracture path parallel to the GB plane, $\sum_{i,j}^{i \neq j}\text{BO}[\text{X}_i,\text{X}_j]$ is the bond order of the bond that exists between X$_i$, X$_j$, where X$_i$, X$_j$, X$_k$, X$_l$ are the atoms that are electronically bonded in sites $i$, $j$, $k$, $l$ respectively. The exact definition and derivation of the bond order is provided in the cited paper \cite{manzIntroducingDDEC6Atomic2017}. X$_1$, X$_2$ represent the $i,j$ and the $k,l$ pairings in any order. X$_i$ and X$_j$ atoms reside entirely within the supercell created (i.e., bonds exist wholly within the cell), whereas X$_k$ and X$_l$ represent atom pairs where only one of X$_k$ and X$_l$ resides in the cell (i.e., possess bonds passing outside of the original cell into a neighbouring image). $z_\text{CP}$ is the \textit{z} coordinate of the cleavage plane. Larger values of ANSBO indicate greater strength of interfacial cohesion. 
\\\\
The ANSBO and cohesion quantities were evaluated for all cleavage planes parallel to the GB plane, and within 3\ \AA\ of the segregated solutes. No cleavage quantities were evaluated between atoms on the same layer. Atoms were considered to be on the same layer where the distance in "z" direction is less than 0.1\ \AA.
\subsubsection{Rice-Wang Rigid-Grain-Separation framework}
We computed the Rice-Wang work of separation in the Rigid-Grain-Separation (RGS) scheme across all structures retained after the aforementioned duplicate filtering process. The $\text{W}_{\text{sep}}^{\text{RGS}}$ were evaluated for all cleavage planes parallel to the GB plane, and within 3\ \AA\ of the segregated solutes, in the same manner as that considered for the bond order cohesion evaluation. In this manner, we search for the weakest cleavage plane in proximity of the solute. The work of separation in the rigid-grain separation framework ($\text{W}_{\rm{sep}}^{\rm{RGS}}$) was calculated by:
\begin{equation}
	\text{W}_{\rm{sep}}^{\rm{RGS}} = (\text{E}_\text{GB-sep} - \text{E}_\text{{GB}}) / \text{A}\quad.
\end{equation}
Here, $\text{E}_\text{GB-sep}$ is the total energy of the cell containing a cleaved GB structure (with/without segregants) and $\text{E}_\text{{GB}}$ is the total energy of the corresponding non-cleaved structure. We emphasise that the atomic positions of the cleaved cell were \textit{not} relaxed. We did not compute works of separation with relaxed surfaces, as is common in studies utilising Rice-Wang theory. The atoms were cleaved with 6 \AA of vacuum separating the slabs.
\section{Results and Discussion}
\subsection{Segregation}
\begin{figure}[h!]
	\centering
	\includegraphics[width=\linewidth]{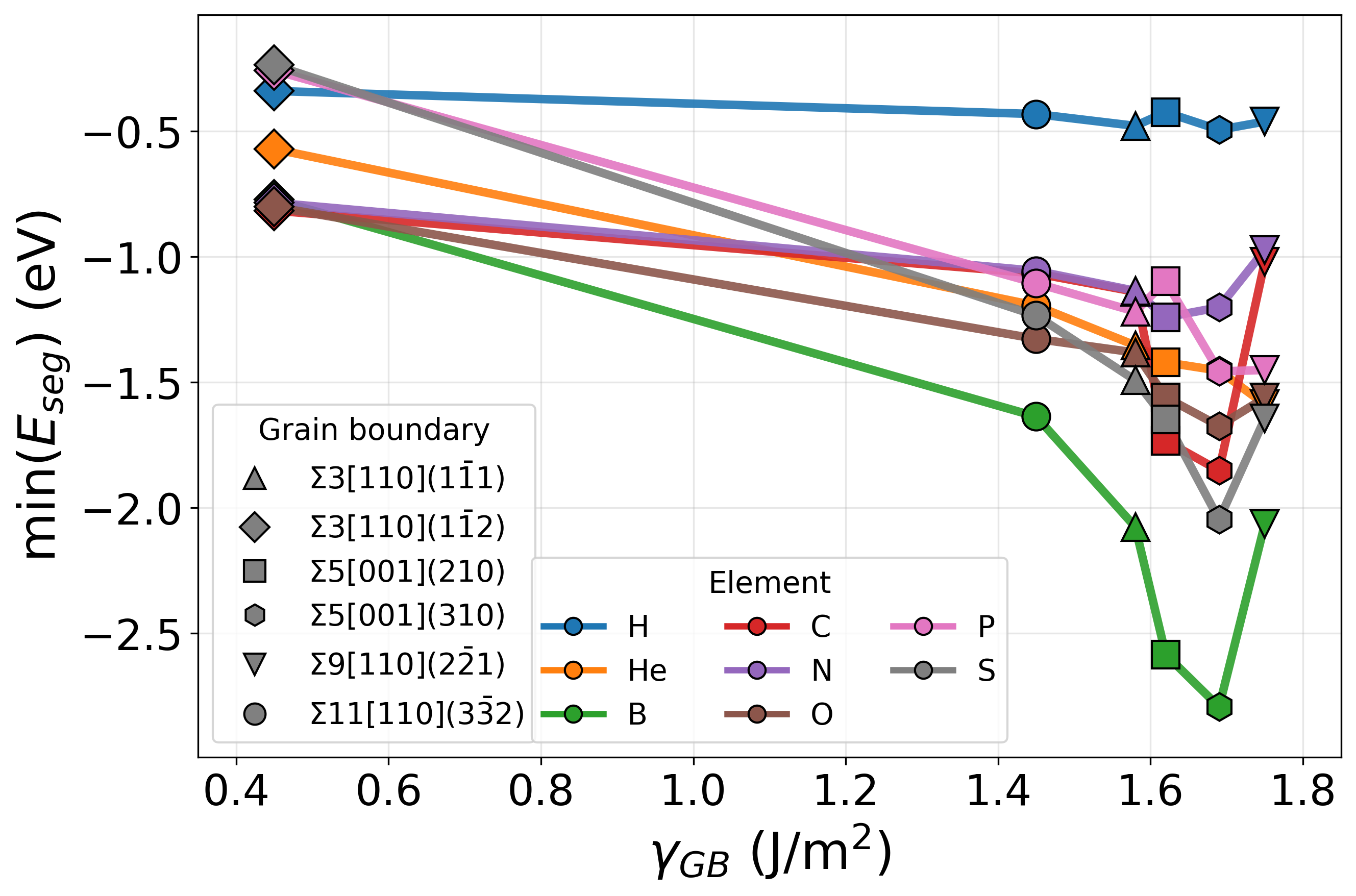}
	\caption{The segregation energies at the strongest trapping sites are plotted against the GB energies of the pure Fe interfaces.}
	\label{fig:min_eseg_vs_gb_energy}
\end{figure}

The minimum segregation energies calculated at each GB of each of the elements is tabulated in Table \ref{tab:GB_SegEnergy_Results}. The minimum segregation energy for a given element at a given GB is the most negative (strongest binding) value across all sampled sites at that boundary. Boron experiences the strongest segregation out of the solutes, followed by O, S, C, He, P and N which have relatively similar segregation energies across the GBs, with H exhibiting the weakest segregation tendencies of the solutes studied. It is notable that H has a similar minimum segregation energy at the twin-like $\Sigma3[110](1\bar{1}2)$ as the other GBs. This is unlike the other elements, which all experience weaker segregation trapping at the twin in comparison with the other GBs. Phosphorus and S experience only very weak segregation at the twin.
\\\\
In Fig. \ref{fig:min_eseg_vs_gb_energy}, we plot the segregation energies of each element against the GB energy of the pure Fe GBs. It is often asserted that differences in grain boundary energies can be used as a heuristic for predicting segregation energies, with claims that "special" lower energy GBs exhibit lower strength of segregation trapping \cite{bechtleGrainboundaryEngineeringMarkedly2009}. Here, it is shown that grain boundary energy is, at best, a very weak predictor of the strength of trapping. We emphasise that we do not doubt the characterisation studies performed in such experiments, merely cautioning against the \textit{interpretation} that the low energy nature of the GBs is cause of such low segregation tendencies, rather than the fact that it is the principally twin nature that is the reason. These arguments typically suggest that higher energy GBs serve as stronger segregation traps for solutes and impurities. However, in Fig. \ref{fig:min_eseg_vs_gb_energy} we show that such assertions are not well-supported by our data across 6 CSL GBs. For H, there is no relationship between the GB energy and the maximum strength of segregation binding.

\begin{table}[h!]
	\centering
	\begin{tabular}{lccccccccccc}
		\hline
		&  &  & \multicolumn{8}{c}{min(E$_\text{seg}$) (eV) / site type}\\
		GB & \begin{tabular}[c]{@{}c@{}}\# int.\\ sites\end{tabular} & \begin{tabular}[c]{@{}c@{}}\# sub.\\ sites\end{tabular} & H & He & B & C & N & O & P & S \\
		\hline
		\textbf{$\Sigma3[110](1\bar{1}1)$} & 78 & 9 & -0.48 & -1.35 & -2.08 & -1.14 & -1.14 & -1.38 & -1.22 & -1.49 \\
		& & & \textcolor{siteint}{\textit{int}} & \textcolor{siteint}{\textit{int}} & \textcolor{siteint}{\textit{int}} & \textcolor{siteint}{\textit{int}} & \textcolor{siteint}{\textit{int}} & \textcolor{siteint}{\textit{int}} & \textcolor{sitesub}{\textit{sub}} & \textcolor{sitesub}{\textit{sub}} \\[2pt]
		\textbf{$\Sigma3[110](1\bar{1}2)$} & 58 & 7 & -0.34 & -0.57 & -0.77 & -0.82 & -0.78 & -0.80 & -0.26 & -0.24 \\
		& & & \textcolor{siteint}{\textit{int}} & \textcolor{sitesub}{\textit{sub}} & \textcolor{siteint}{\textit{int}} & \textcolor{siteint}{\textit{int}} & \textcolor{siteint}{\textit{int}} & \textcolor{siteint}{\textit{int}} & \textcolor{sitesub}{\textit{sub}} & \textcolor{sitesub}{\textit{sub}} \\[2pt]
		\textbf{$\Sigma5[001](210)$} & 106 & 7 & -0.42 & -1.42 & -2.58 & -1.73 & -1.24 & -1.56 & -1.10 & -1.65 \\
		& & & \textcolor{siteint}{\textit{int}} & \textcolor{sitesub}{\textit{sub}} & \textcolor{siteint}{\textit{int}} & \textcolor{siteint}{\textit{int}} & \textcolor{siteint}{\textit{int}} & \textcolor{siteint}{\textit{int}} & \textcolor{siteint}{\textit{int}} & \textcolor{siteint}{\textit{int}} \\[2pt]
		\textbf{$\Sigma5[001](310)$} & 144 & 5 & -0.49 & -1.45 & -2.79 & -1.85 & -1.20 & -1.67 & -1.46 & -2.05 \\
		& & & \textcolor{siteint}{\textit{int}} & \textcolor{sitesub}{\textit{sub}} & \textcolor{siteint}{\textit{int}} & \textcolor{siteint}{\textit{int}} & \textcolor{siteint}{\textit{int}} & \textcolor{siteint}{\textit{int}} & \textcolor{siteint}{\textit{int}} & \textcolor{siteint}{\textit{int}} \\[2pt]
		\textbf{$\Sigma9[110](2\bar{2}1)$} & 103 & 14 & -0.46 & -1.59 & -2.06 & -1.02 & -0.97 & -1.56 & -1.45 & -1.64 \\
		& & & \textcolor{siteint}{\textit{int}} & \textcolor{sitemix}{\textit{mix}} & \textcolor{sitesub}{\textit{sub}} & \textcolor{siteint}{\textit{int}} & \textcolor{sitemix}{\textit{mix}} & \textcolor{sitemix}{\textit{mix}} & \textcolor{sitemix}{\textit{mix}} & \textcolor{sitesub}{\textit{sub}} \\[2pt]
		\textbf{$\Sigma11[110](3\bar{3}2)$} & 60 & 12 & -0.43 & -1.19 & -1.64 & -1.07 & -1.06 & -1.33 & -1.11 & -1.23 \\
		& & & \textcolor{siteint}{\textit{int}} & \textcolor{siteint}{\textit{int}} & \textcolor{siteint}{\textit{int}} & \textcolor{siteint}{\textit{int}} & \textcolor{siteint}{\textit{int}} & \textcolor{siteint}{\textit{int}} & \textcolor{sitemix}{\textit{mix}} & \textcolor{siteint}{\textit{int}} \\
		\hline
	\end{tabular}
	\caption{Number of sampled starting interstitial and substitutional sites (before duplicate removal), and minimum $E_\mathrm{seg}$ (eV) for each element at each grain boundary. The site type of the minimum-energy site is indicated: \textit{int} = interstitial starting position, \textit{sub} = substitutional, \textit{mix} = duplicate sites found from both starting position types.}
	\label{tab:GB_SegEnergy_Results}
\end{table}

\begin{figure}[h!]
	\centering
	\begin{subfigure}{0.48\linewidth}
		\centering
		\includegraphics[width=\linewidth]{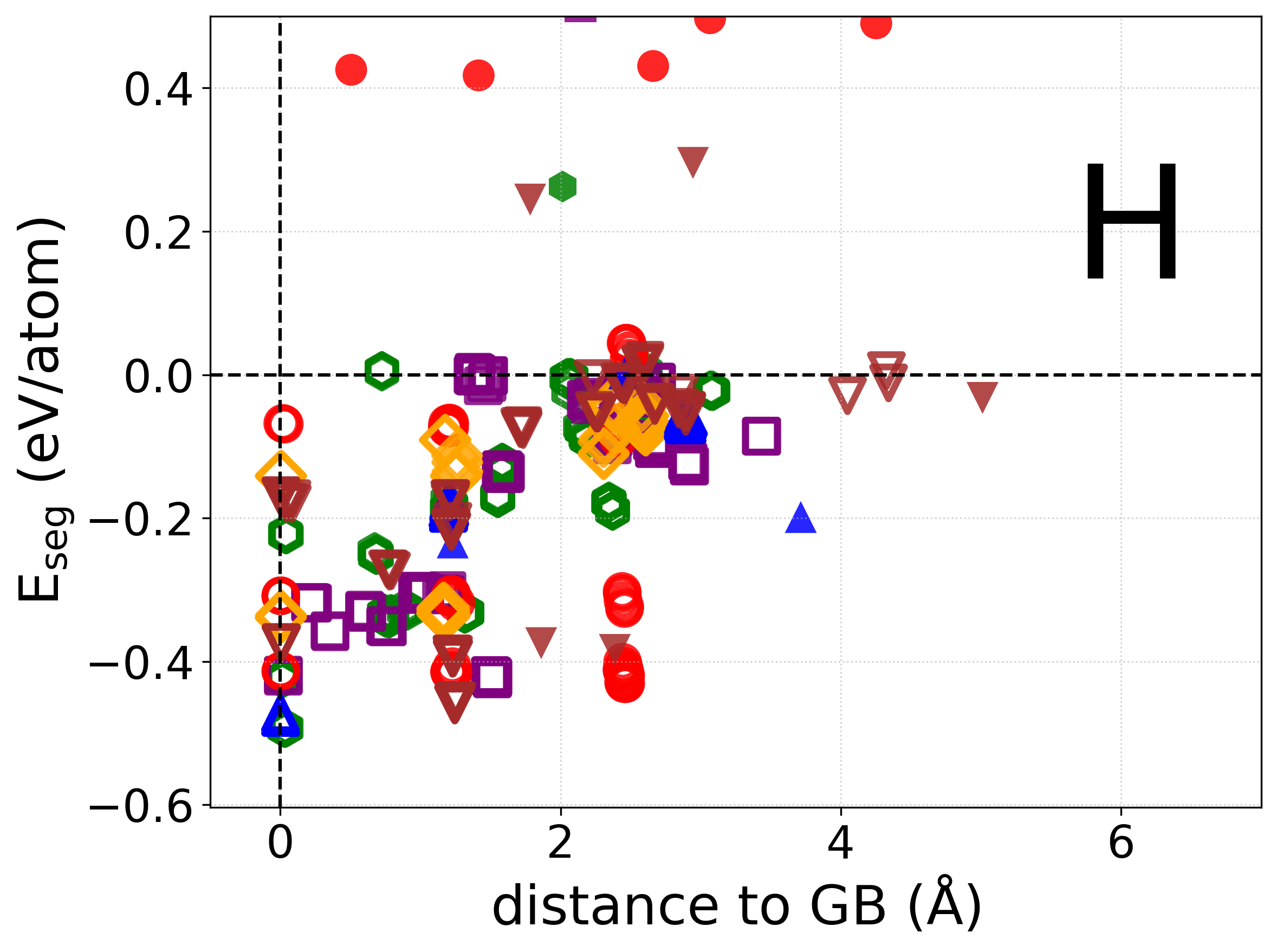}
		\caption{}
		\label{fig:Eseg_dist_GB_H}
	\end{subfigure}\hfill
	\begin{subfigure}{0.48\linewidth}
		\centering
		\includegraphics[width=\linewidth]{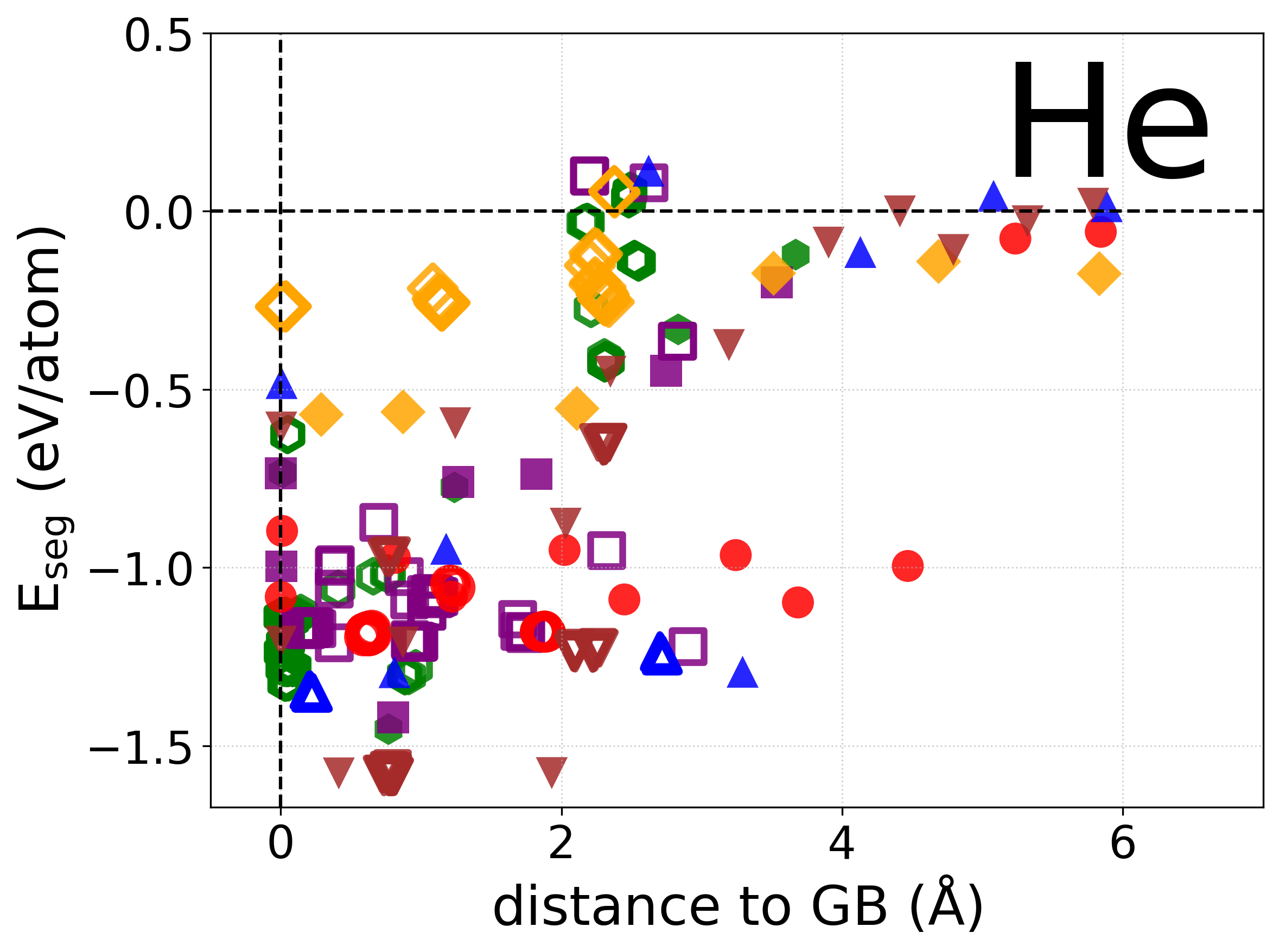}
		\caption{}
		\label{fig:Eseg_dist_GB_He}
	\end{subfigure}
	
	\begin{subfigure}{0.48\linewidth}
		\centering
		\includegraphics[width=\linewidth]{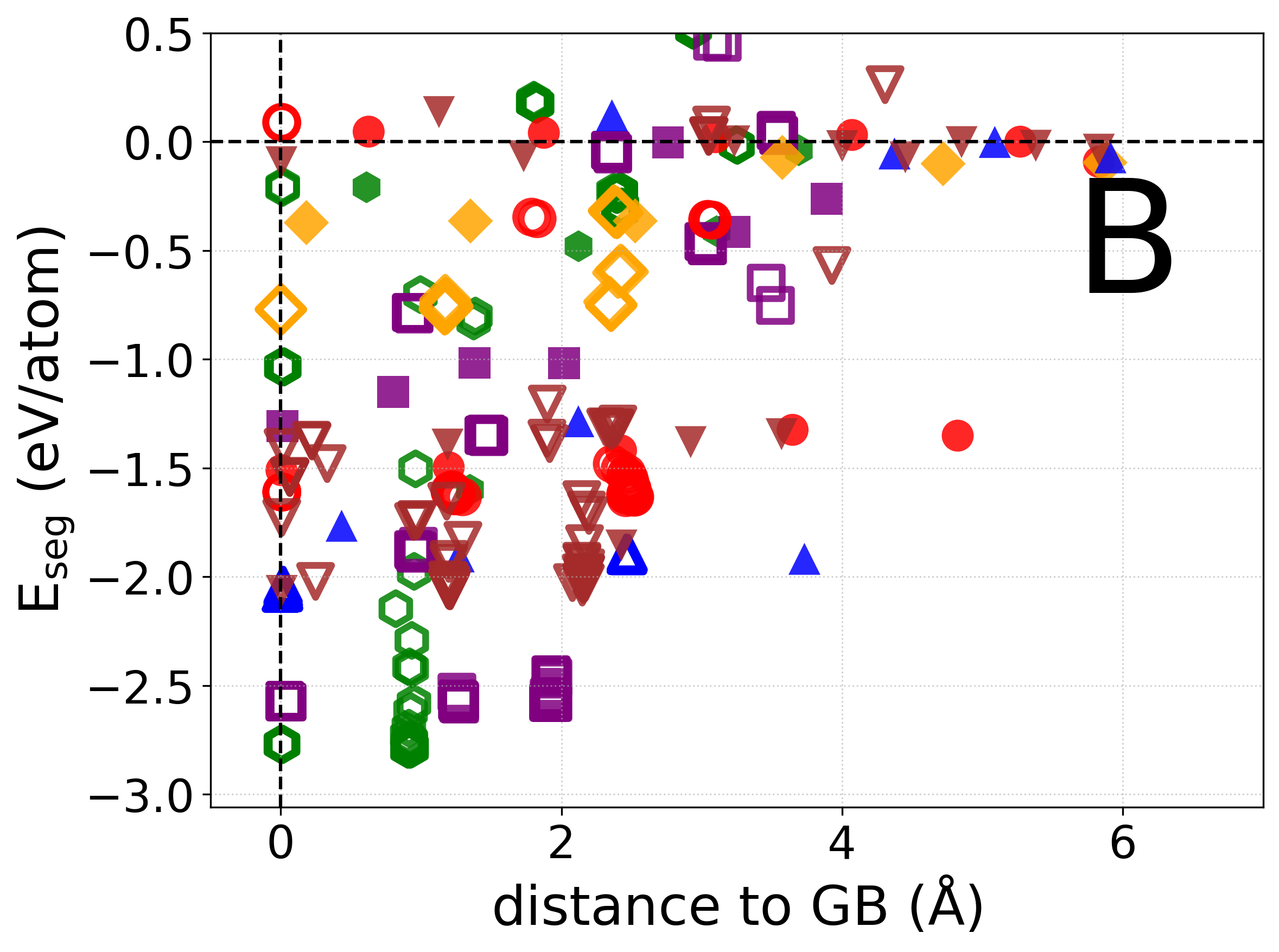}
		\caption{}
		\label{fig:Eseg_dist_GB_B}
	\end{subfigure}\hfill
	\begin{subfigure}{0.48\linewidth}
		\centering
		\includegraphics[width=\linewidth]{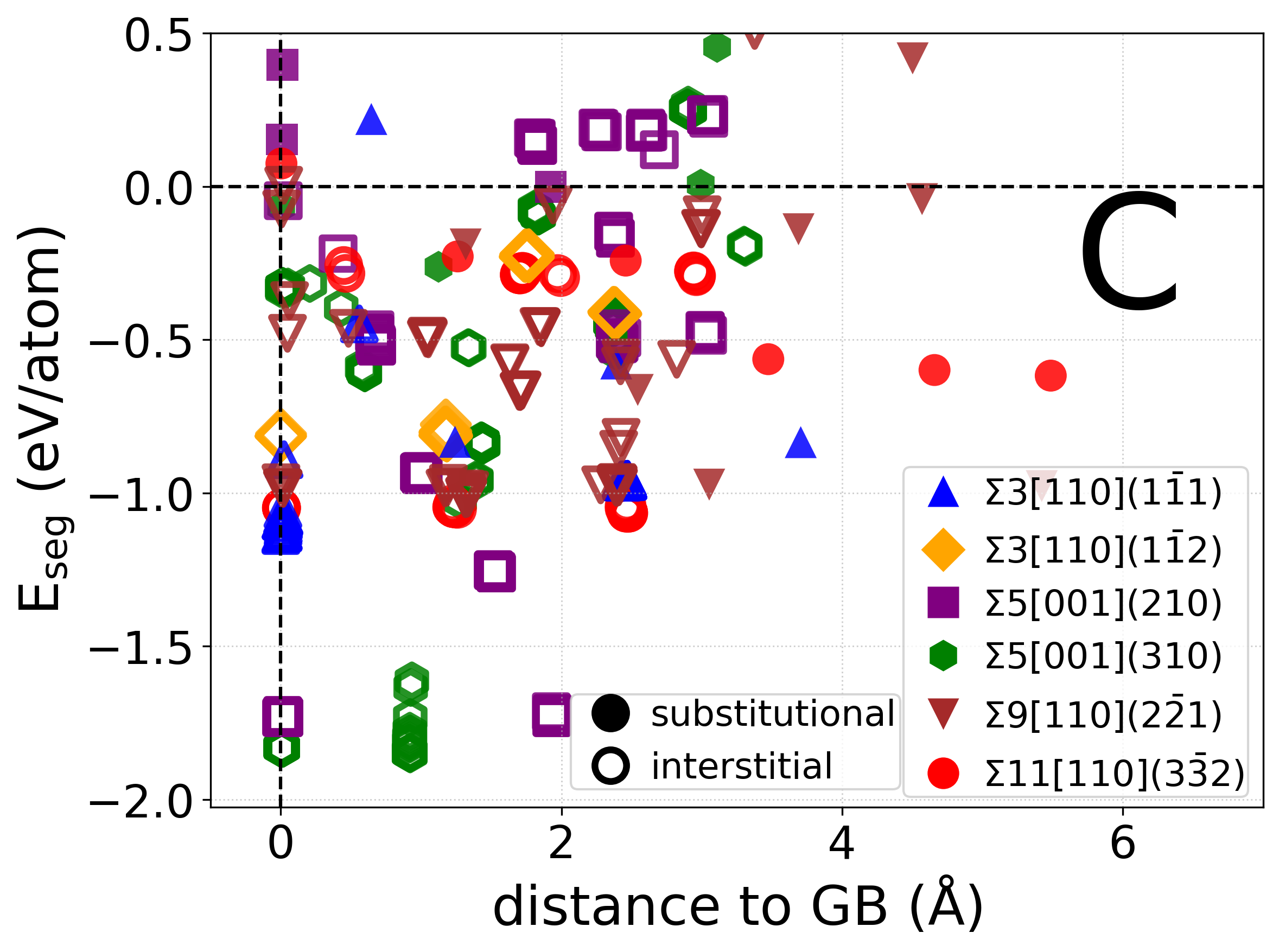}
		\caption{}
		\label{fig:Eseg_dist_GB_C}
	\end{subfigure}
	
	\begin{subfigure}{0.48\linewidth}
		\centering
		\includegraphics[width=\linewidth]{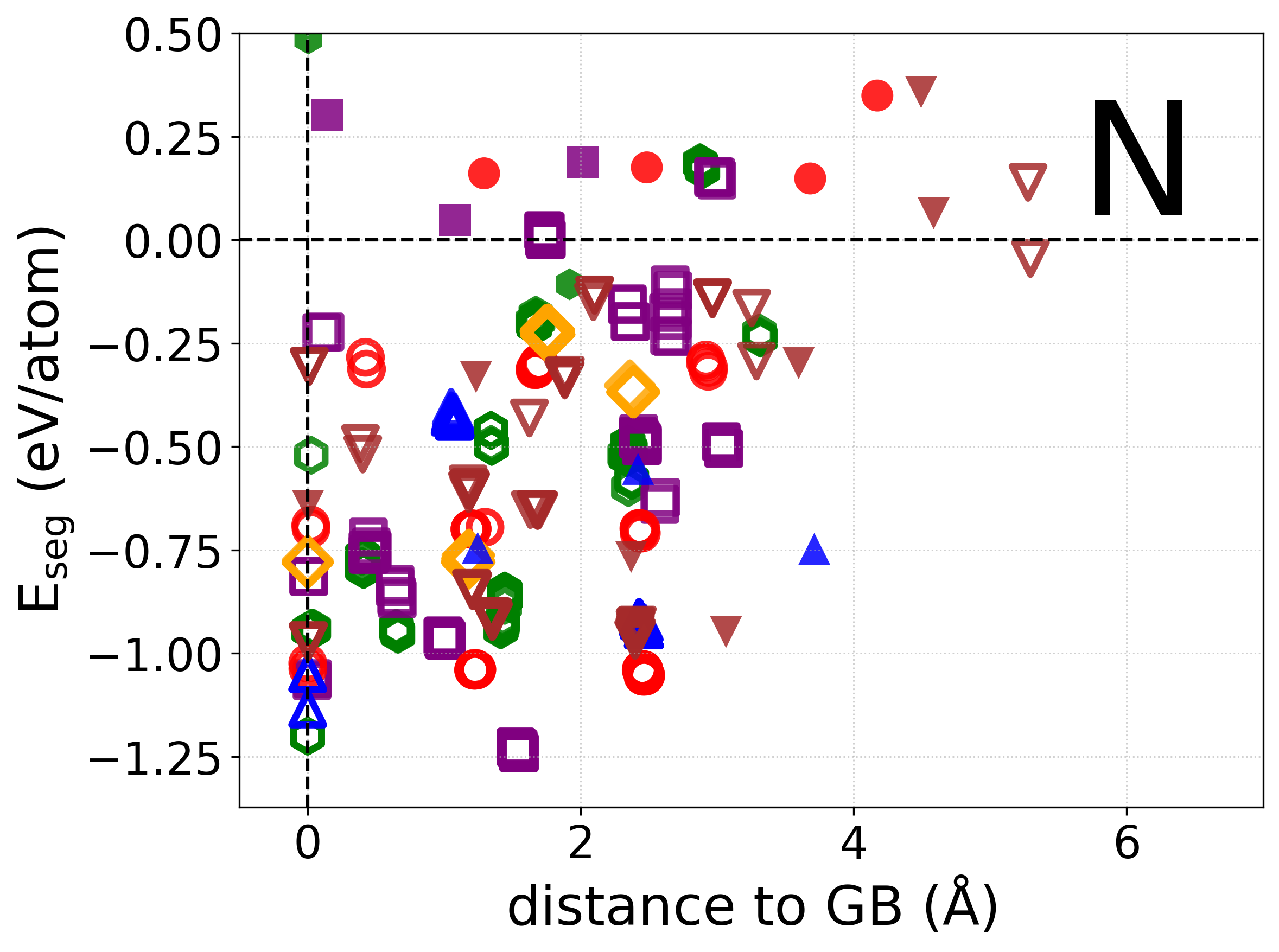}
		\caption{}
		\label{fig:Eseg_dist_GB_N}
	\end{subfigure}\hfill
	\begin{subfigure}{0.48\linewidth}
		\centering
		\includegraphics[width=\linewidth]{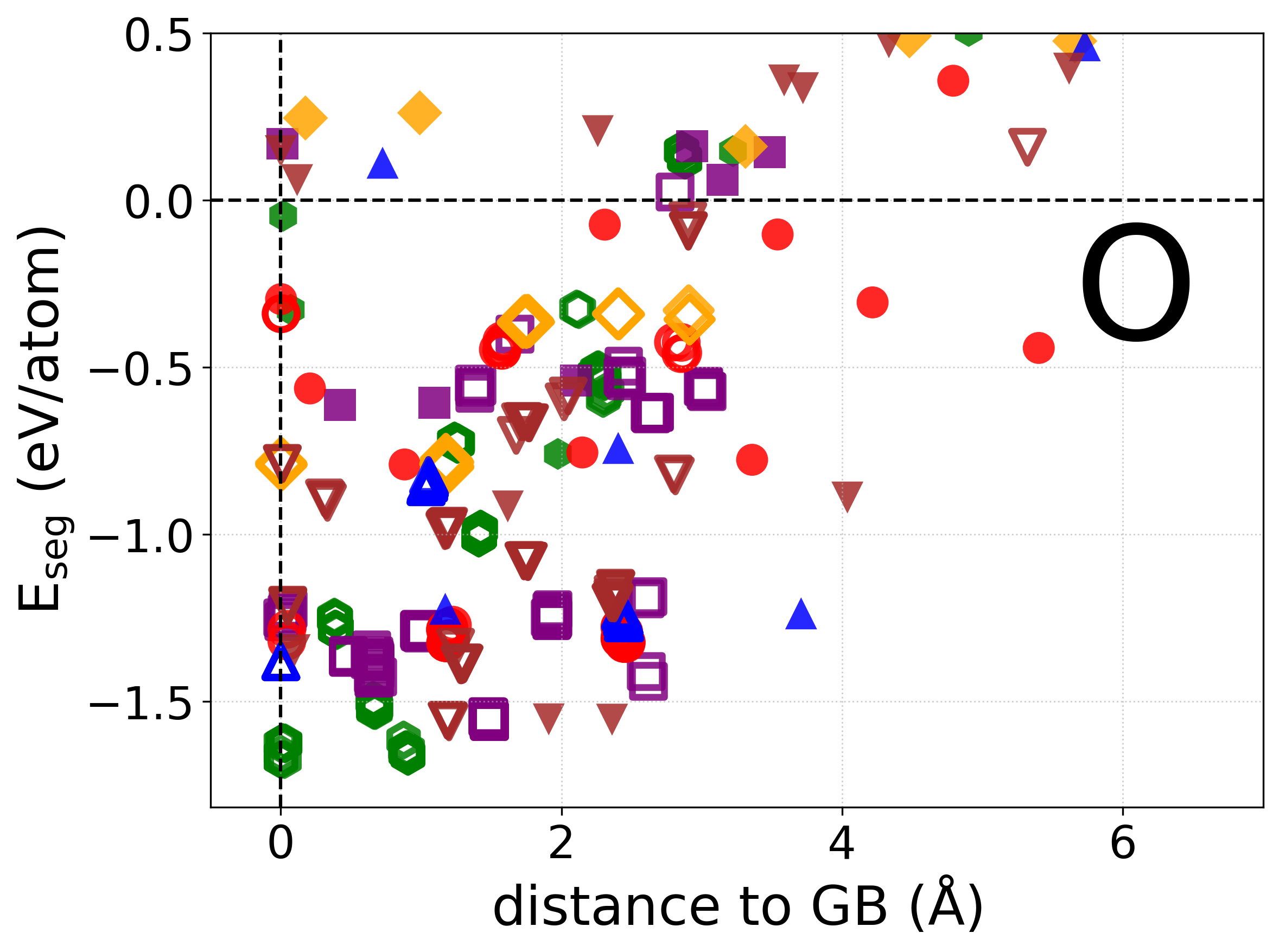}
		\caption{}
		\label{fig:Eseg_dist_GB_O}
	\end{subfigure}
	
	\begin{subfigure}{0.48\linewidth}
		\centering
		\includegraphics[width=\linewidth]{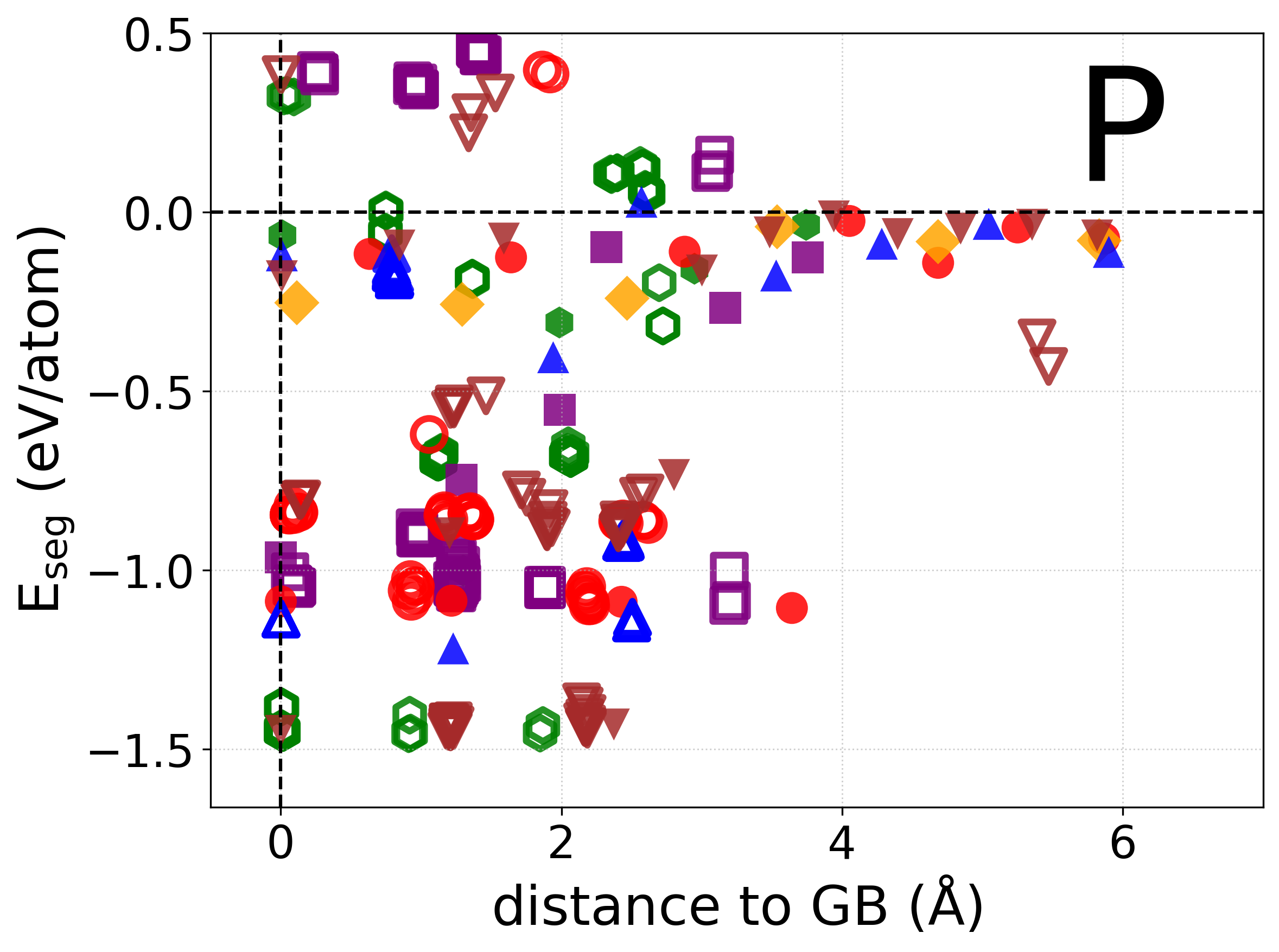}
		\caption{}
		\label{fig:Eseg_dist_GB_P}
	\end{subfigure}\hfill
	\begin{subfigure}{0.48\linewidth}
		\centering
		\includegraphics[width=\linewidth]{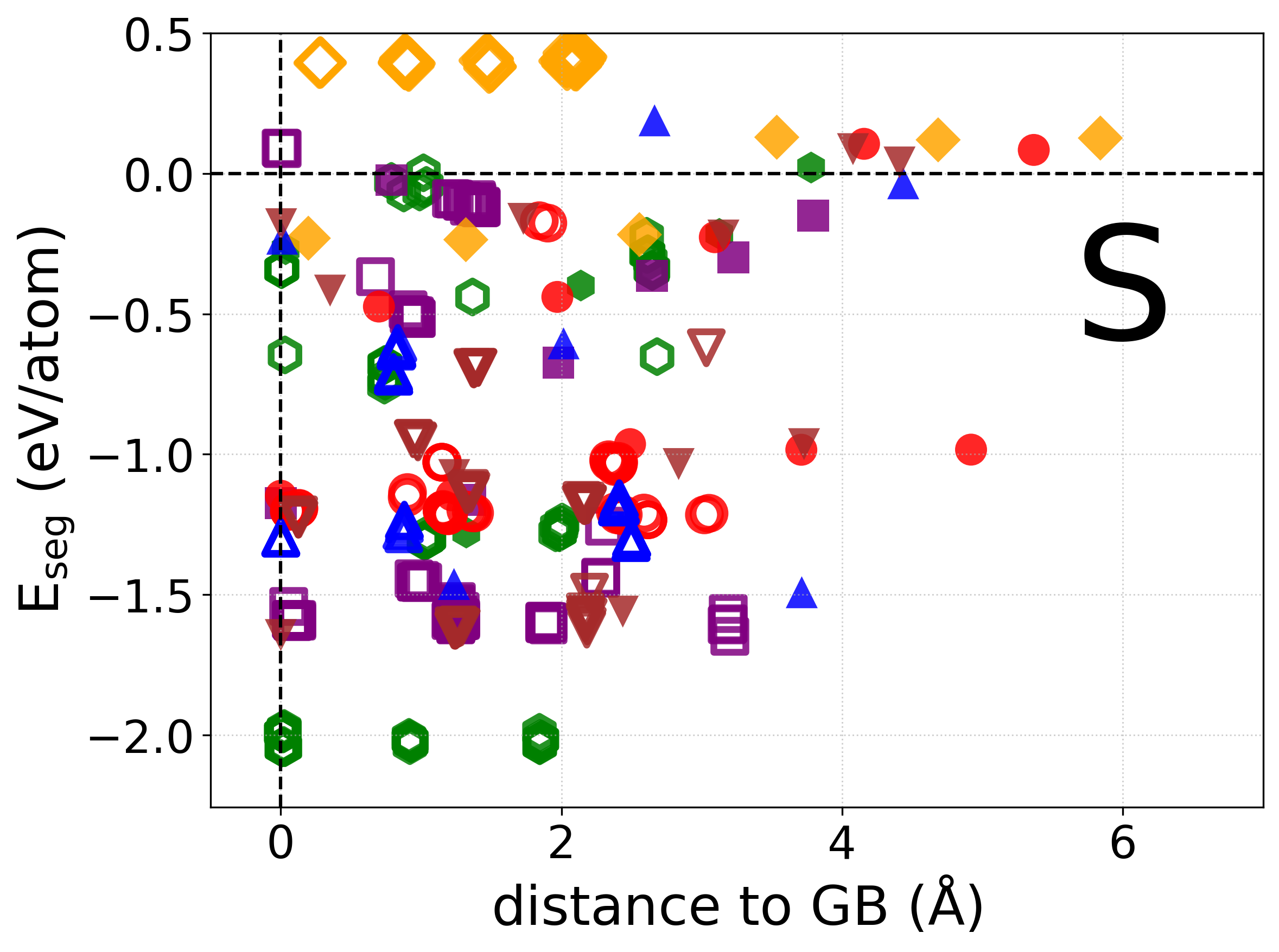}
		\caption{}
		\label{fig:Eseg_dist_GB_S}
	\end{subfigure}
	
	\caption{Comparison of the distance from the grain boundary versus segregation energy for solutes in the same grain boundary: (\ref{fig:Eseg_dist_GB_H}) H; (\ref{fig:Eseg_dist_GB_He}) He; (\ref{fig:Eseg_dist_GB_B}) B; (\ref{fig:Eseg_dist_GB_C}) C; (\ref{fig:Eseg_dist_GB_N}) N; (\ref{fig:Eseg_dist_GB_O}) O; (\ref{fig:Eseg_dist_GB_P}) P; (\ref{fig:Eseg_dist_GB_S}) S. Each panel shows the correlation between the distance of the solute’s most favorable site from the grain boundary plane and its minimum segregation energy, \(E_{\mathrm{seg}}\).}
	\label{fig:Eseg_dist_GB_all}
\end{figure}

In Table \ref{tab:eseg_literature_comparison}, we compare our minimum segregation energies with available DFT literature values at matching grain boundaries. Relatively good agreement is achieved for most solutes. Here we note the very large segregation trapping calculated by Mirzaev et al. \cite{mirzaevInitioModellingInteraction2016} at the $\Sigma5[001](210)$ for H of $-0.81$ eV compared to our value of $-0.42$ eV. We conclude that this value is likely erroneous as we do not observe such large segregation energies in any of the GBs we have studied, making this a distinct outlier and exclude it from the comparison. A prior work by Wachowicz et al. \cite{wachowiczEffectImpuritiesStructural2011} have studied similar solutes here, but as their study utilised a different, less favourable bulk reference state i.e. substitutional O, N, C in bcc Fe, their study reports significantly shifted segregation energies which overestimate the strength of trapping of these solutes at the GB \cite{wachowiczEffectImpuritiesStructural2011}. As such, we do not compare their values to ours in Table \ref{tab:eseg_literature_comparison}.
\begin{table}[h!]
	\centering
	\small
	\setlength{\tabcolsep}{4pt}
	\renewcommand{\arraystretch}{1.2}
	\begin{tabular}{ll r p{8.5cm}}
		\toprule
		Solute & GB & \begin{tabular}[c]{@{}r@{}}This work\\(eV)\end{tabular} & Literature $E_\mathrm{seg}$ (eV) \\
		\midrule
		\multirow{3}{*}{H} & $\Sigma3[110](1\bar{1}1)$ & $-0.48$ & $-0.39$ \cite{mirzaevInitioModellingInteraction2016}; $-0.43$$^\dagger$ \cite{duFirstprinciplesStudyInteraction2011a}; $-0.47$ \cite{heFirstprinciplesInvestigationEffect2013, kholtobinaEffectAlloyingElements2021};\; $-0.48$ \cite{yamaguchiMobileEffectHydrogen2012} \\
		& $\Sigma5[001](310)$ & $-0.49$ & $-0.43$ \cite{mirzaevInitioModellingInteraction2016} \\
		\midrule
		He & $\Sigma3[110](1\bar{1}1)$ & $-1.35$ & $-1.23$$^\dagger$ \cite{suzudoAtomisticModelingHe2013} \\
		& $\Sigma5[001](310)$ & $-1.45$ & $-1.43$ \cite{zhangFirstprinciplesStudyHe2010} \\
		\midrule
		B  & $\Sigma3[110](1\bar{1}1)$ & $-2.08$ & $-1.96$ \cite{kholtobinaEffectAlloyingElements2021}; $-2.10$$^\dagger$ \cite{wachowiczEffectImpuritiesStructural2011}; $-1.99$$^\dagger$ \cite{yamaguchiFirstprinciplesStudyGrain2011a} \\
		\midrule
		\multirow{4}{*}{C} & $\Sigma3[110](1\bar{1}1)$ & $-1.14$ & $-1.04$ \cite{kholtobinaEffectAlloyingElements2021}; $-1.15$$^\dagger$ \cite{yamaguchiFirstprinciplesStudyGrain2011a} \\
		& $\Sigma3[110](1\bar{1}2)$ & $-0.82$ & $-0.67$ \cite{wangFirstprinciplesStudyCarbon2016a} \\
		& $\Sigma5[001](210)$ & $-1.73$ & $-1.63$ \cite{wangFirstprinciplesStudyCarbon2016a} \\
		& $\Sigma5[001](310)$ & $-1.85$ & $-1.77$ \cite{wangFirstprinciplesStudyCarbon2016a} \\
		\midrule
		N  & $\Sigma3[110](1\bar{1}1)$ & $-1.14$ & $-0.95$ \cite{kholtobinaEffectAlloyingElements2021} \\
		\midrule
		\multirow{2}{*}{P} & $\Sigma3[110](1\bar{1}1)$ & $-1.22$ & $-1.15$$^\dagger$ \cite{yamaguchiFirstprinciplesStudyGrain2011};\; $-1.07$$^\dagger$ \cite{yamaguchiDecohesionIronGrain2007};\; $-1.2$$^\dagger$ {\cite{itoElectronicOriginGrain2020}} \\
		& $\Sigma5[001](310)$ & $-1.46$ & $-1.5$$^\dagger$ \cite{cernySegregationPhosphorusSilicon2024} \\
		\midrule
		\multirow{1}{*}{S} & $\Sigma3[110](1\bar{1}1)$ & $-1.49$ & $-1.46$$^\dagger$ \cite{yamaguchiDecohesionIronGrain2007}; $-1.01$$^\dagger$ \cite{yamaguchiFirstprinciplesStudyGrain2011} \\
		\bottomrule
	\end{tabular}
	\caption{Comparison of minimum segregation energies from this study with DFT literature values at matching grain boundaries. $^\dagger$Values estimated from published figures, small errors expected from author re-extraction from figure interpretation.}
	\label{tab:eseg_literature_comparison}
\end{table}
\\\\
We plot the segregation energies as a function of distance away from the GBs in Fig. \ref{fig:Eseg_dist_GB_all}. This distance is measured as the distance of the relaxed solute to the initial position of the GB, as it is not possible to cleanly distinguish the GB plane after relaxation in many cases. In general, the strongest sites for binding occur within 2.5\ \AA\ of the interface. Note that there are some large segregation energies that exist past the range of observed structural distortion ($\approx$ 3\ \AA\ of the interface). This can be attributed to giant relaxations that the solutes induce, warping the interface plane towards the solute atom, effectively inducing a shift of the GB plane. This phenomenon can be observed in even larger cells (see the Supplementary Information of \cite{maiPhosphorusTransitionMetal2023}), and as such we deem them physical, and not artefacts of the size of the cells that we use here.
\\\\

\begin{figure}[h!]
	\centering
	\begin{subfigure}{0.48\linewidth}
		\centering
		\includegraphics[width=\linewidth]{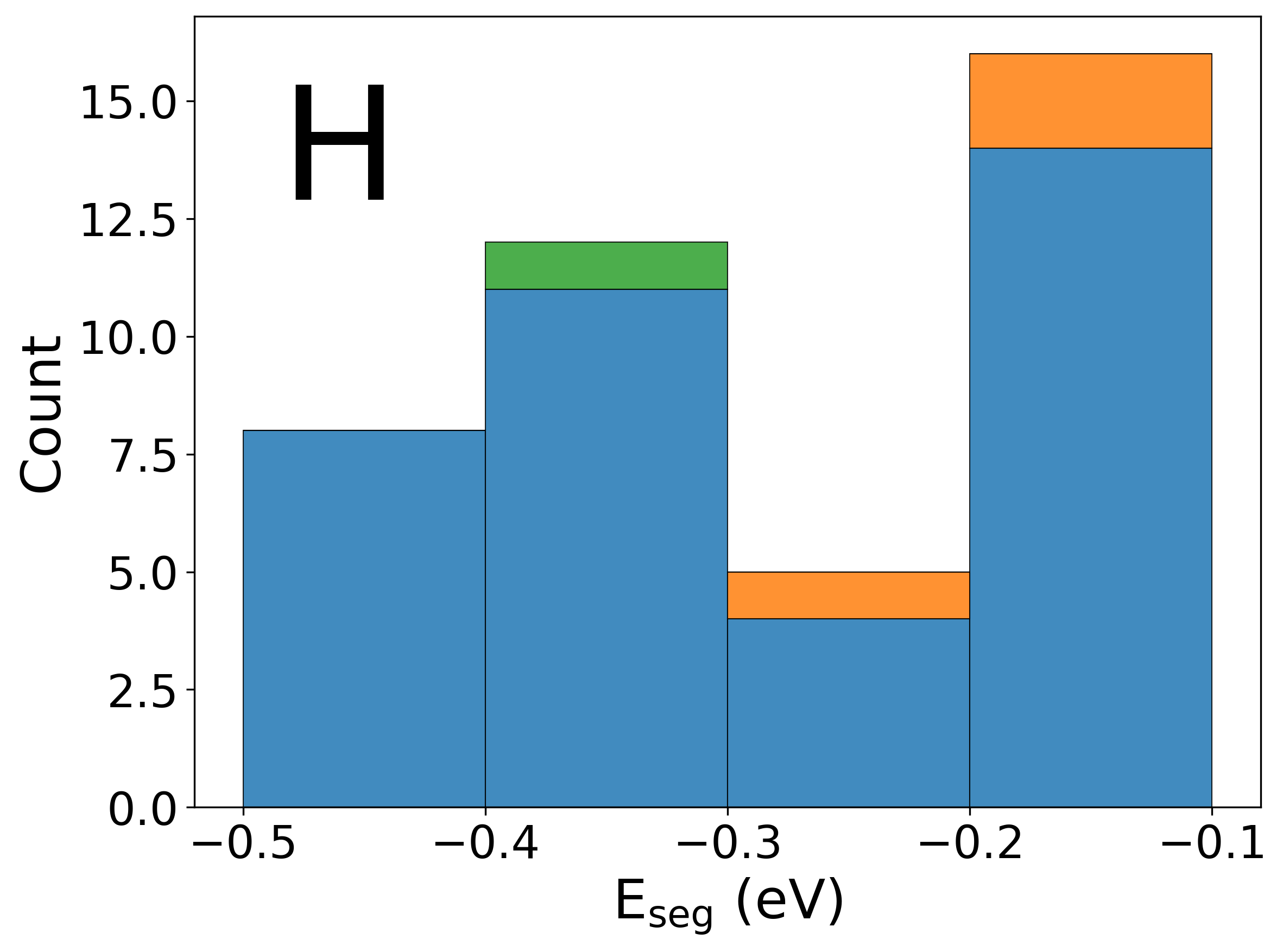}
		\caption{}
		\label{fig:Eseg_hist_H}
	\end{subfigure}\hfill
	\begin{subfigure}{0.48\linewidth}
		\centering
		\includegraphics[width=\linewidth]{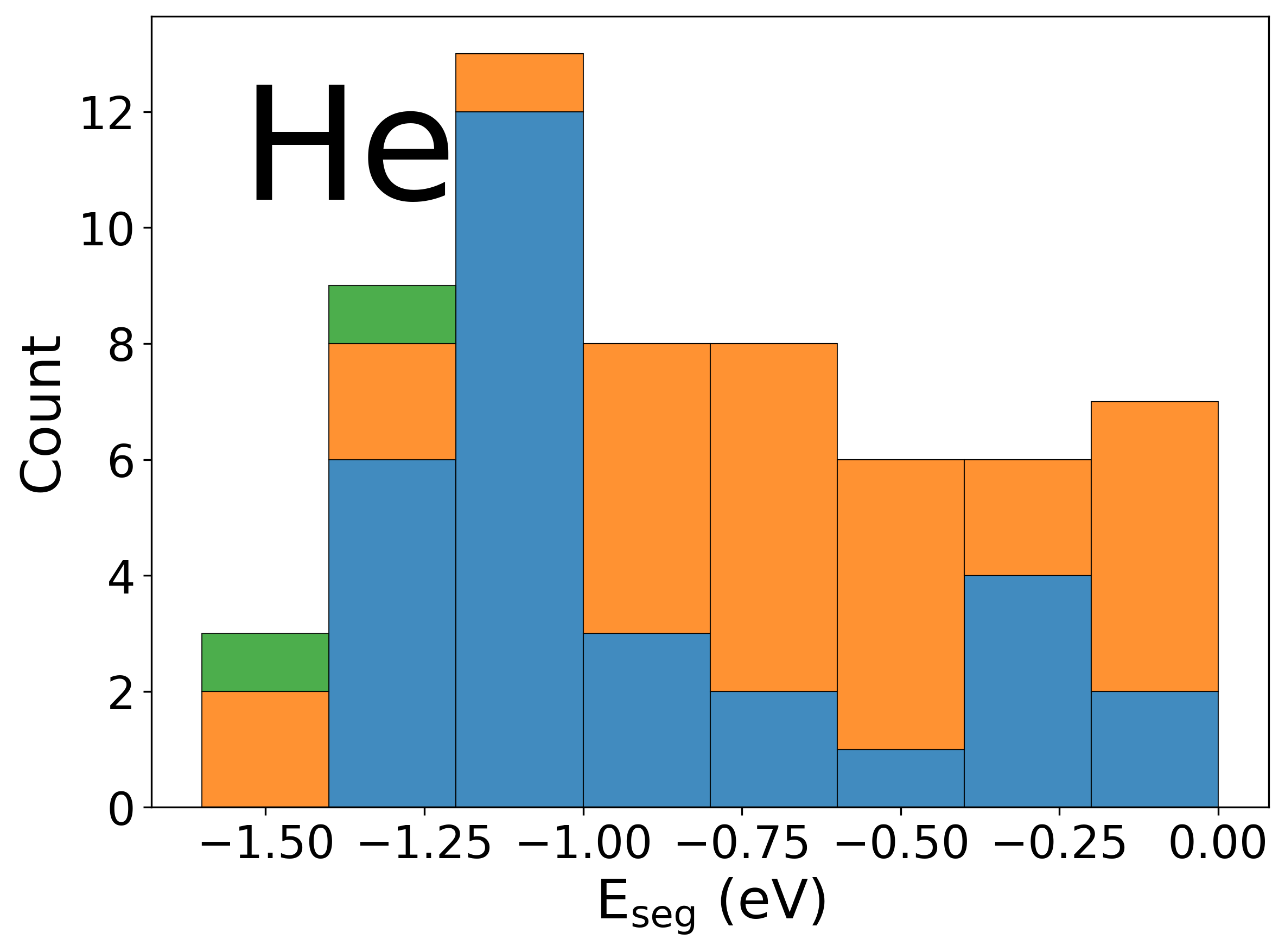}
		\caption{}
		\label{fig:Eseg_hist_He}
	\end{subfigure}
	\begin{subfigure}{0.48\linewidth}
		\centering
		\includegraphics[width=\linewidth]{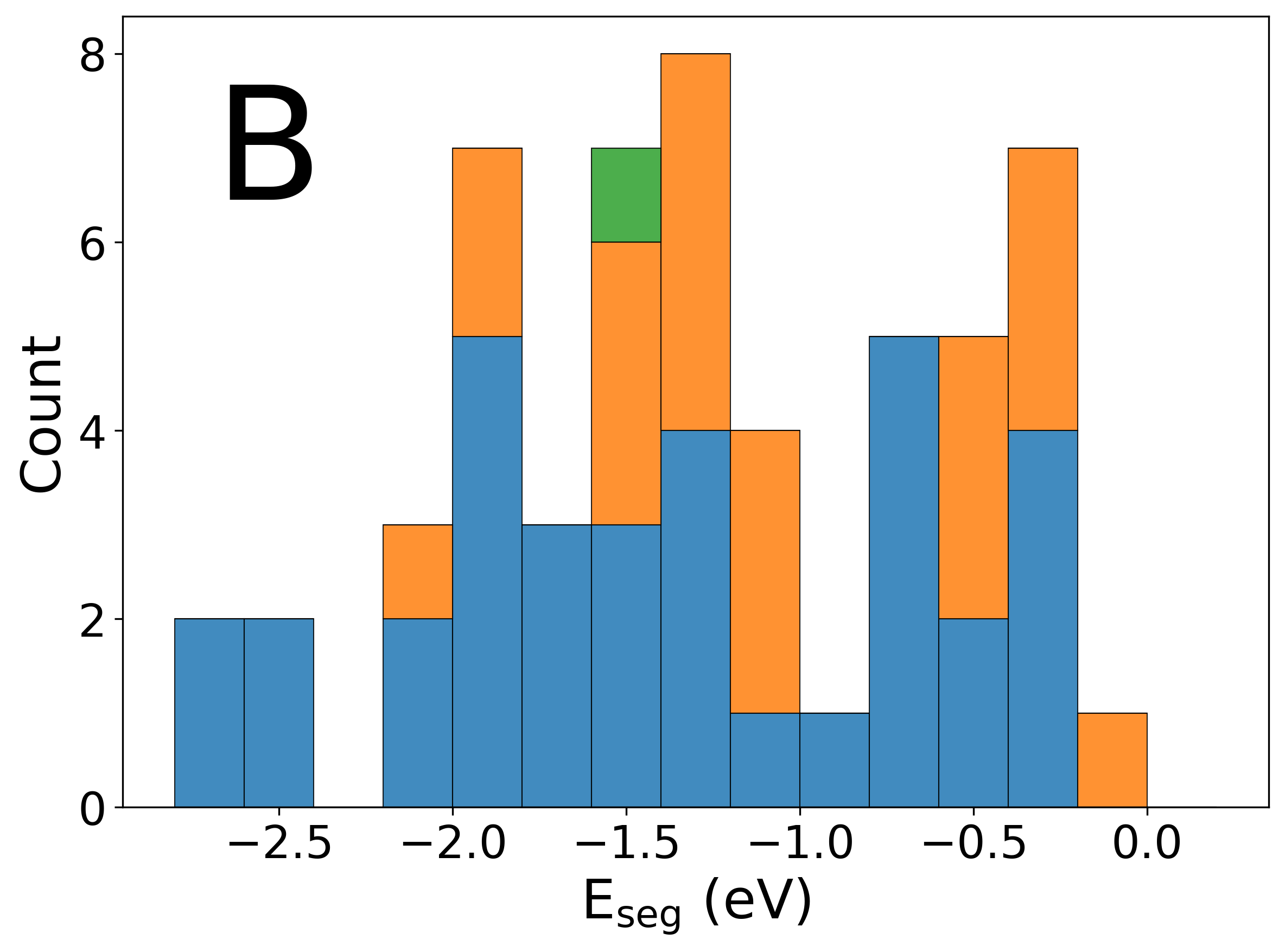}
		\caption{}
		\label{fig:Eseg_hist_B}
	\end{subfigure}\hfill
	\begin{subfigure}{0.48\linewidth}
		\centering
		\includegraphics[width=\linewidth]{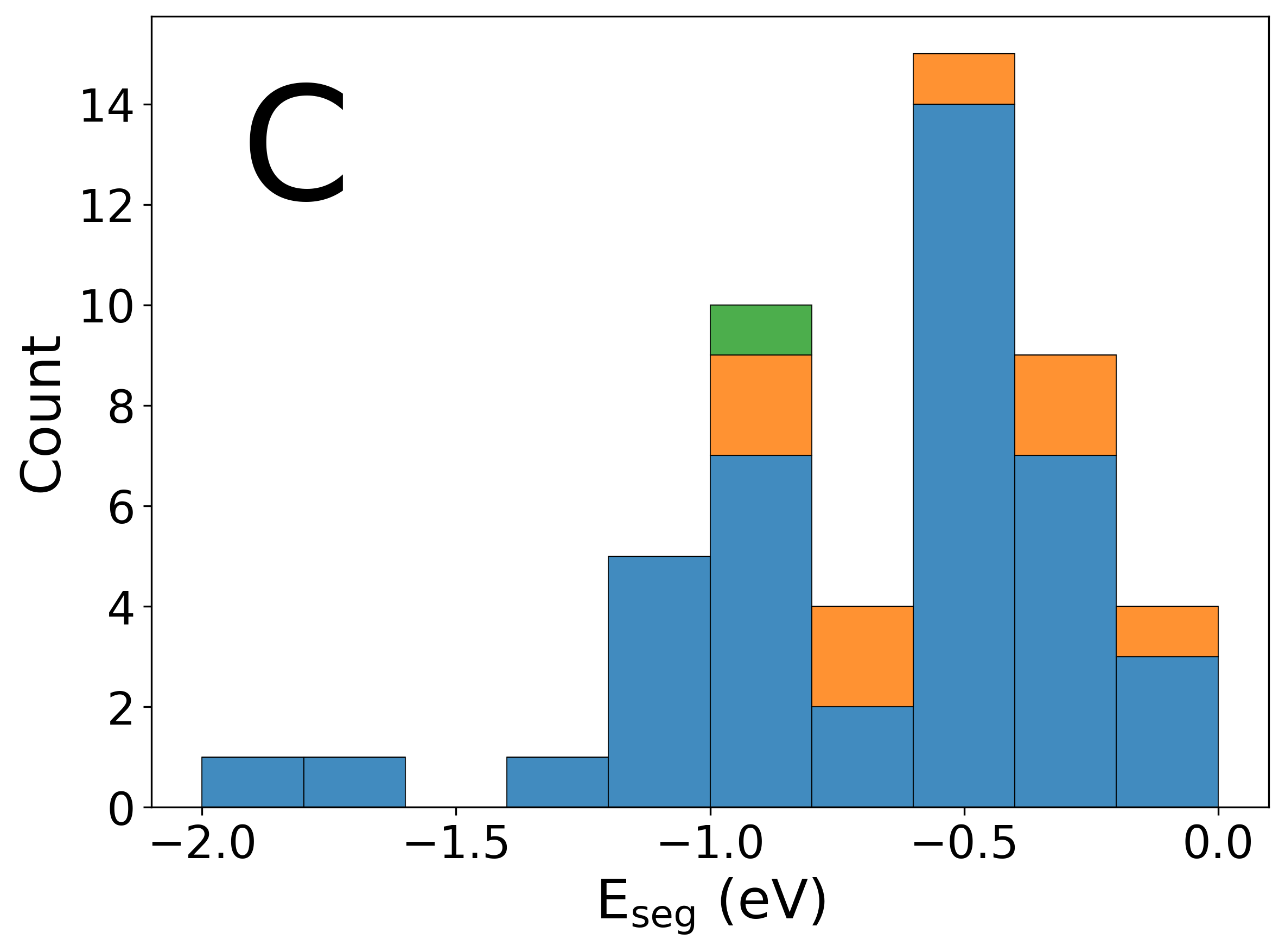}
		\caption{}
		\label{fig:Eseg_hist_C}
	\end{subfigure}
	\begin{subfigure}{0.48\linewidth}
		\centering
		\includegraphics[width=\linewidth]{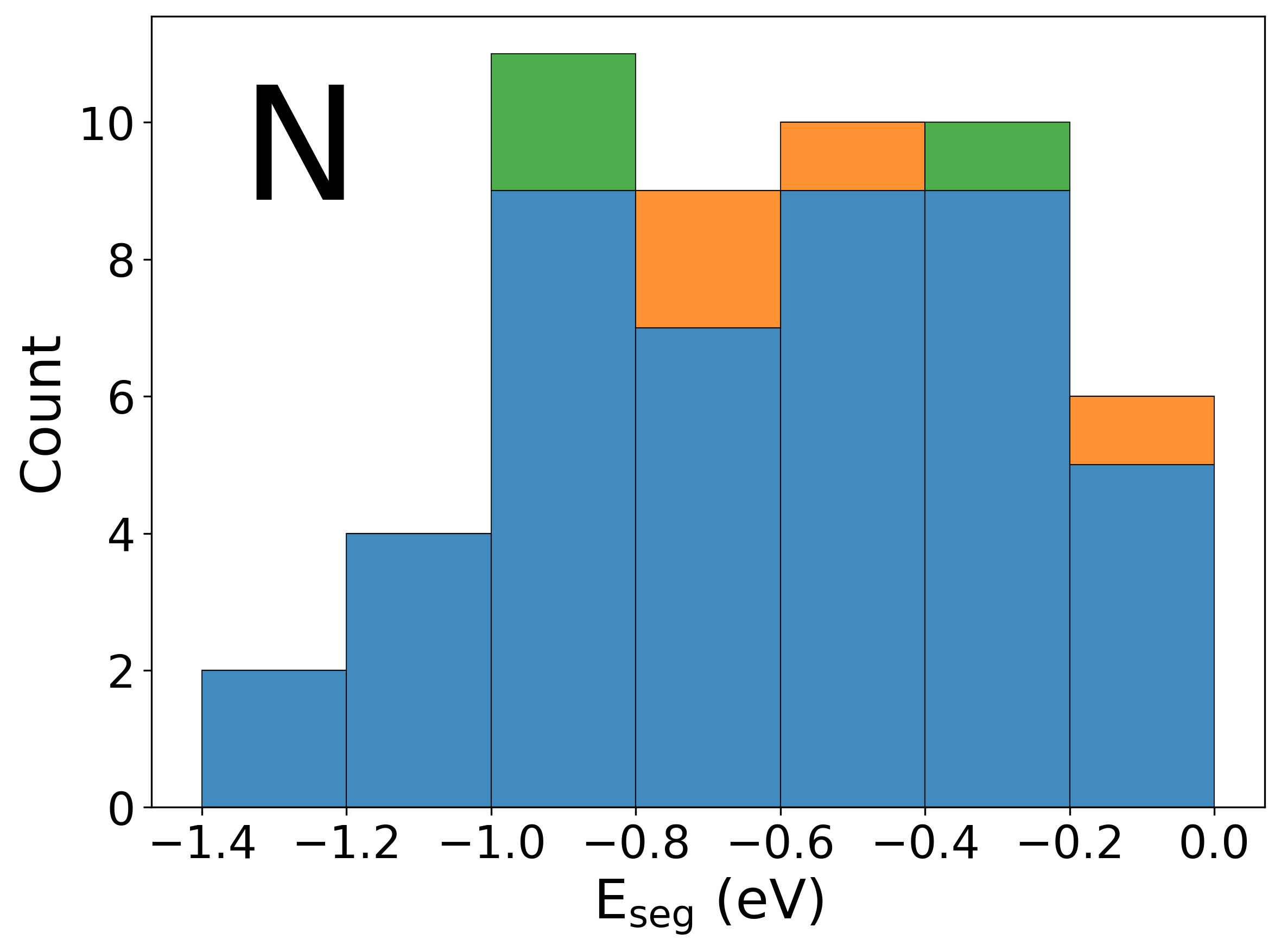}
		\caption{}
		\label{fig:Eseg_hist_N}
	\end{subfigure}\hfill
	\begin{subfigure}{0.48\linewidth}
		\centering
		\includegraphics[width=\linewidth]{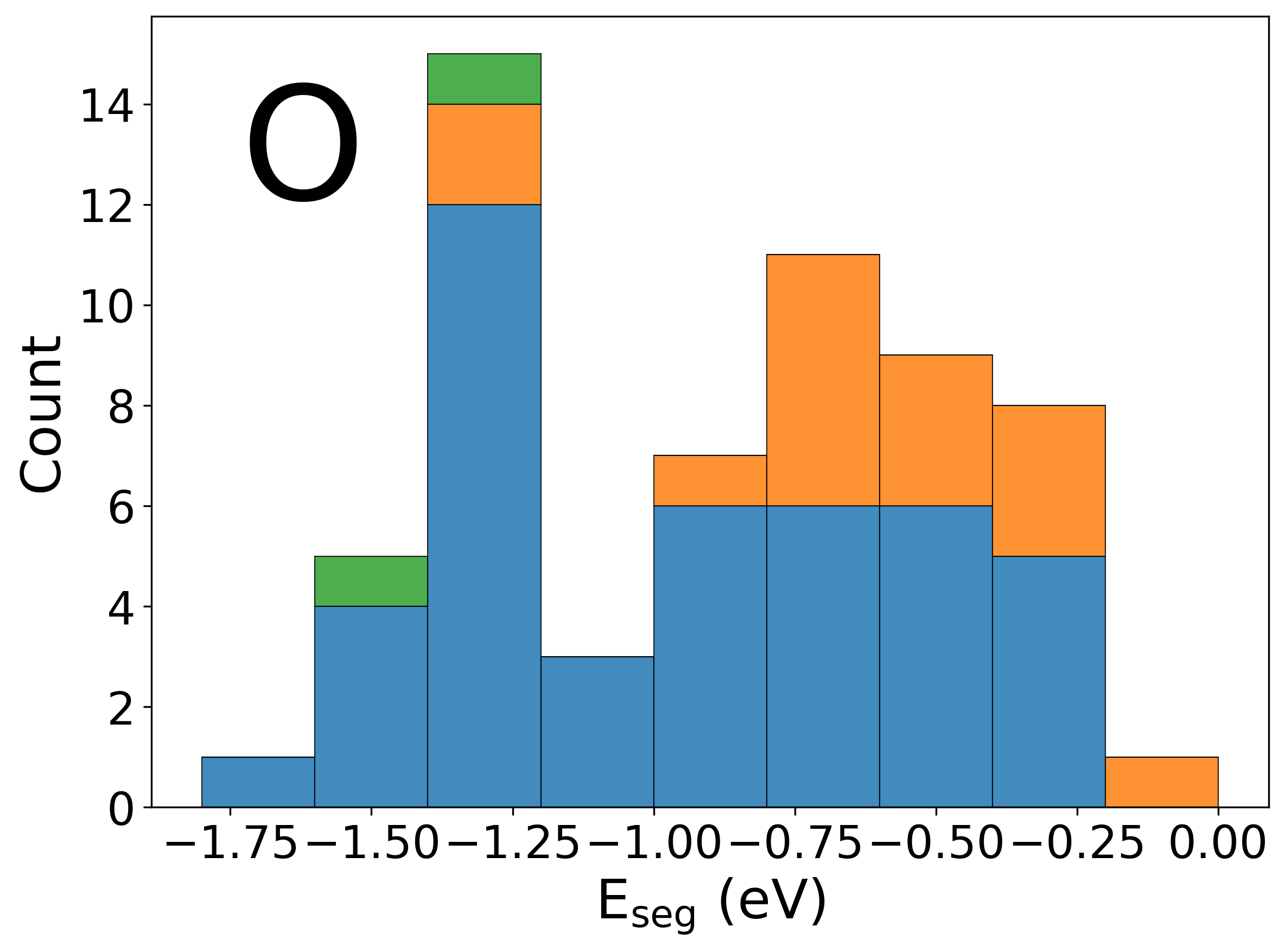}
		\caption{}
		\label{fig:Eseg_hist_O}
	\end{subfigure}
	\begin{subfigure}{0.48\linewidth}
		\centering
		\includegraphics[width=\linewidth]{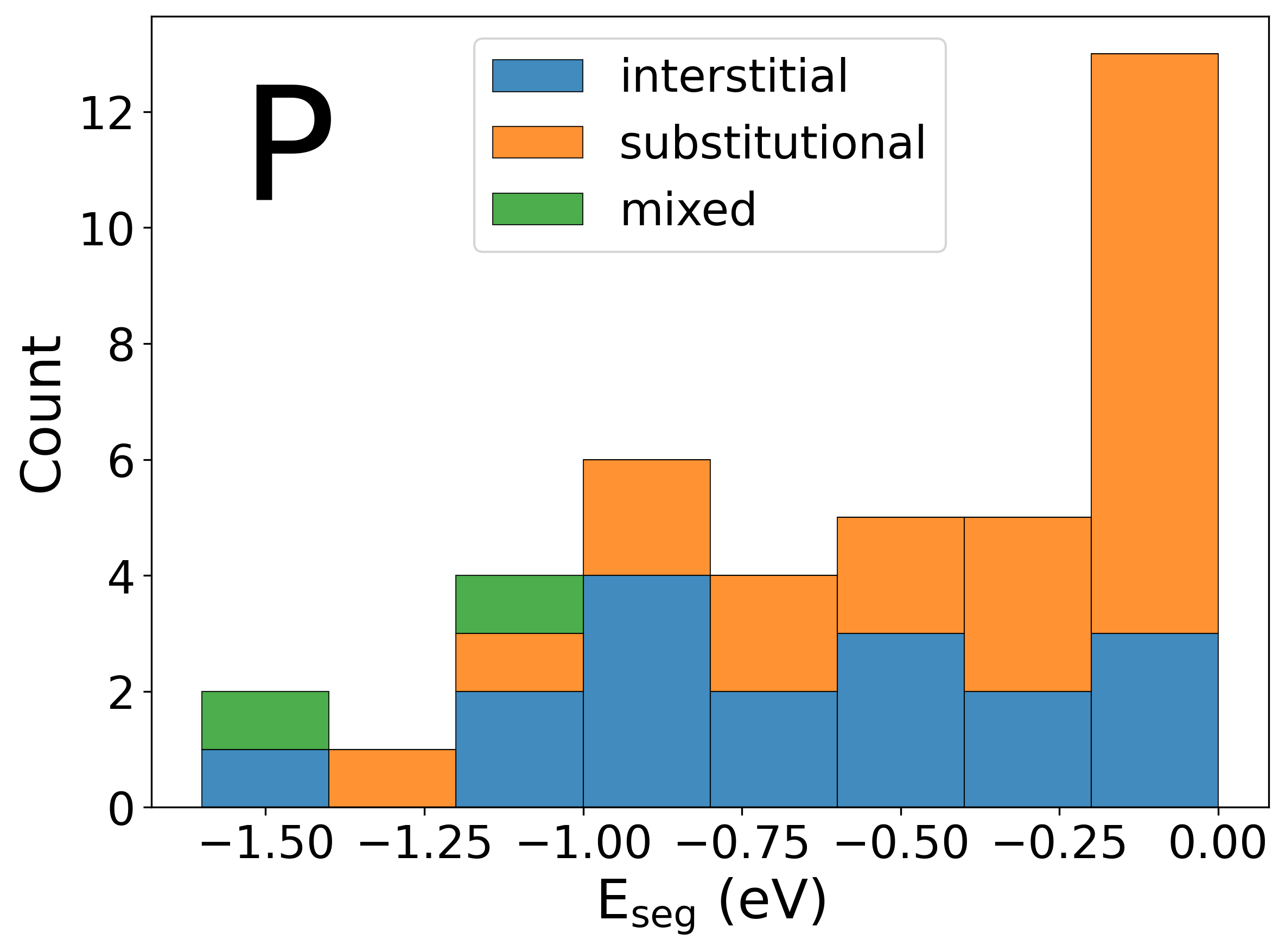}
		\caption{}
		\label{fig:Eseg_hist_P}
	\end{subfigure}\hfill
	\begin{subfigure}{0.48\linewidth}
		\centering
		\includegraphics[width=\linewidth]{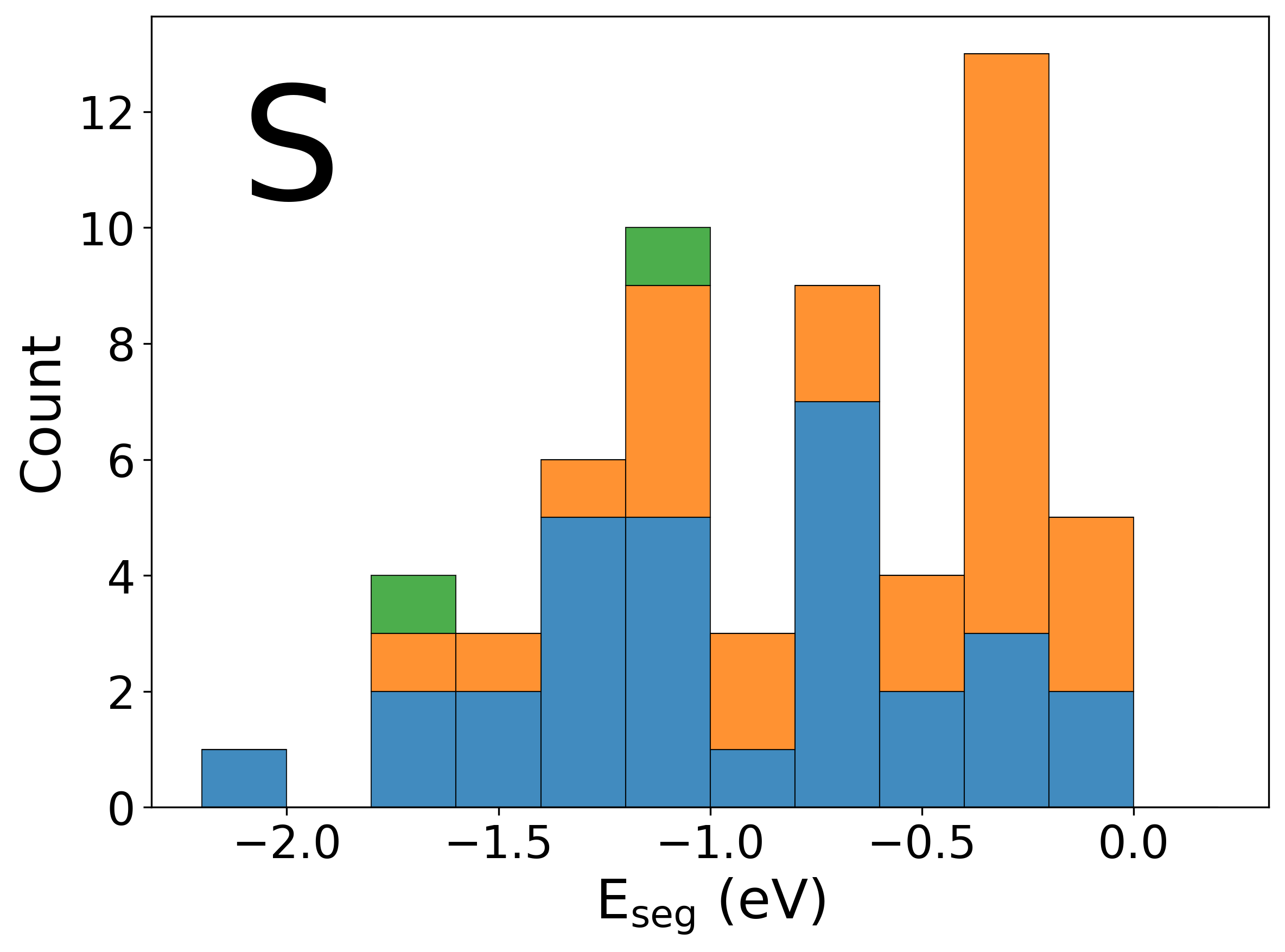}
		\caption{}
		\label{fig:Eseg_hist_S}
	\end{subfigure}
	\caption{Distributions of segregation energies of the unique sites (after SOAP-based duplicate removal, $E_\mathrm{seg} < -0.1$~eV) for each element after our duplicate removal procedure in our study: (\ref{fig:Eseg_hist_H}) H; (\ref{fig:Eseg_hist_He}) He; (\ref{fig:Eseg_hist_B}) B; (\ref{fig:Eseg_hist_C}) C; (\ref{fig:Eseg_hist_N}) N; (\ref{fig:Eseg_hist_O}) O; (\ref{fig:Eseg_hist_P}) P; (\ref{fig:Eseg_hist_S}) S. Bars are coloured by the starting site type: interstitial (blue), substitutional (orange), and mixed (green).}
	\label{fig:Eseg_histograms}
\end{figure}

The distributions of segregation energies for the unique sites retained after our duplicate filtering process are shown in Fig. \ref{fig:Eseg_histograms}. The histograms are split by starting site type: interstitial (blue), substitutional (orange), and mixed (green, where both starting types relaxed to the same final configuration). The segregation spectra vary considerably between elements. H exhibits a narrow distribution concentrated near $-0.2$ to $-0.5$ eV, while B and S show broad spectra extending to $-2.8$ and $-2.0$ eV respectively. The relative proportions of site types vary: H, B, C, N, O are predominantly interstitial, while He, P, S have significant substitutional contributions.
\\\\
No clear correlation is observed in segregation trapping at either smaller or larger sites for most solutes. There is a preference for a specific volume that results in deepest trapping for some solutes, such as B, C, N, O. For these solutes, distinct "U" shapes/wells exist in the segregation energy plots as a function of volume. It is important to note that this site volume does not inform how strongly a solute is trapped; but rather informs a rough lower bound for the segregation energy. For P, S, He and H, there is considerable scatter and this relationship is not observed. There is some remarkable similarity in the shapes of the plots for P and S. One of these features is the clear preference for small solute sites at 9\ \AA$^3$\ in the $\Sigma$5[001](210) and the $\Sigma$5[001](310) GBs, with similarities elsewhere, indicating that P and S likely prefer similar kinds of sites.
\\ 
\begin{figure}[h!]
	\centering
	\begin{subfigure}{0.48\linewidth}
		\centering
		\includegraphics[width=\linewidth]{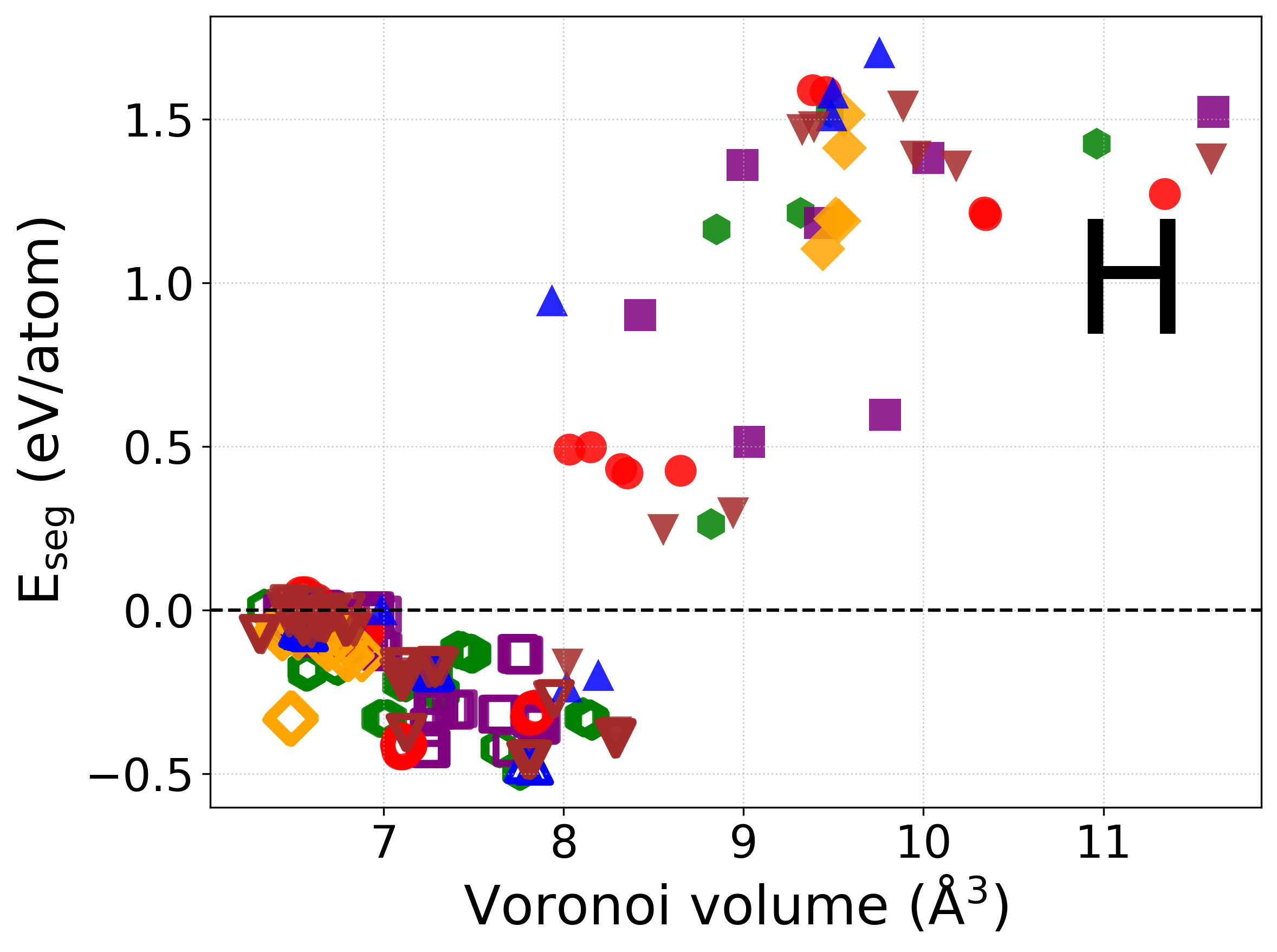}
		\caption{}
		\label{fig:VorSeg_H}
	\end{subfigure}\hfill
	\begin{subfigure}{0.48\linewidth}
		\centering
		\includegraphics[width=\linewidth]{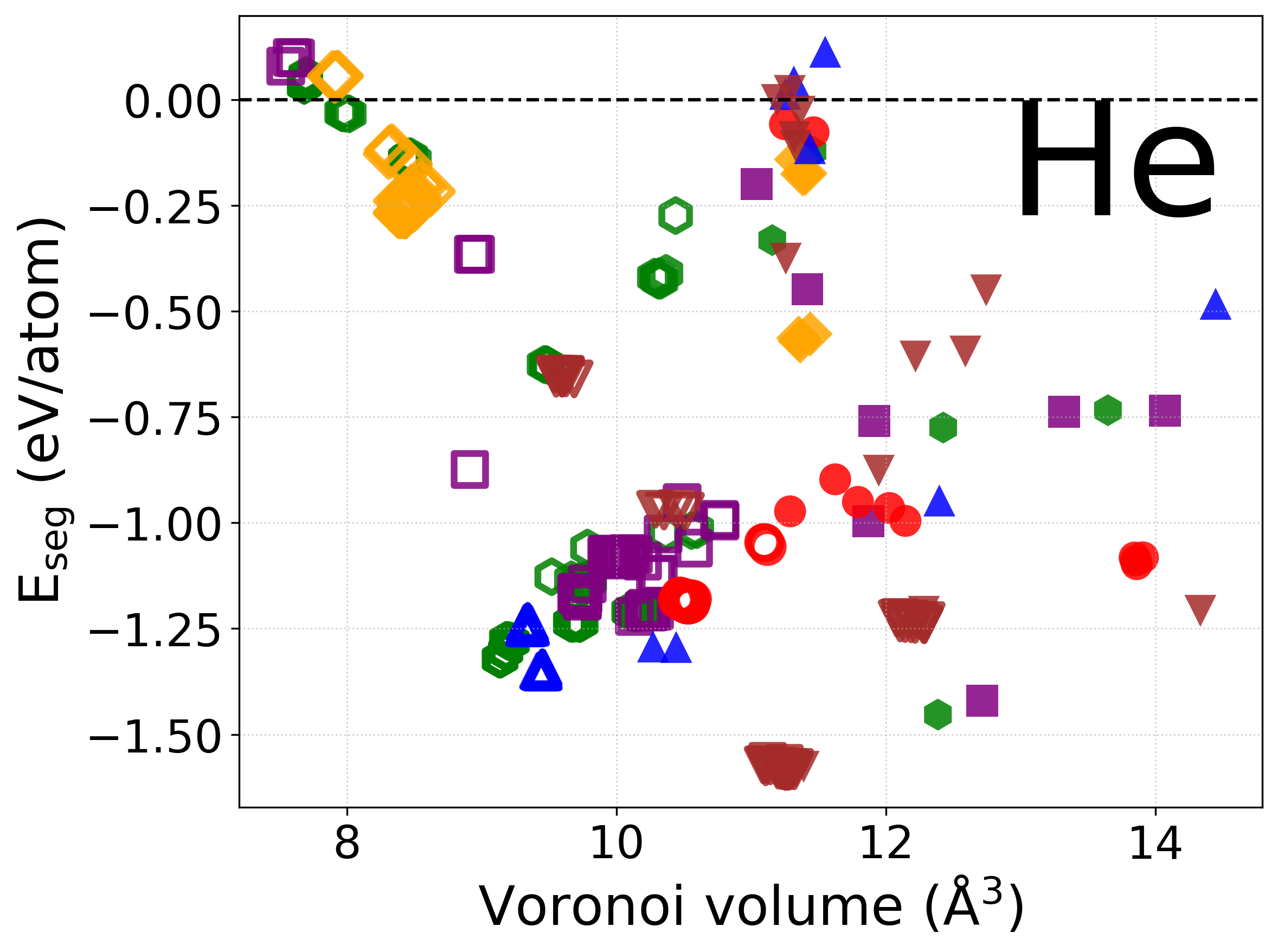}
		\caption{}
		\label{fig:VorSeg_He}
	\end{subfigure}
	\begin{subfigure}{0.48\linewidth}
		\centering
		\includegraphics[width=\linewidth]{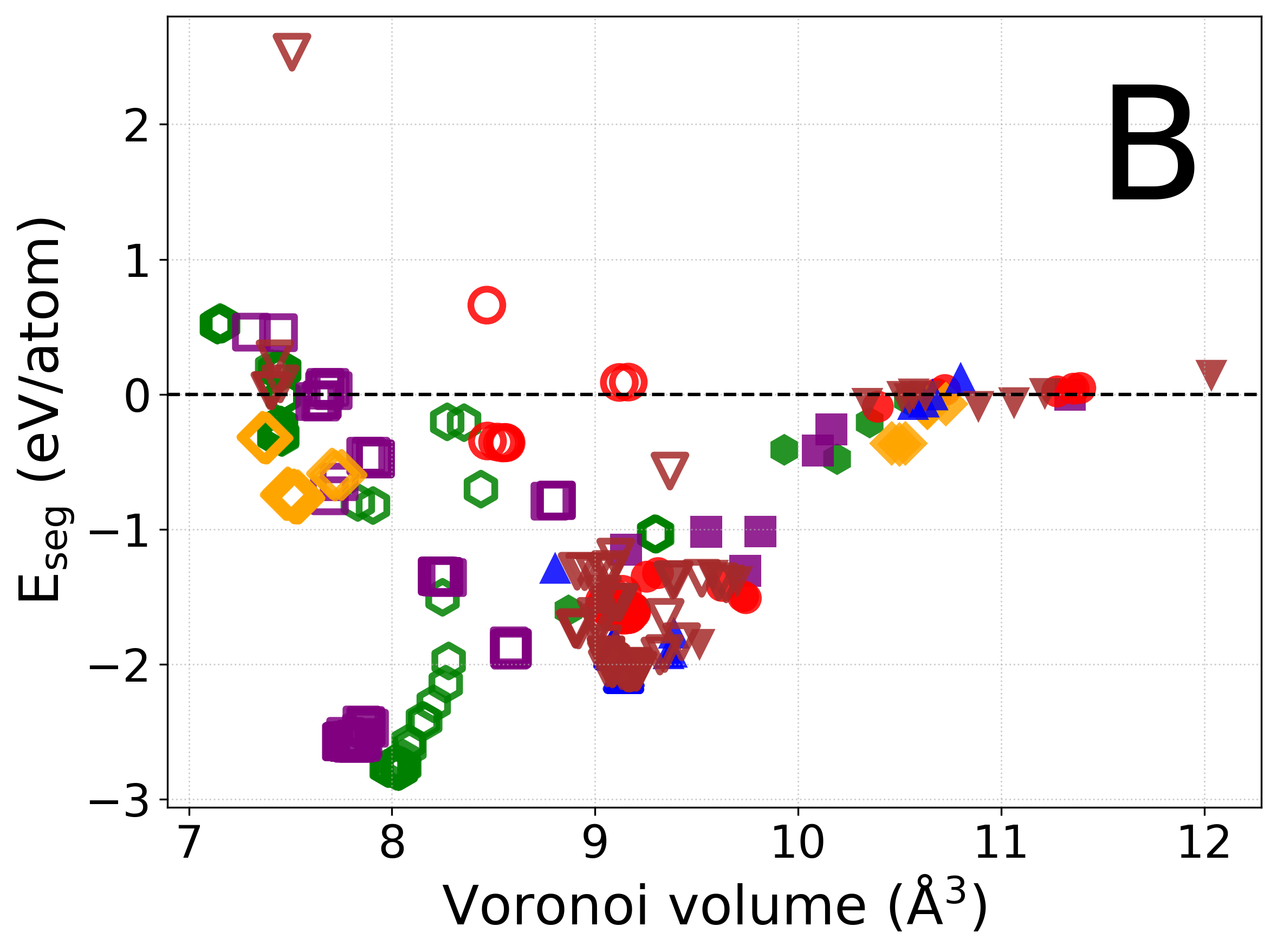}
		\caption{}
		\label{fig:VorSeg_B}
	\end{subfigure}\hfill
	\begin{subfigure}{0.48\linewidth}
		\centering
		\includegraphics[width=\linewidth]{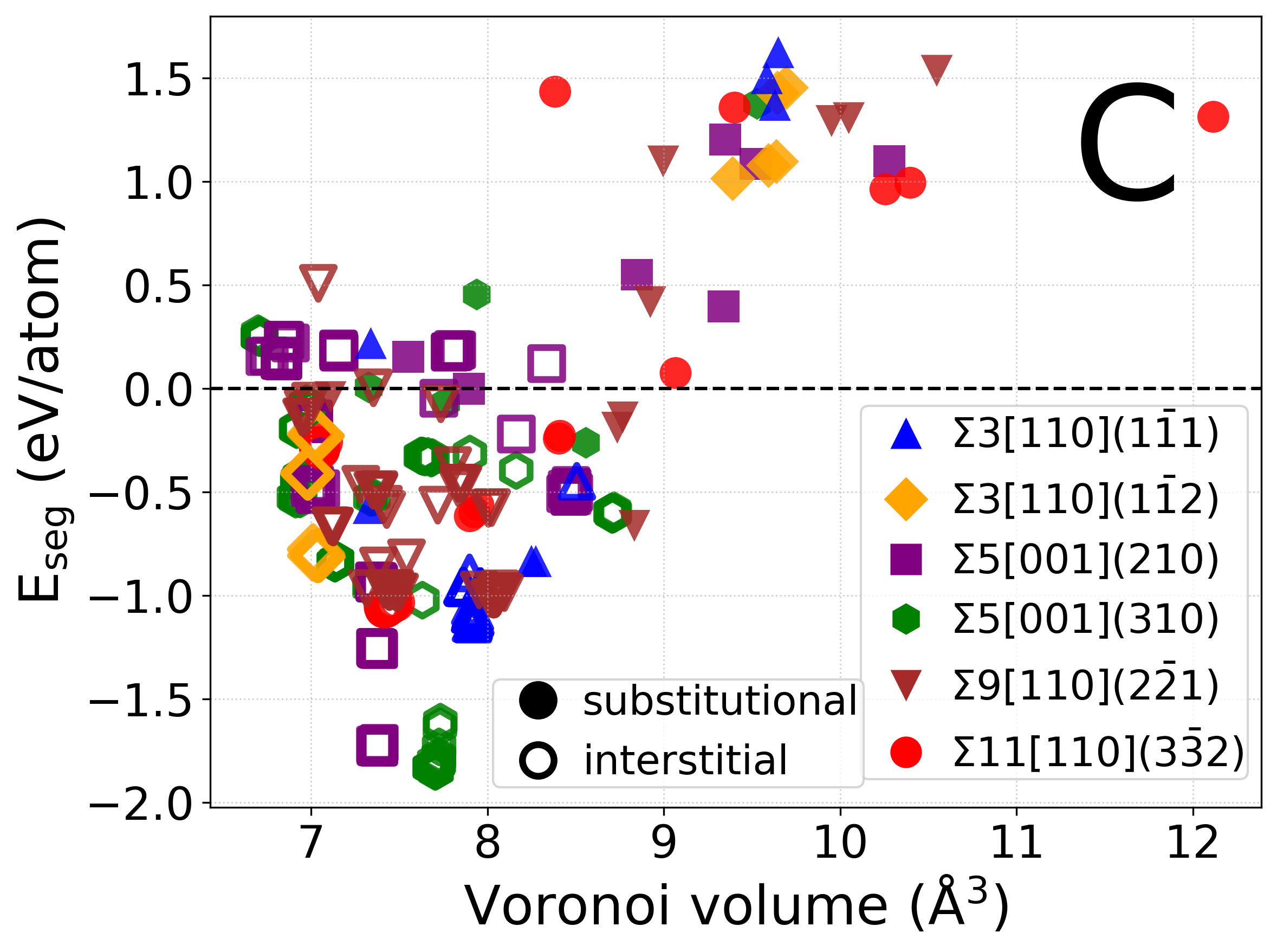}
		\caption{}
		\label{fig:VorSeg_C}
	\end{subfigure}
	\begin{subfigure}{0.48\linewidth}
		\centering
		\includegraphics[width=\linewidth]{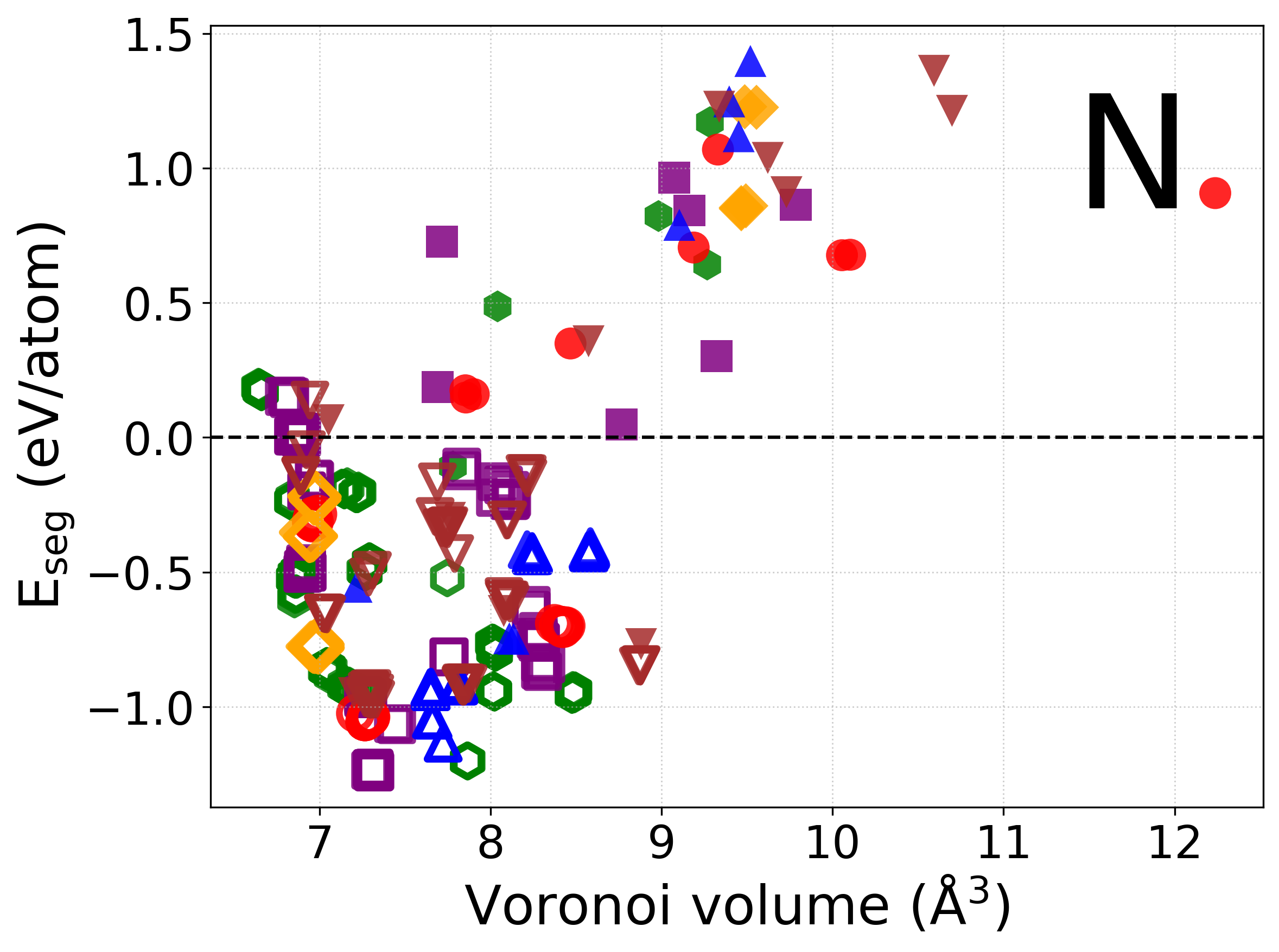}
		\caption{}
		\label{fig:VorSeg_N}
	\end{subfigure}\hfill
	\begin{subfigure}{0.48\linewidth}
		\centering
		\includegraphics[width=\linewidth]{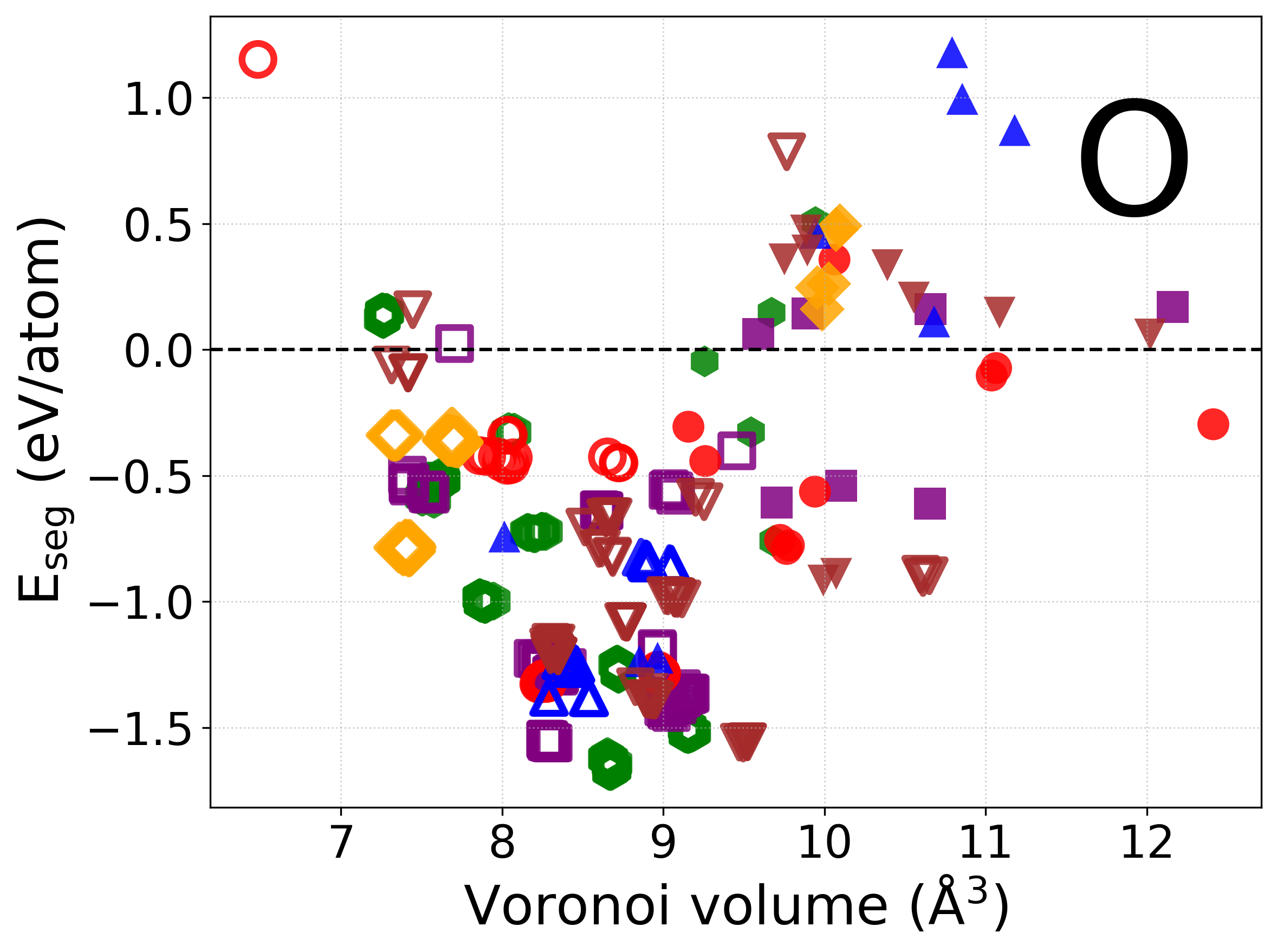}
		\caption{}
		\label{fig:VorSeg_O}
	\end{subfigure}
	\begin{subfigure}{0.48\linewidth}
		\centering
		\includegraphics[width=\linewidth]{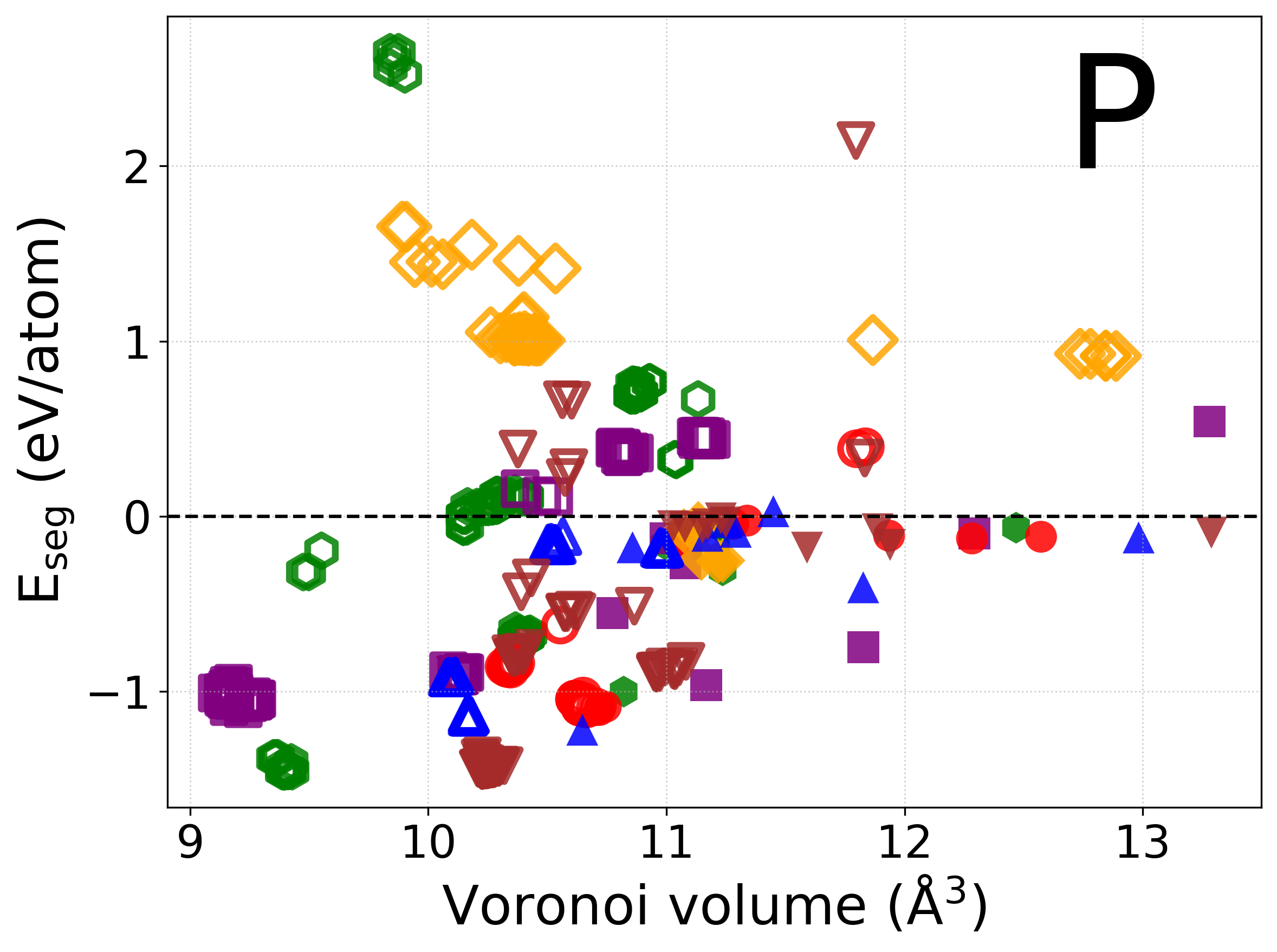}
		\caption{}
		\label{fig:VorSeg_P}
	\end{subfigure}\hfill
	\begin{subfigure}{0.48\linewidth}
		\centering
		\includegraphics[width=\linewidth]{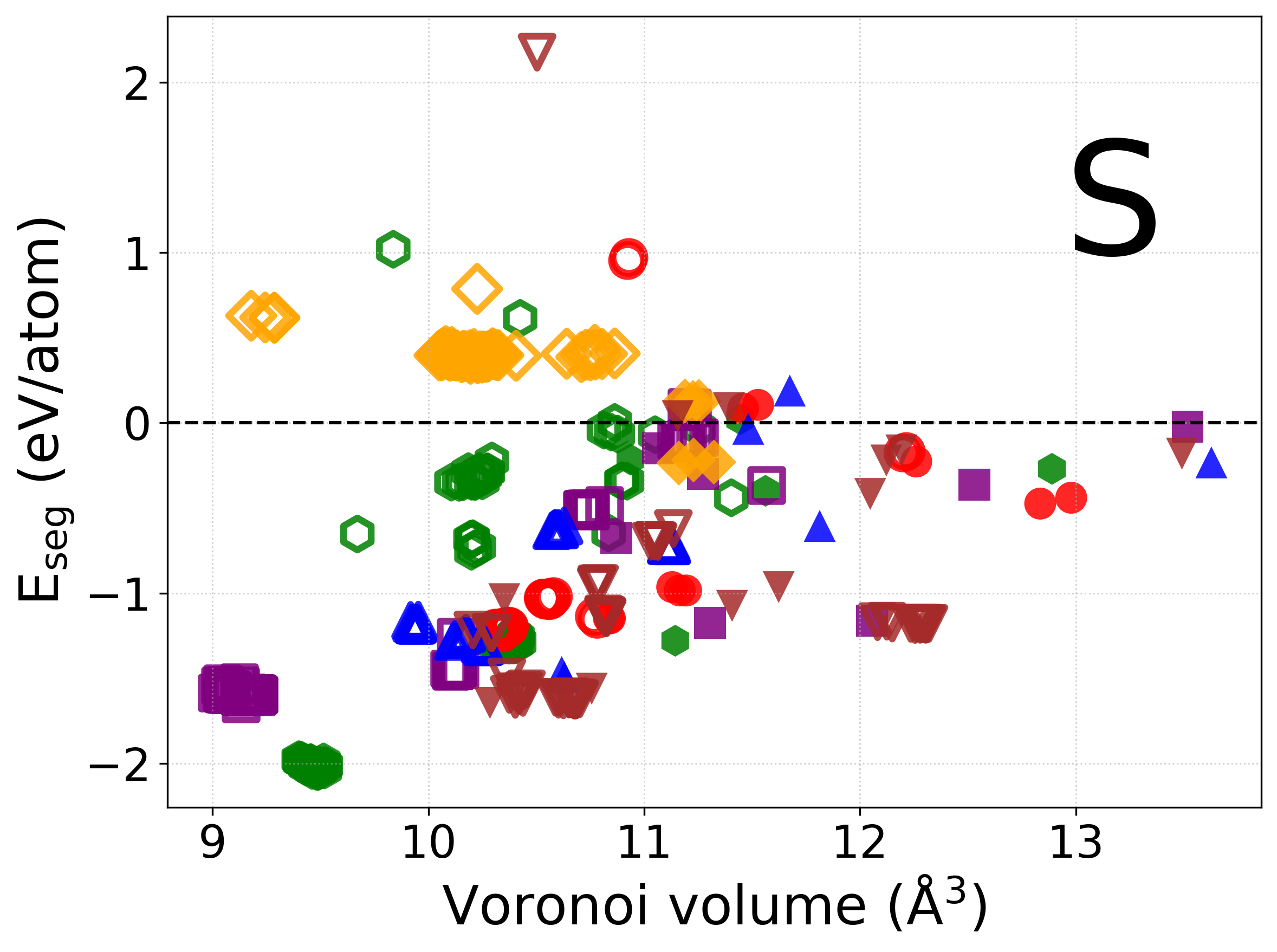}
		\caption{}
		\label{fig:VorSeg_S}
	\end{subfigure}
	
	\caption{The final relaxed segregation energy is plotted against the corresponding Voronoi volume of that site for each element: (\ref{fig:VorSeg_H}) H, (\ref{fig:VorSeg_He}) He, (\ref{fig:VorSeg_B}) B, (\ref{fig:VorSeg_C}) C, (\ref{fig:VorSeg_N}) N, (\ref{fig:VorSeg_O}) O, (\ref{fig:VorSeg_P}) P and (\ref{fig:VorSeg_S}) S.}
	\label{fig:VoronoiSegregation_All}
\end{figure}

\begin{figure}[h!]
	\centering
	\begin{subfigure}{0.48\linewidth}
		\centering
		\includegraphics[width=\linewidth]{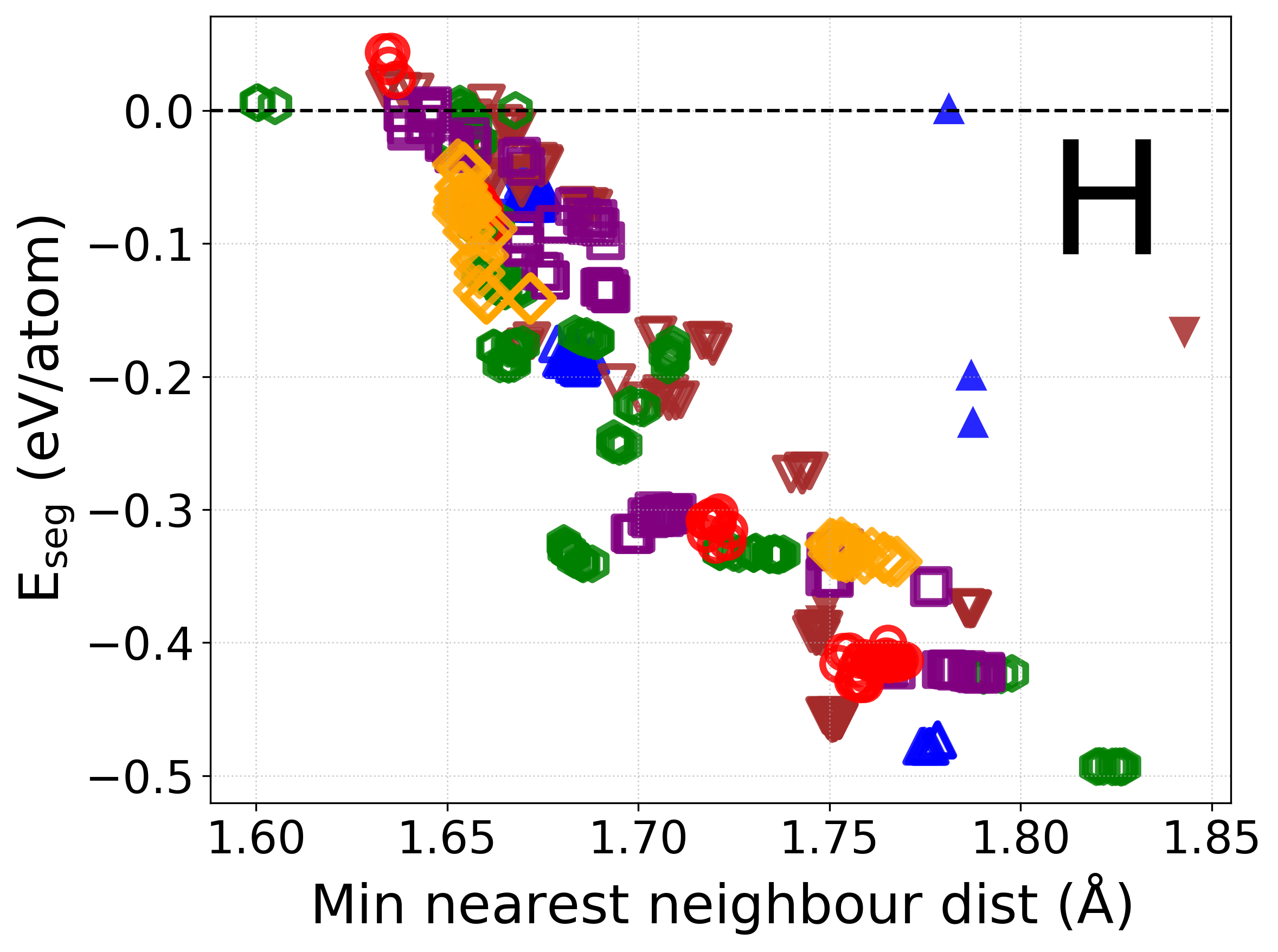}
		\caption{}
		\label{fig:NNdistSeg_H}
	\end{subfigure}\hfill
	\begin{subfigure}{0.48\linewidth}
		\centering
		\includegraphics[width=\linewidth]{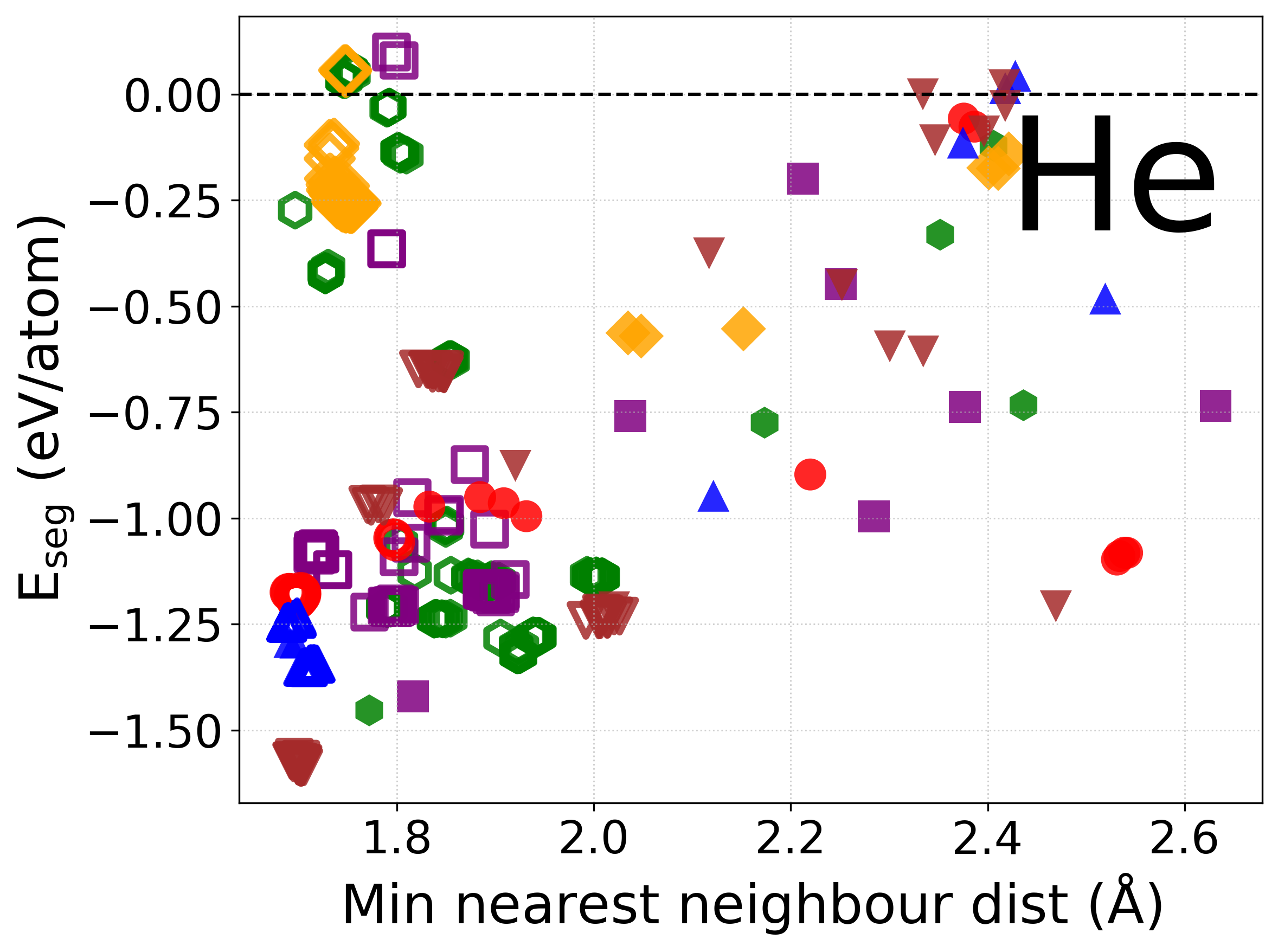}
		\caption{}
		\label{fig:NNdistSeg_He}
	\end{subfigure}
	\begin{subfigure}{0.48\linewidth}
		\centering
		\includegraphics[width=\linewidth]{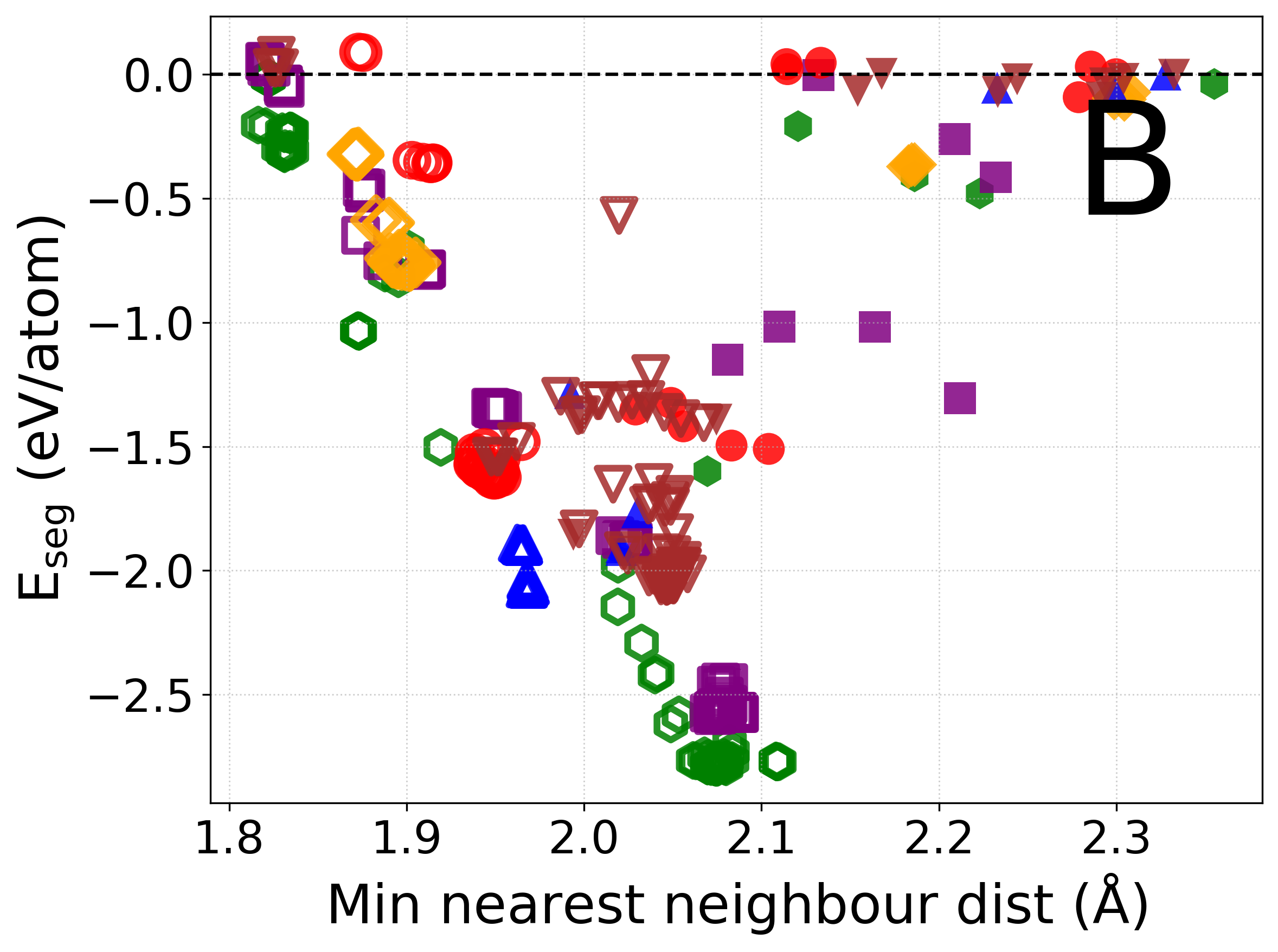}
		\caption{}
		\label{fig:NNdistSeg_B}
	\end{subfigure}\hfill
	\begin{subfigure}{0.48\linewidth}
		\centering
		\includegraphics[width=\linewidth]{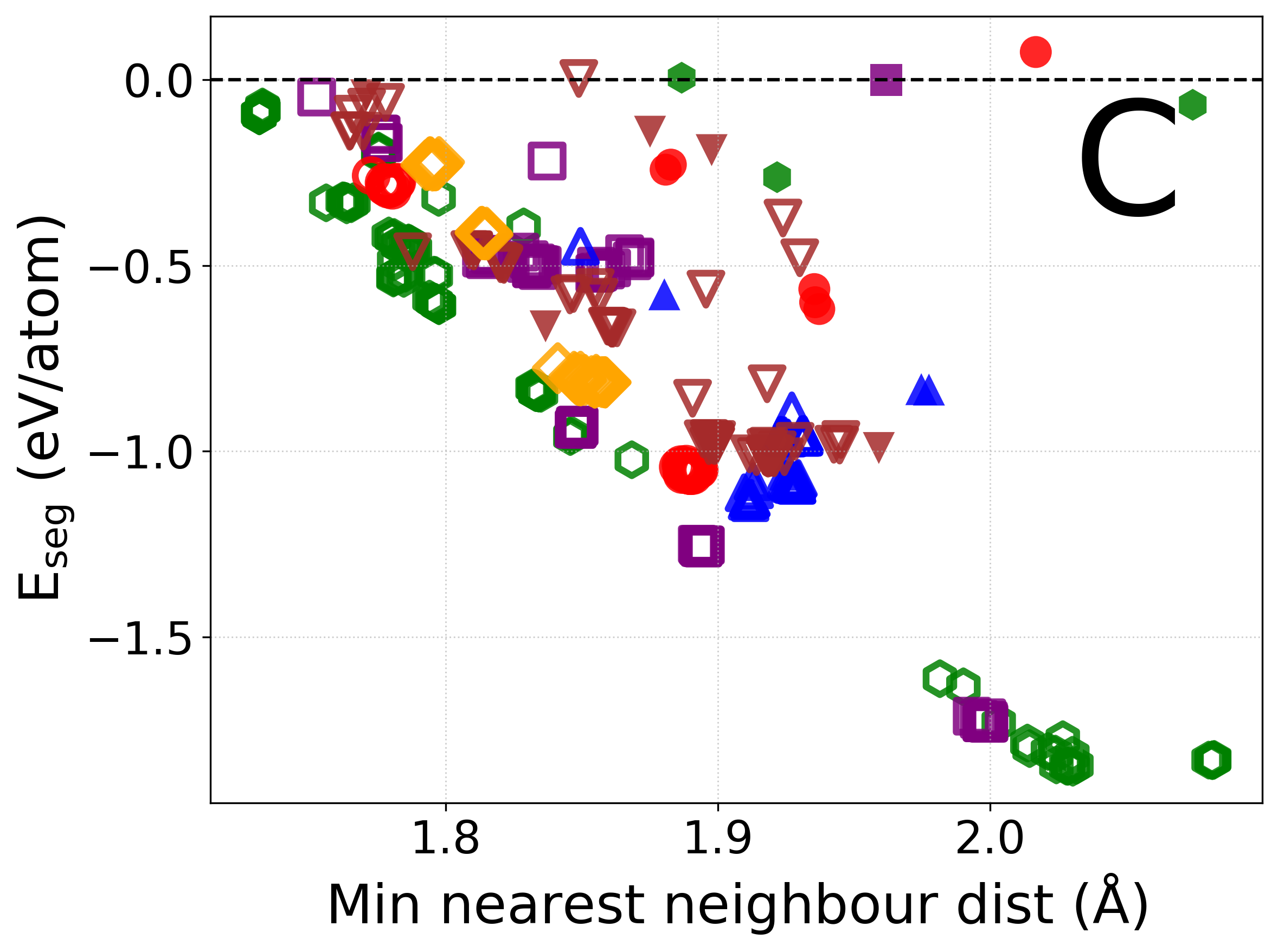}
		\caption{}
		\label{fig:NNdistSeg_C}
	\end{subfigure}
	\begin{subfigure}{0.48\linewidth}
		\centering
		\includegraphics[width=\linewidth]{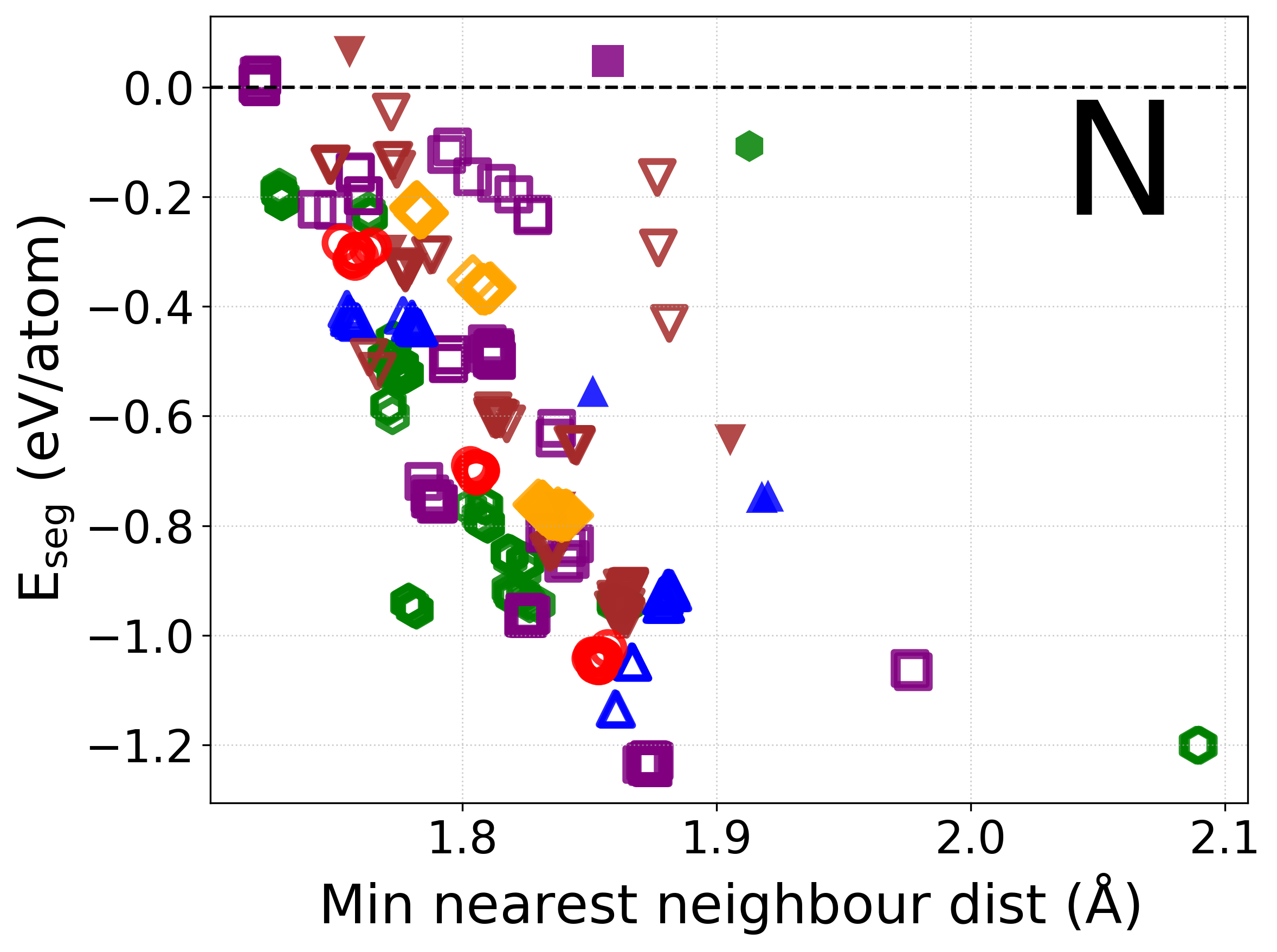}
		\caption{}
		\label{fig:NNdistSeg_N}
	\end{subfigure}\hfill
	\begin{subfigure}{0.48\linewidth}
		\centering
		\includegraphics[width=\linewidth]{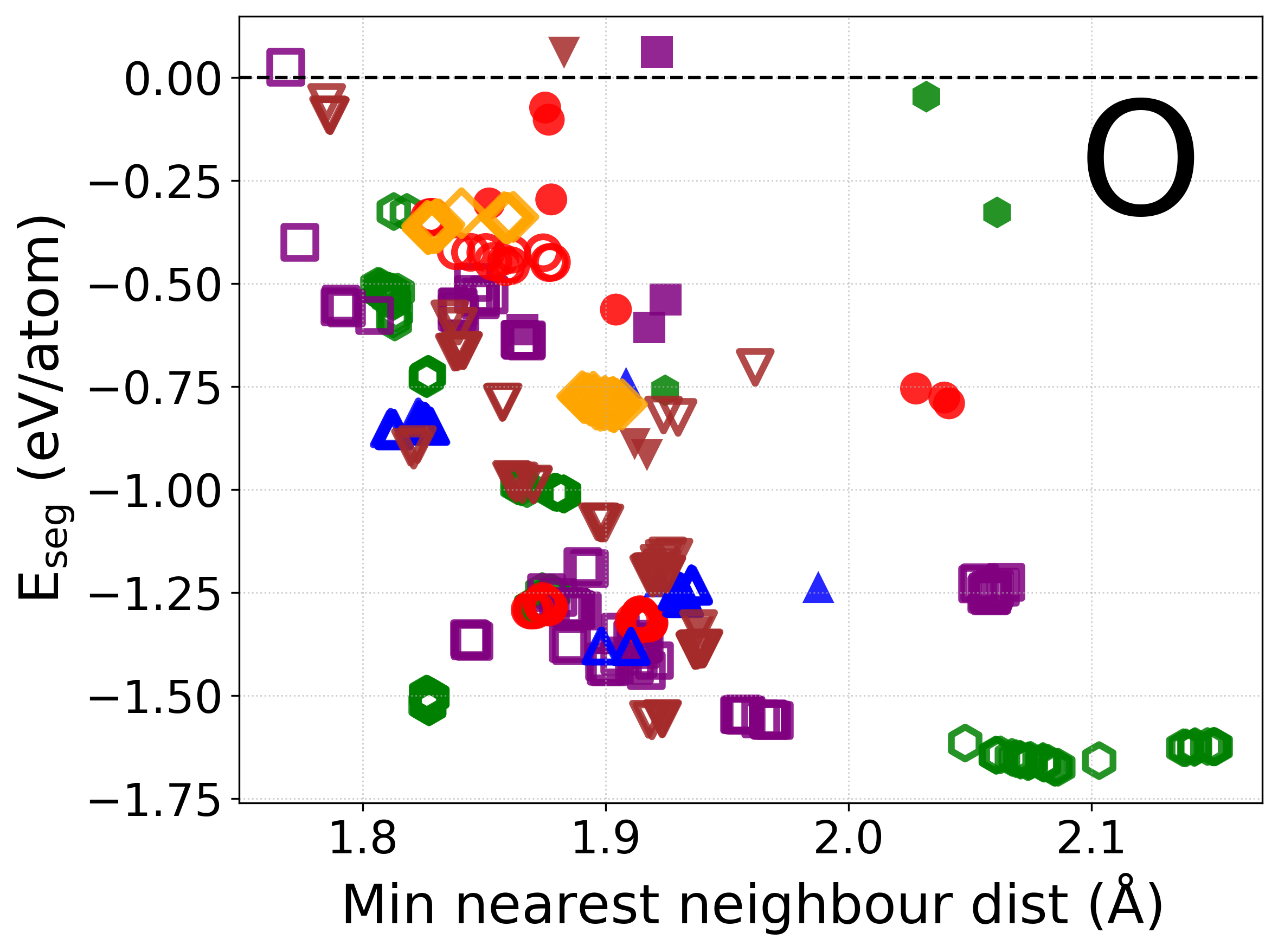}
		\caption{}
		\label{fig:NNdistSeg_O}
	\end{subfigure}
	\begin{subfigure}{0.48\linewidth}
		\centering
		\includegraphics[width=\linewidth]{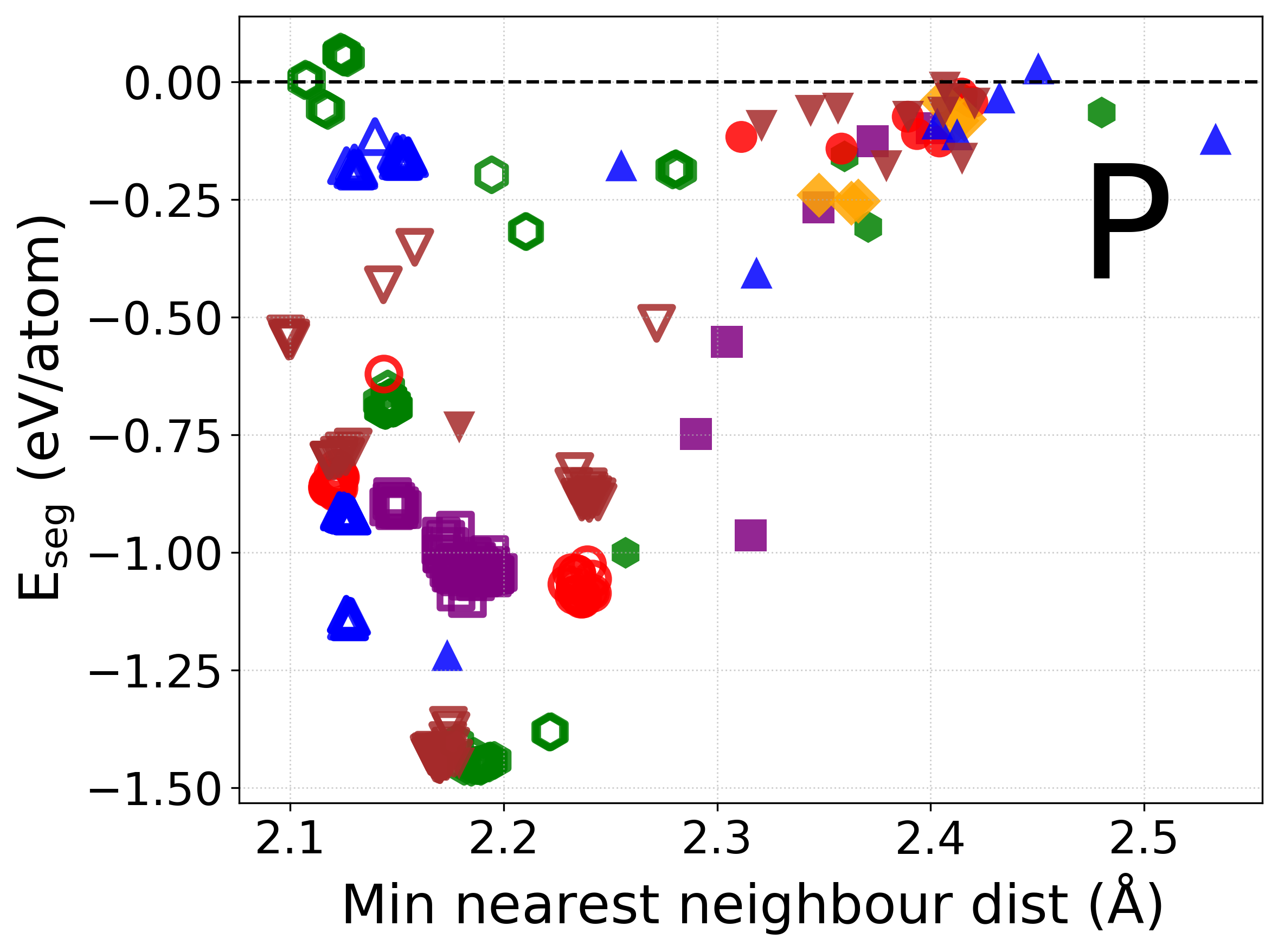}
		\caption{}
		\label{fig:NNdistSeg_P}
	\end{subfigure}\hfill
	\begin{subfigure}{0.48\linewidth}
		\centering
		\includegraphics[width=\linewidth]{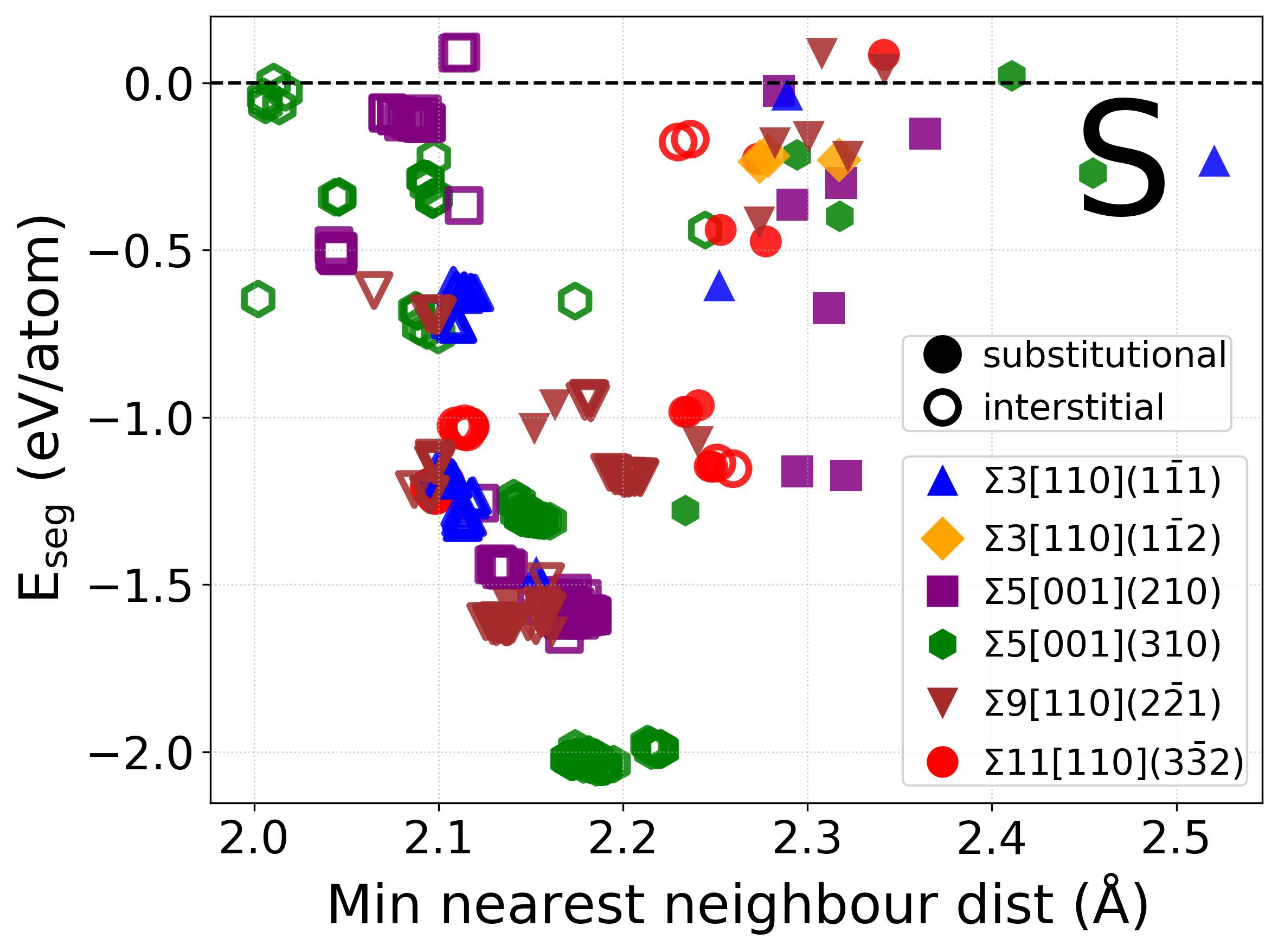}
		\caption{}
		\label{fig:NNdistSeg_S}
	\end{subfigure}
	
	\caption{Nearest‐neighbor distance vs.\ relaxed segregation energy comparison for solutes in the same GB: (\ref{fig:NNdistSeg_H}) H; (\ref{fig:NNdistSeg_He}) He; (\ref{fig:NNdistSeg_B}) B; (\ref{fig:NNdistSeg_C}) C; (\ref{fig:NNdistSeg_N}) N; (\ref{fig:NNdistSeg_O}) O; (\ref{fig:NNdistSeg_P}) P; (\ref{fig:NNdistSeg_S}) S. Each panel plots the nearest‐neighbor distance of that element’s preferred site against its minimum segregation energy \(E_{\rm seg}\).}
	\label{fig:NNdistSeg_All}
\end{figure}
In the literature, it is often posited that higher levels of structural disorder are responsible for segregation phenomena at defects. Here, we refine the argument in the case of the light elements, that it is instead the ability to accommodate strain in the form of an increase in the minimum solute-host atomic distance (hereby referred to as nearest-neighbour distance) to the favoured bond length of the solute. To visualise this, we have plotted the minimum distance from the solute to the nearest-neighbour in the final relaxed state of each studied configuration in Fig. \ref{fig:NNdistSeg_All}. The minimum distance between atoms proves to be clearly correlated to the lower bound of segregation energies than the Voronoi volume. Notably, this relationship is not present for He, which does not form bonds. So, one can interpret that the maximum strength of segregation binding for these solutes (excl. He) to be a function of the minimum atom-atom bonding length, indicating that strain accommodation in the shortest bonds is a dominant feature in determining the energetic favourability of a segregation site at GBs.
\\\\
To quantify this purported relationship between the lower bound of achievable segregation energies and minimum nearest neighbour distance we performed a binning exercise for each element. For elements which exhibit a U-shaped energy well, we restricted the analysis to the descending branch by automatically identifying the bin containing the deepest $E_\mathrm{seg}$ and use its right edge for a cutoff. We identified the lower envelope of segregation energies via partitioning them into 12 equally spaced bins and extracted the minimum $E_\mathrm{seg}$ within each bin. Computing the Spearman rank correlations on each of the binned minima yields correlations ($\rho$), which are reported for each element in \ref{tab:nn_dist_correlation}. All elements (excluding He) yield strong negative correlations ($\rho \leq -0.64$), confirming our observation above, wherein the lower bound is controlled by the 1st nearest neighbour distance.

\begin{table}[h!]
	\centering
	\begin{tabular}{lccccc}
		\hline
		Element & Cutoff (\AA) & N & min($E_\mathrm{seg}$) (eV) & $\rho_\mathrm{full}$ & $\rho_\mathrm{envelope}$ \\
		\hline
		H  & 1.84 & 551 & $-$0.49 & $-$0.88 & $-$0.92 \\
		He & 1.77 & 302 & $-$1.59 & $+$0.54 & $+$0.82 \\
		B  & 2.09 & 486 & $-$2.79 & $-$0.90 & $-$0.96 \\
		C  & 2.05 & 484 & $-$1.85 & $-$0.80 & $-$0.93 \\
		N  & 1.91 & 511 & $-$1.24 & $-$0.79 & $-$0.93 \\
		O  & 2.09 & 511 & $-$1.67 & $-$0.53 & $-$0.64 \\
		P  & 2.21 & 337 & $-$1.46 & $-$0.62 & $-$0.90 \\
		S  & 2.22 & 400 & $-$2.05 & $-$0.72 & $-$0.95 \\
		\hline
	\end{tabular}
	\caption{Spearman rank correlation between 1st nearest-neighbour distance (i.e. solute to nearest neighbouring Fe atom) and segregation energy on the descending limb of the energy well. The cutoff is automatically determined as the right edge of the bin containing the deepest $E_\mathrm{seg}$. $\rho_\mathrm{full}$: correlation over all sites below the cutoff. $\rho_\mathrm{envelope}$: correlation of the binned lower-bound minima.}
	\label{tab:nn_dist_correlation}
\end{table}

To check whether the correlation is truly driven by the shortest bond or reflects a longer-range effect, we repeated the analysis for the $k$-th nearest-neighbour distance with $k = 2,\ldots,6$ (Supplementary Information, Sec.~S3, Table~S4). The correlation persists at the 2nd nearest-neighbour distance with comparable strength for H, C, N and O (Spearman $\rho$ typically within $\sim 0.1$ of the 1st-neighbour value), reflecting that the 1st and 2nd neighbours are both equally influential on the lower bound segregation energy. Beyond the 3rd nearest-neighbour, the correlation weakens substantially, and by the 5th--6th nearest-neighbour it is effectively lost for all elements. This suggests that the controlling quantity is the strain state of the immediate shortest bonds rather than any longer-range elastic field, and supports our interpretation of segregation binding as governed by accommodation of the shortest solute--host bond(s).
\\\\
From this, we can infer that rather than light elements preference for being where volume is initially available to host them, the favourable sites are instead those that can relax to maximise the nearest neighbour bond distance. This is an important distinction; although voids may initially offer a larger minimum distance to a neighbouring atom, the surrounding atomic structure may be rigid; meaning that little may be afforded in the way of strain release via atomic relaxation. Instead, a local environment involving structures which are "soft", i.e. those which allow large atomic displacements/distortions to facilitate the release of strain energy in the shortest bonds, are preferred by these smaller elements overall. We do not discuss further here on how GBs may be identified a-priori as "soft", but acknowledge this is an interesting avenue for future research.

\subsection{On site sampling strategies for light elements}
\label{sec:site_sampling}
A typical practitioner's approach to identifying favourable segregation sites for light elements is to site the atom at the largest possible starting volume at the GB and then applying a relaxation. This is often achieved by applying a Voronoi tesselation on the GB structure, identifying sites which lie on the polyhedra vertices and then taking the sites which have the largest Voronoi volume to calculate the segregation energy \cite{scheiberImpactSegregationEnergy2021}. With this heuristic approach, it is usually assumed that sites which have the largest initial volume are the most favourable for segregation. The underlying physical reasoning is that interstitials would prefer to be where there is sufficient additional volume to accommodate their presence, reducing strain energy. This assumption is often employed to avoid the extensive surveys and incurring expensive computational costs as done in this work.
\\\\
Here, we explicitly investigate the validity of this assumption, and present the results in Table \ref{tab:interstitial-voronoi-regret}. The table presents the energy differential of the site computed with the initially largest Voronoi volume compared to the minimum segregation energy actually computed from our candidate structures. Importantly, we find that this limited sampling approach can frequently cause studies to miss the most favourable segregation sites for light elements. 
\begin{table}[ht]
	\centering
	\small
	\setlength{\tabcolsep}{5pt}
	\renewcommand{\arraystretch}{1.15}
	\begin{tabular}{lccccccccc}
		\toprule
		GB & $N_\mathrm{sites}$ & H & He & B & C & N & O & P & S \\
		\midrule
		\textbf{$\Sigma3[110](1\bar{1}1)$}  & 54  & \textcolor{blue}{0.00} & \textcolor{blue}{0.00} & \textcolor{blue}{0.00} & \textcolor{blue}{0.00} & \textcolor{blue}{0.00} & \textcolor{blue}{0.01} & \textcolor{blue}{0.00} & \textcolor{blue}{0.01} \\
		\textbf{$\Sigma3[110](1\bar{1}2)$}  & 24  & \textcolor{blue}{0.01} & \textcolor{blue}{0.00} & \textcolor{blue}{0.00} & \textcolor{blue}{0.00} & \textcolor{blue}{0.00} & \textcolor{blue}{0.02} & \cellcolor{red!19}0.09 & \cellcolor{red!53}0.25 \\
		\textbf{$\Sigma5[001](210)$}        & 98  & \textcolor{blue}{0.02} & \cellcolor{red!15}0.07 & \cellcolor{red!11}0.05 & \textcolor{blue}{0.03} & \cellcolor{red!100}0.47 & \cellcolor{red!73}0.34 & \cellcolor{red!14}0.07 & \cellcolor{red!11}0.05 \\
		\textbf{$\Sigma5[001](310)$}        & 91  & \textcolor{blue}{0.04} & \cellcolor{red!20}0.09 & \textcolor{blue}{0.02} & \cellcolor{red!96}0.45 & \textcolor{blue}{0.00} & \cellcolor{red!14}0.07 & \textcolor{blue}{0.00} & \cellcolor{red!15}0.07 \\
		\textbf{$\Sigma9[110](2\bar{2}1)$}  & 101 & \textcolor{blue}{0.00} & \textcolor{blue}{0.00} & \textcolor{blue}{0.00} & \textcolor{blue}{0.03} & \cellcolor{red!12}0.06 & \cellcolor{red!37}0.17 & \textcolor{blue}{0.01} & \cellcolor{red!87}0.41 \\
		\textbf{$\Sigma11[110](3\bar{3}2)$} & 40  & \textcolor{blue}{0.02} & \textcolor{blue}{0.03} & \textcolor{blue}{0.02} & \textcolor{blue}{0.02} & \textcolor{blue}{0.02} & \textcolor{blue}{0.04} & \cellcolor{red!49}0.23 & \textcolor{blue}{0.02} \\
		\bottomrule
	\end{tabular}
	\caption{The "regret" that is associated with heuristically selecting the interstitial site with the largest unrelaxed Voronoi volume compared to the actual minimum segregation energy computed at a grain boundary. Regret is computed as $E_{\mathrm{seg}}(V_{\mathrm{Voronoi}}^{\max}) - \min(E_{\mathrm{seg}})$. Entries are regret in eV relative to the true minimum segregation energy interstitial site within each grain boundary and element. Values below 0.05 eV are treated as effectively the minimum and shown in blue. Larger misses are shaded from white to red, normalized to the largest interstitial regret (0.47 eV).}
	\label{tab:interstitial-voronoi-regret}
\end{table}
\\\\
It is tempting, then, to attribute this disagreement to the fact that it should be the \textit{final} relaxed volume of the site that is important, rather than the initial volume. Such agreement would then assert that additional volume is a controlling factor due to the release of strain energy associated with larger volumes to host the additional atoms. This is however, not the case. In the Voronoi volume - segregation energy plots in Fig. \ref{fig:VoronoiSegregation_All}, we can see that the Voronoi volume is not a good predictor of the final segregation energy for these light elements, as there is significant scatter in the data. Therefore, the commonly applied volume criterion and the underlying physical reasoning to rationalise light element segregation are not supported by our data. Instead, we have shown in Fig. \ref{fig:NNdistSeg_All} that the minimum nearest-neighbour distance is a much better predictor of the lower bound of final segregation energy for these light elements. Attempts to find the most favourable segregation sites should then be attributed to finding such "soft" structures which can deform and effectively accommodate the strain energy in the shortest bonds.
\\\\
An obvious next step would be to use the unrelaxed (single-point) segregation energy on the starting structures as an inexpensive pre-screening tool; we show in the Supplementary Information that this also fails, both for the absolute values and for the ranking of a site's energetic favourability. The relaxation energies are substantial, and the unrelaxed ranking only moderately correlates with the relaxed ranking (Spearman $\rho \leq 0.58$ across all elements); the unrelaxed calculation identifies the same strongest site as the relaxed calculation in only 4 of 48 GB--element combinations. Full ionic relaxation is therefore required, and the implication for practitioners is that comprehensive sampling of candidate sites --- rather than \textit{a priori} selection based on pre-relaxation geometric or energetic descriptors is necessary to accurately capture the segregation behaviour of these light-element solutes.
\\\\
MLIP-based pre-screening is an emerging alternative once sufficiently accurate steel-specific MLIPs are available. Here, we stress that it is critical that the \textit{forces} in the MLIPs are accurate; in the absence of accurate forces, relaxations by MLIPs are likely to result in untrustworthy relaxation trajectories, and hence structurally wrong final configurations when compared to DFT.

\subsection{Segregation-cohesion engineering maps}
\begin{figure}[h!]
	\centering
	\begin{subfigure}{0.48\linewidth}
		\centering
		\includegraphics[width=\linewidth]{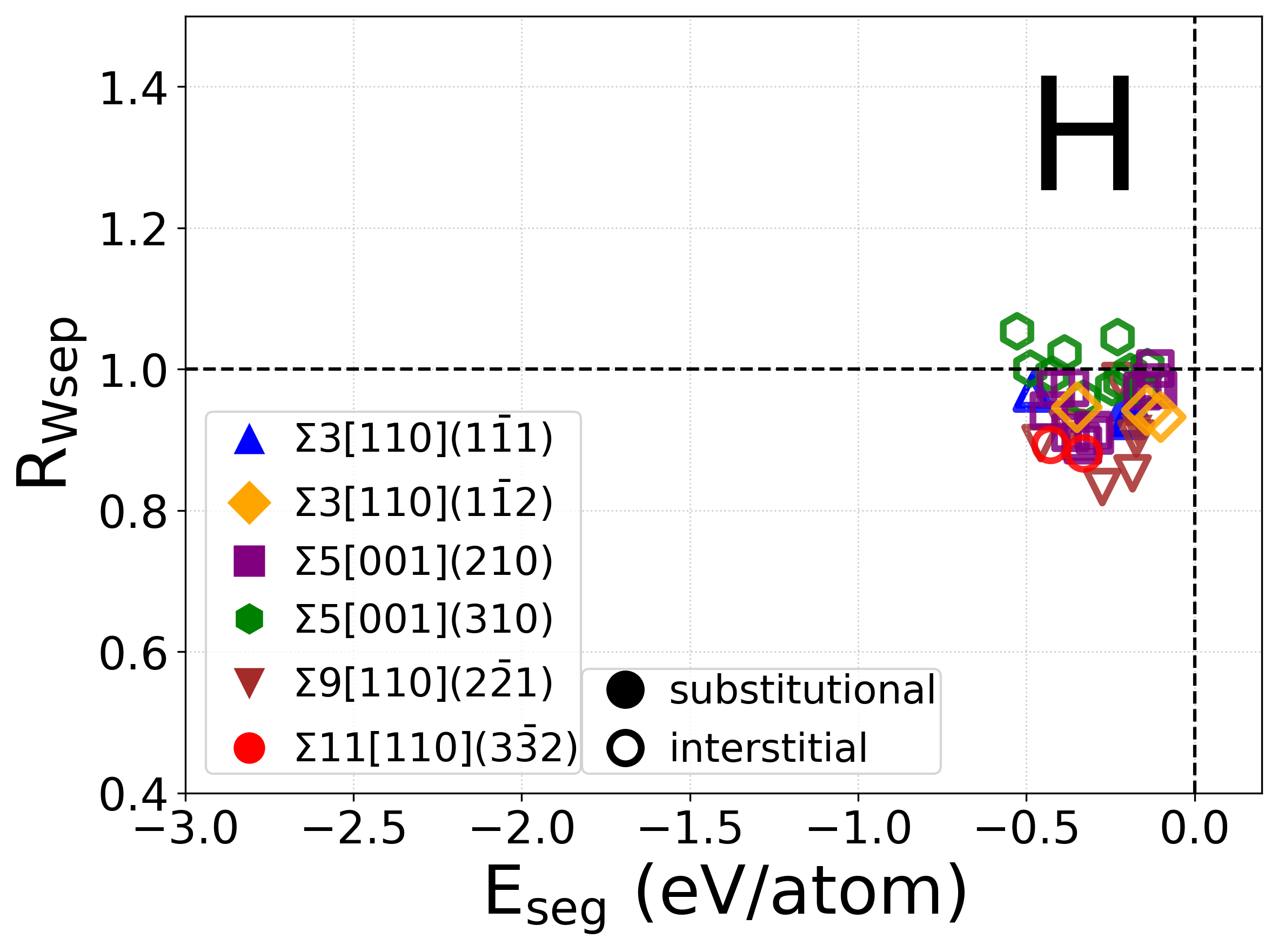}
		\caption{}
		\label{fig:SegEng_Wsep_H}
	\end{subfigure}\hfill
	\begin{subfigure}{0.48\linewidth}
		\centering
		\includegraphics[width=\linewidth]{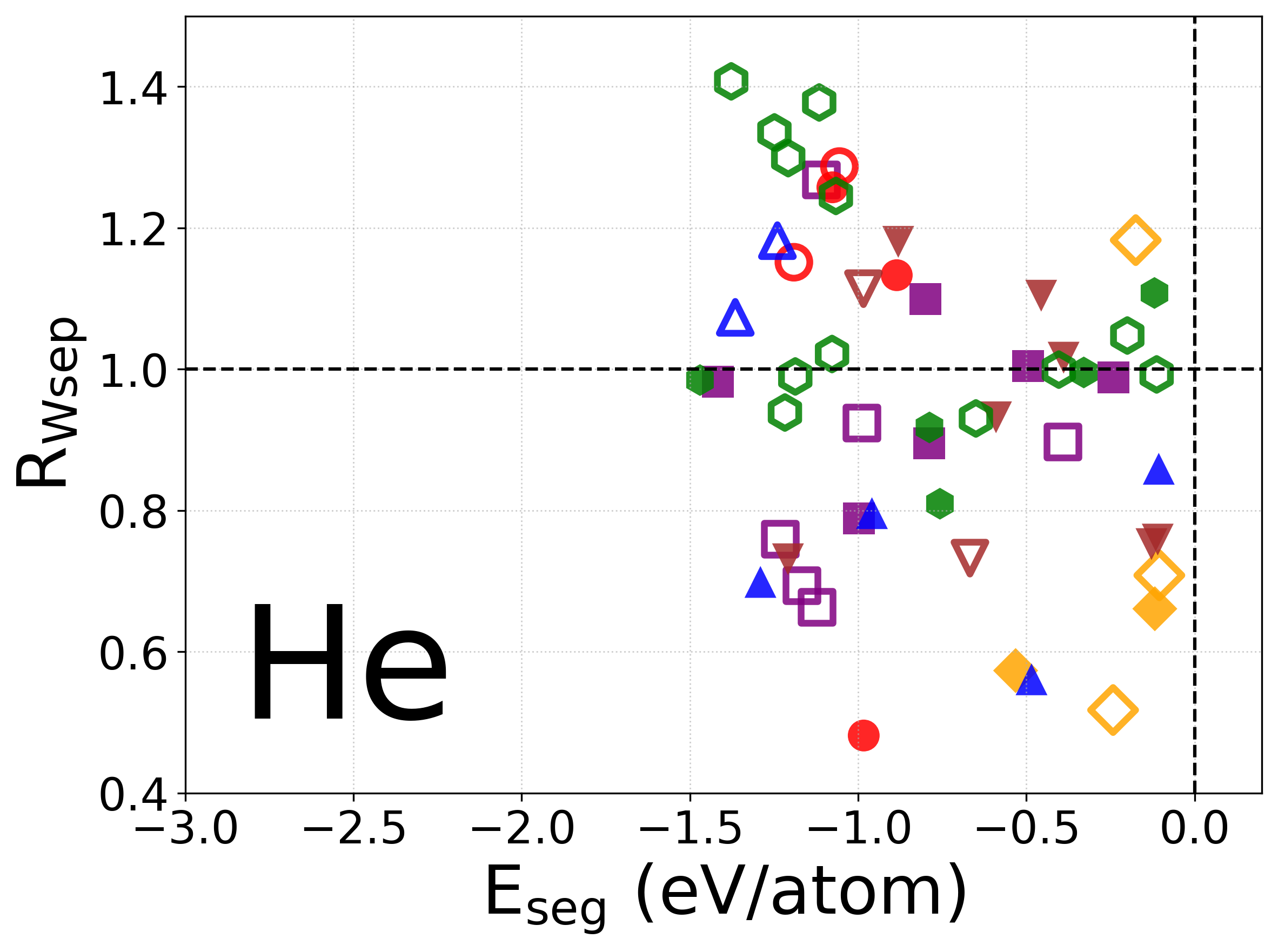}
		\caption{}
		\label{fig:SegEng_Wsep_He}
	\end{subfigure}
	\begin{subfigure}{0.48\linewidth}
		\centering
		\includegraphics[width=\linewidth]{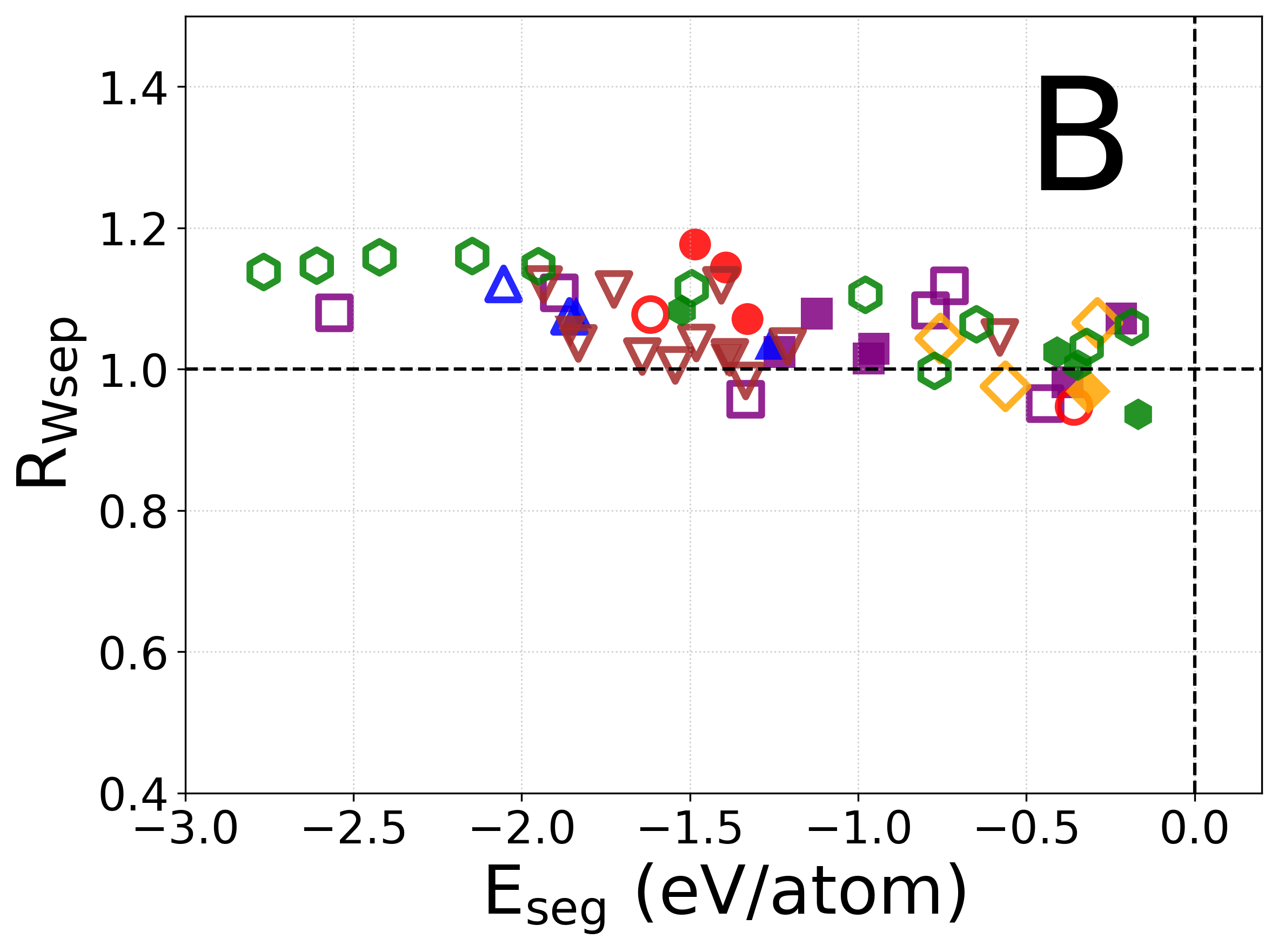}
		\caption{}
		\label{fig:SegEng_Wsep_B}
	\end{subfigure}\hfill
	\begin{subfigure}{0.48\linewidth}
		\centering
		\includegraphics[width=\linewidth]{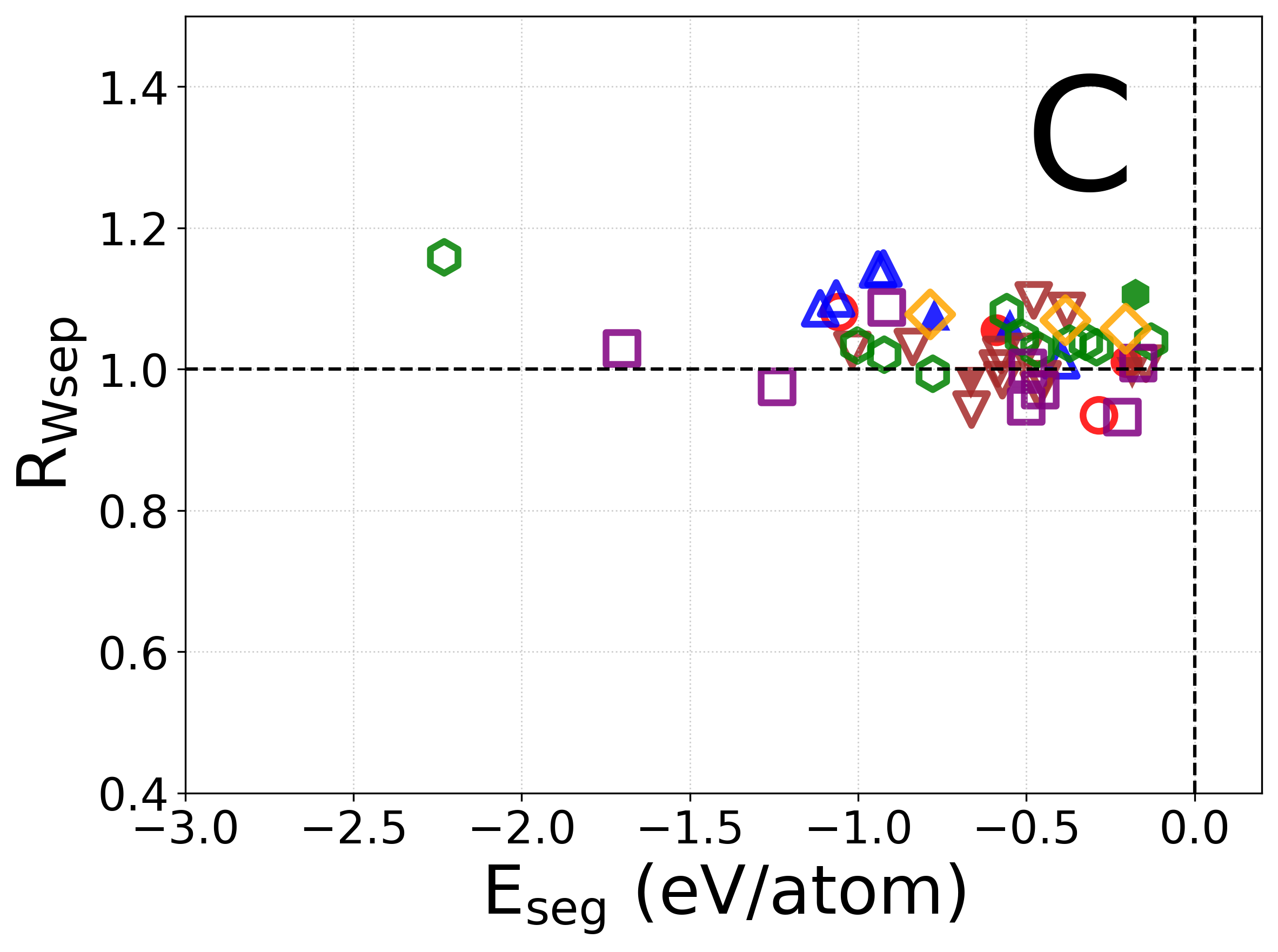}
		\caption{}
		\label{fig:SegEng_Wsep_C}
	\end{subfigure}	
	\begin{subfigure}{0.48\linewidth}
		\centering
		\includegraphics[width=\linewidth]{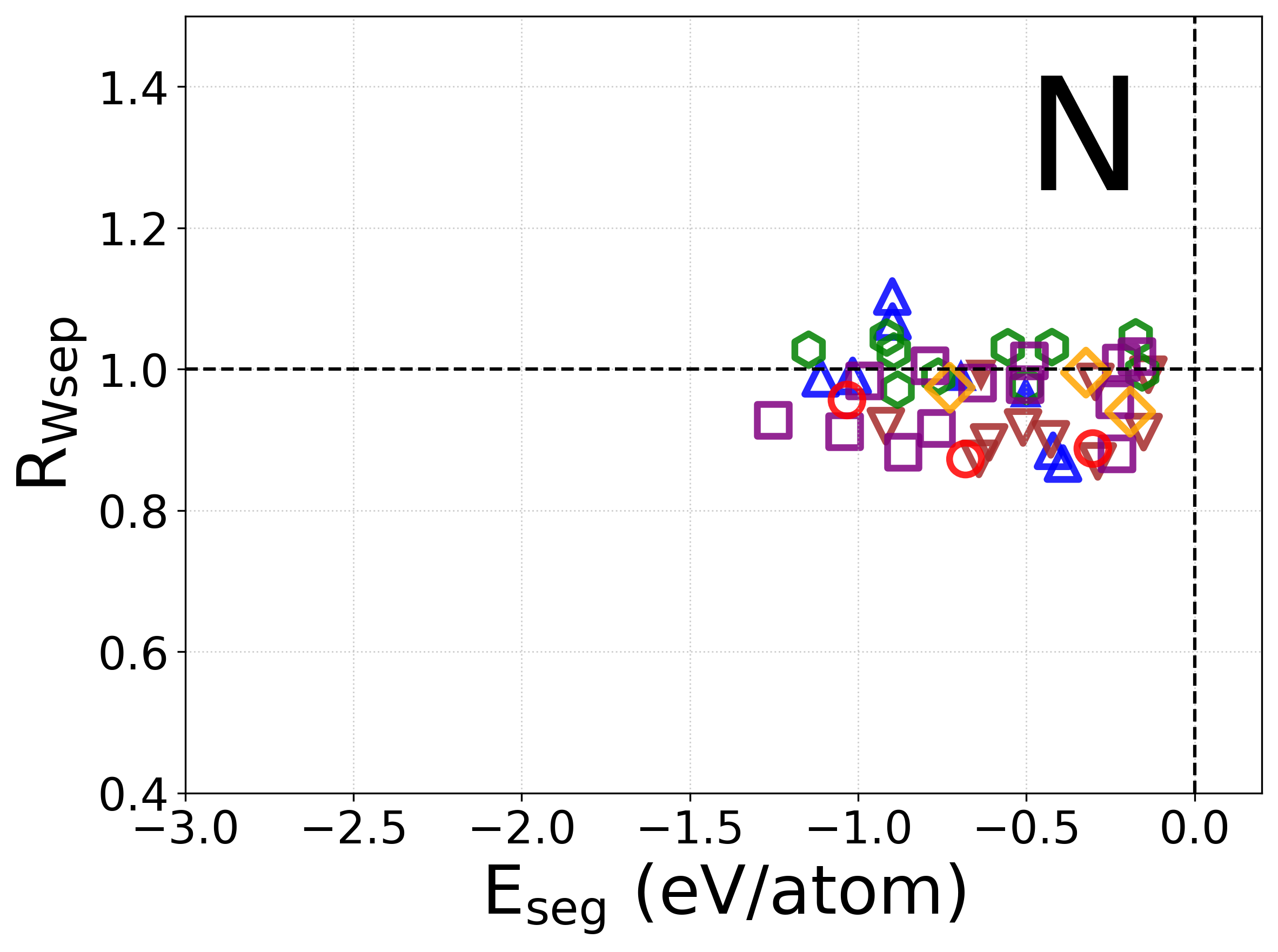}
		\caption{}
		\label{fig:SegEng_Wsep_N}
	\end{subfigure}\hfill
	\begin{subfigure}{0.48\linewidth}
		\centering
		\includegraphics[width=\linewidth]{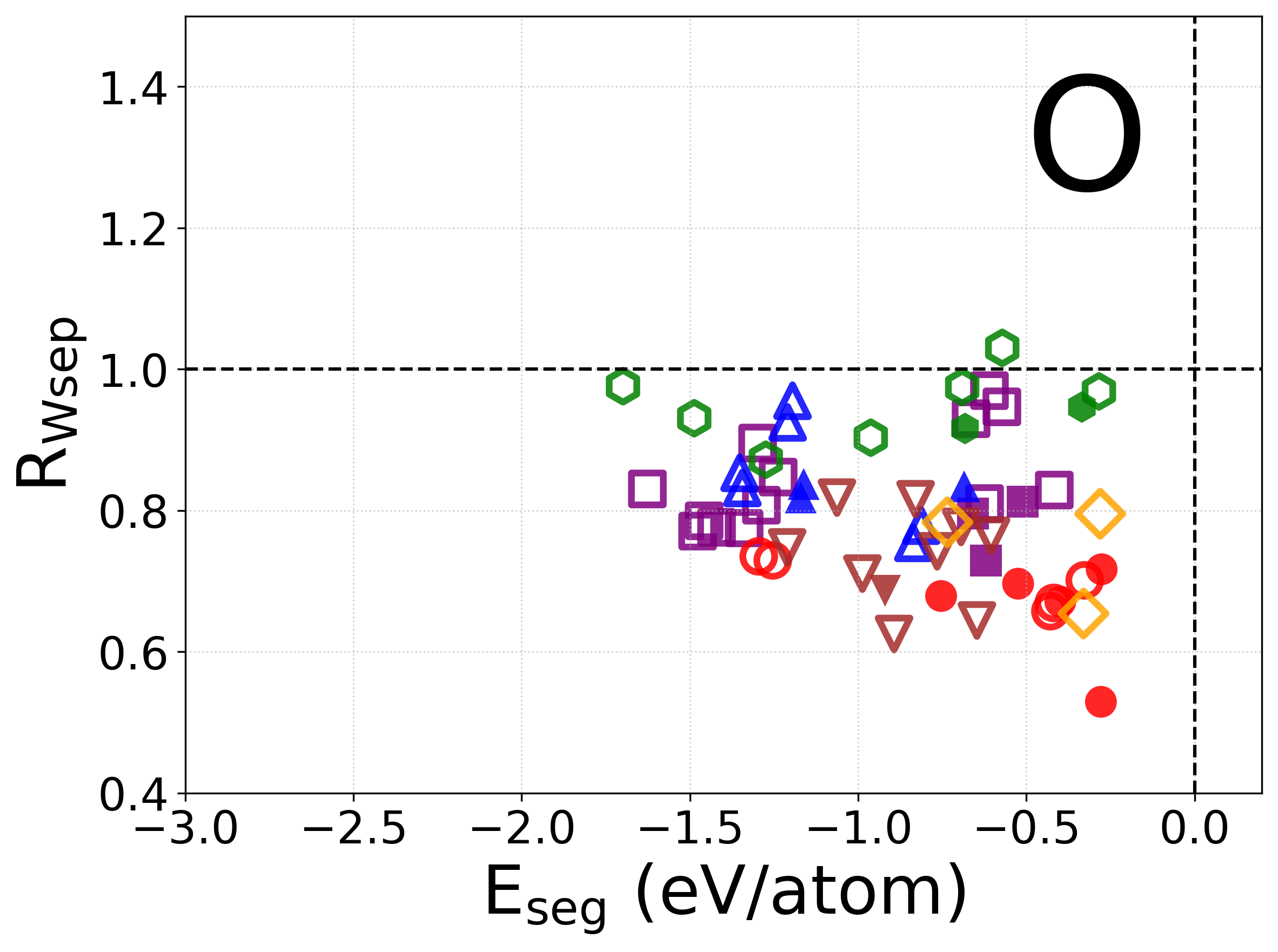}
		\caption{}
		\label{fig:SegEng_Wsep_O}
	\end{subfigure}
	\begin{subfigure}{0.48\linewidth}
		\centering
		\includegraphics[width=\linewidth]{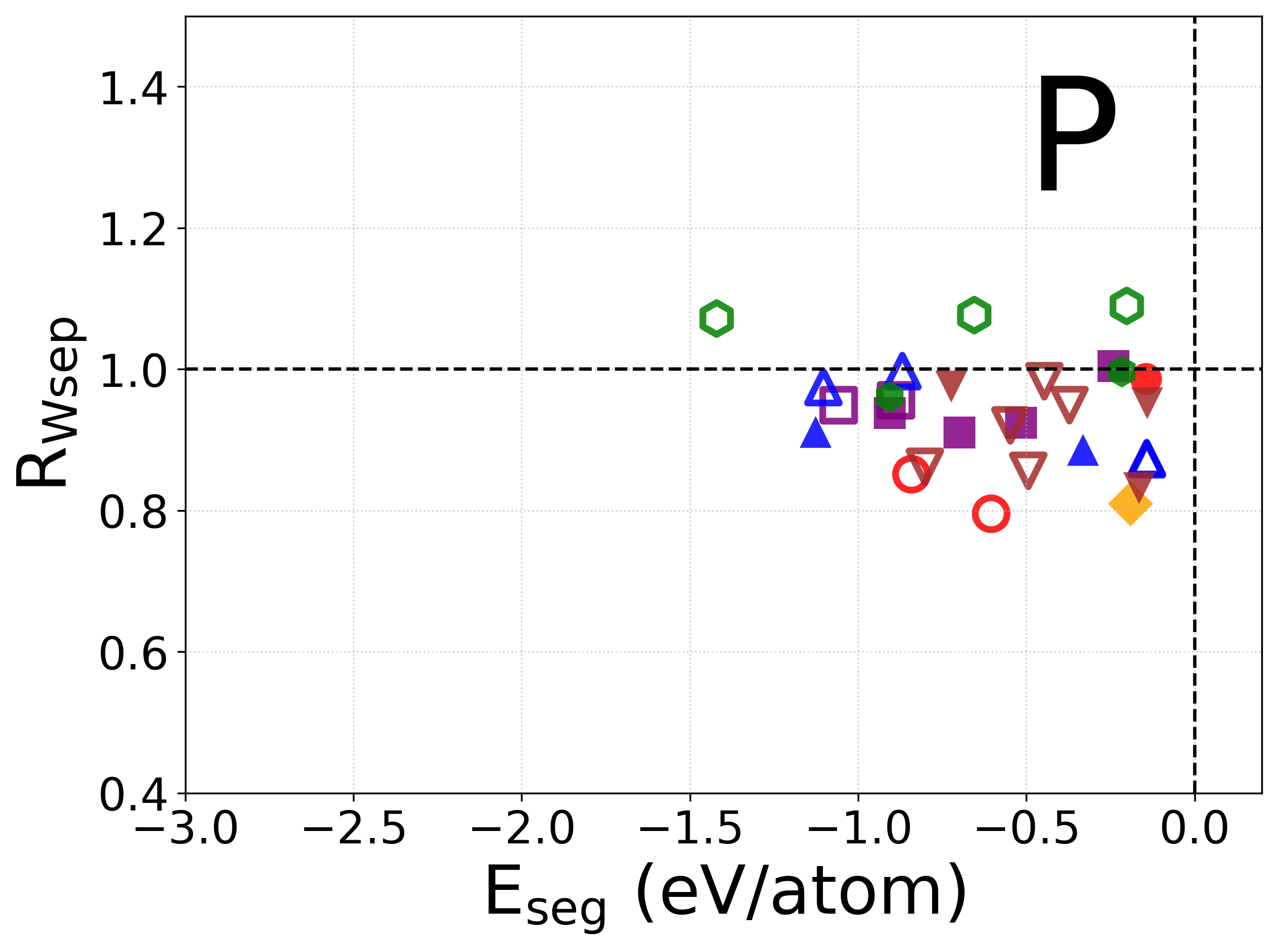}
		\caption{}
		\label{fig:SegEng_Wsep_P}
	\end{subfigure}\hfill
	\begin{subfigure}{0.48\linewidth}
		\centering
		\includegraphics[width=\linewidth]{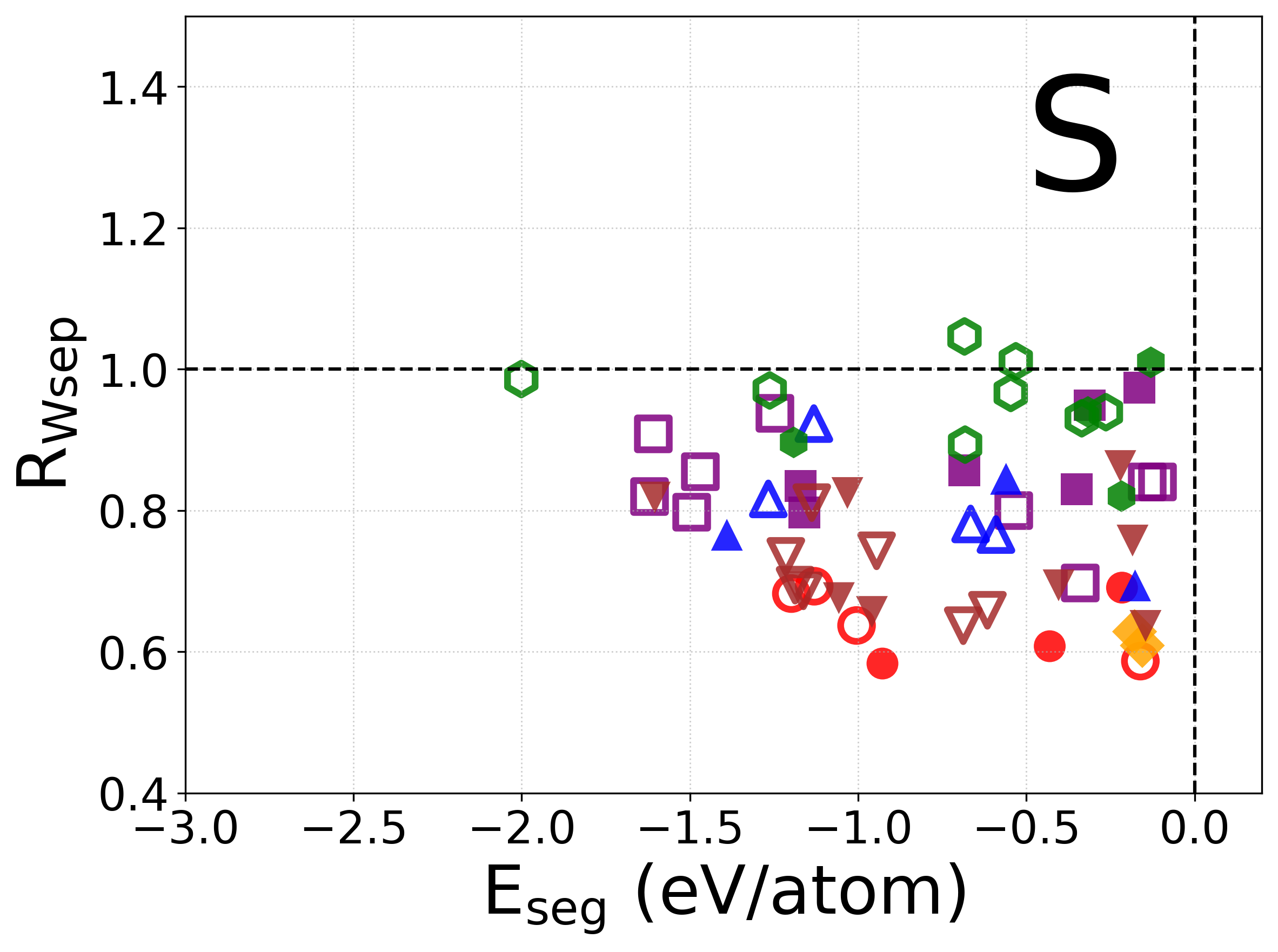}
		\caption{}
		\label{fig:SegEng_Wsep_S}
	\end{subfigure}
	
	\caption{Segregation engineering maps (segregation binding strength vs.\ GB cohesion modifier) for solutes in the same GB: (\ref{fig:SegEng_Wsep_H}) H; (\ref{fig:SegEng_Wsep_He}) He; (\ref{fig:SegEng_Wsep_B}) B; (\ref{fig:SegEng_Wsep_C}) C; (\ref{fig:SegEng_Wsep_N}) N; (\ref{fig:SegEng_Wsep_O}) O; (\ref{fig:SegEng_Wsep_P}) P; (\ref{fig:SegEng_Wsep_S}) S. Each panel shows the minimum segregation energy \(\min(E_{\rm seg})\) plotted against the cohesive strength modifier \(\eta\) at that element’s preferred site.}
	\label{fig:SegregationEngineering_Wsep_All}
\end{figure}

\begin{figure}[h!]
	\centering
	\begin{subfigure}{0.48\linewidth}
		\centering
		\includegraphics[width=\linewidth]{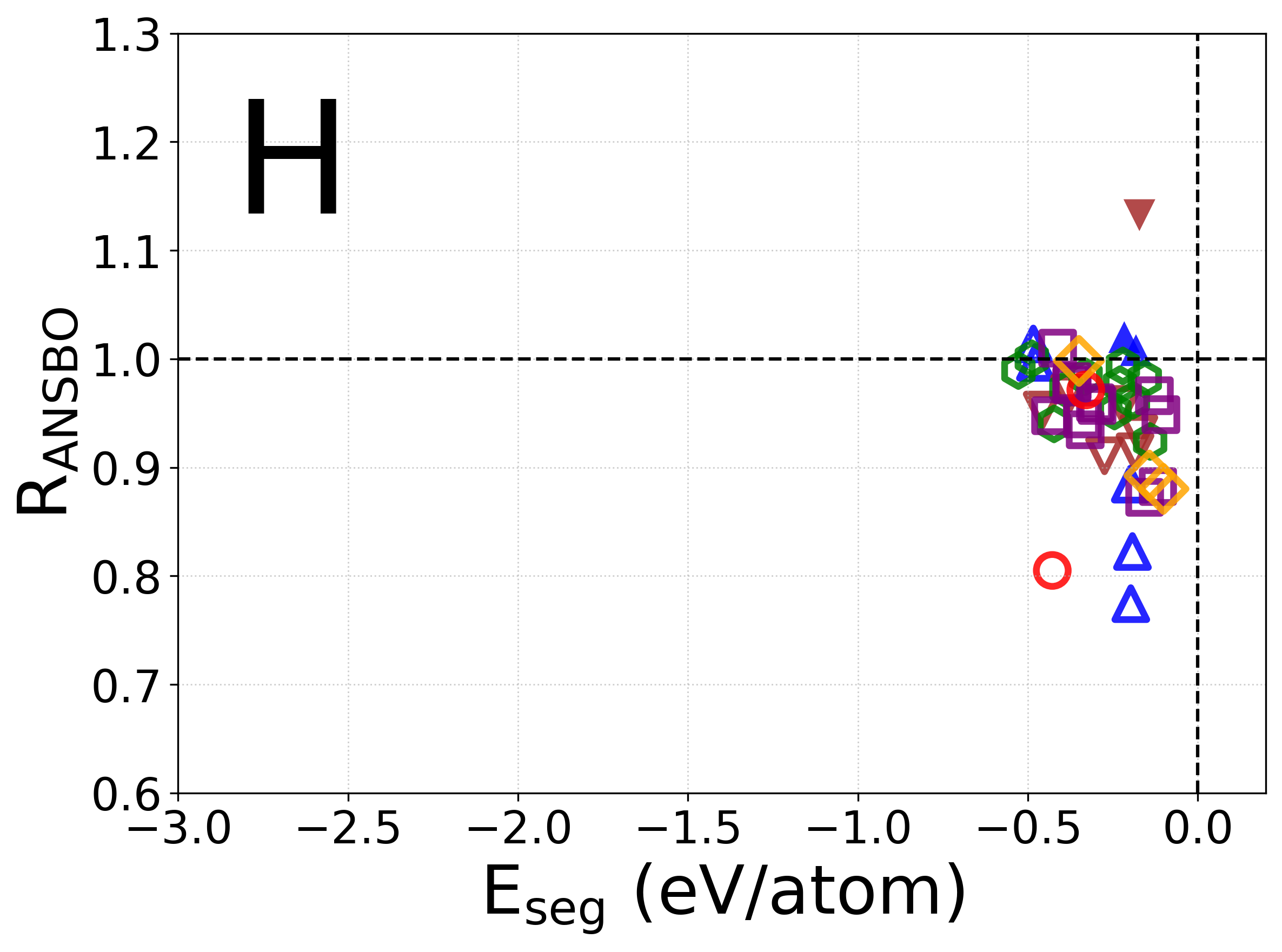}
		\caption{}
		\label{fig:SegEng_ANSBO_H}
	\end{subfigure}\hfill
	\begin{subfigure}{0.48\linewidth}
		\centering
		\includegraphics[width=\linewidth]{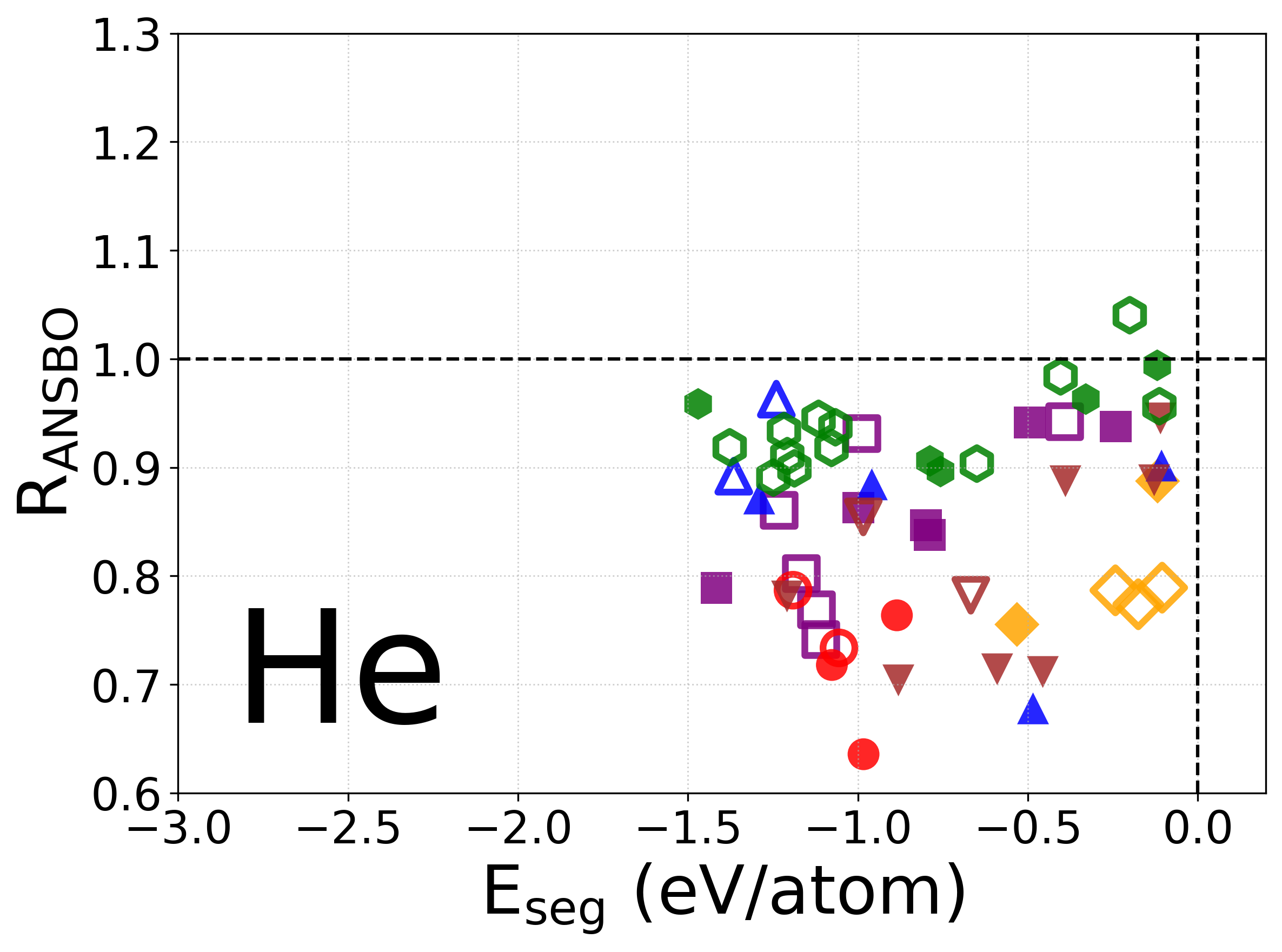}
		\caption{}
		\label{fig:SegEng_ANSBO_He}
	\end{subfigure}
	\begin{subfigure}{0.48\linewidth}
		\centering
		\includegraphics[width=\linewidth]{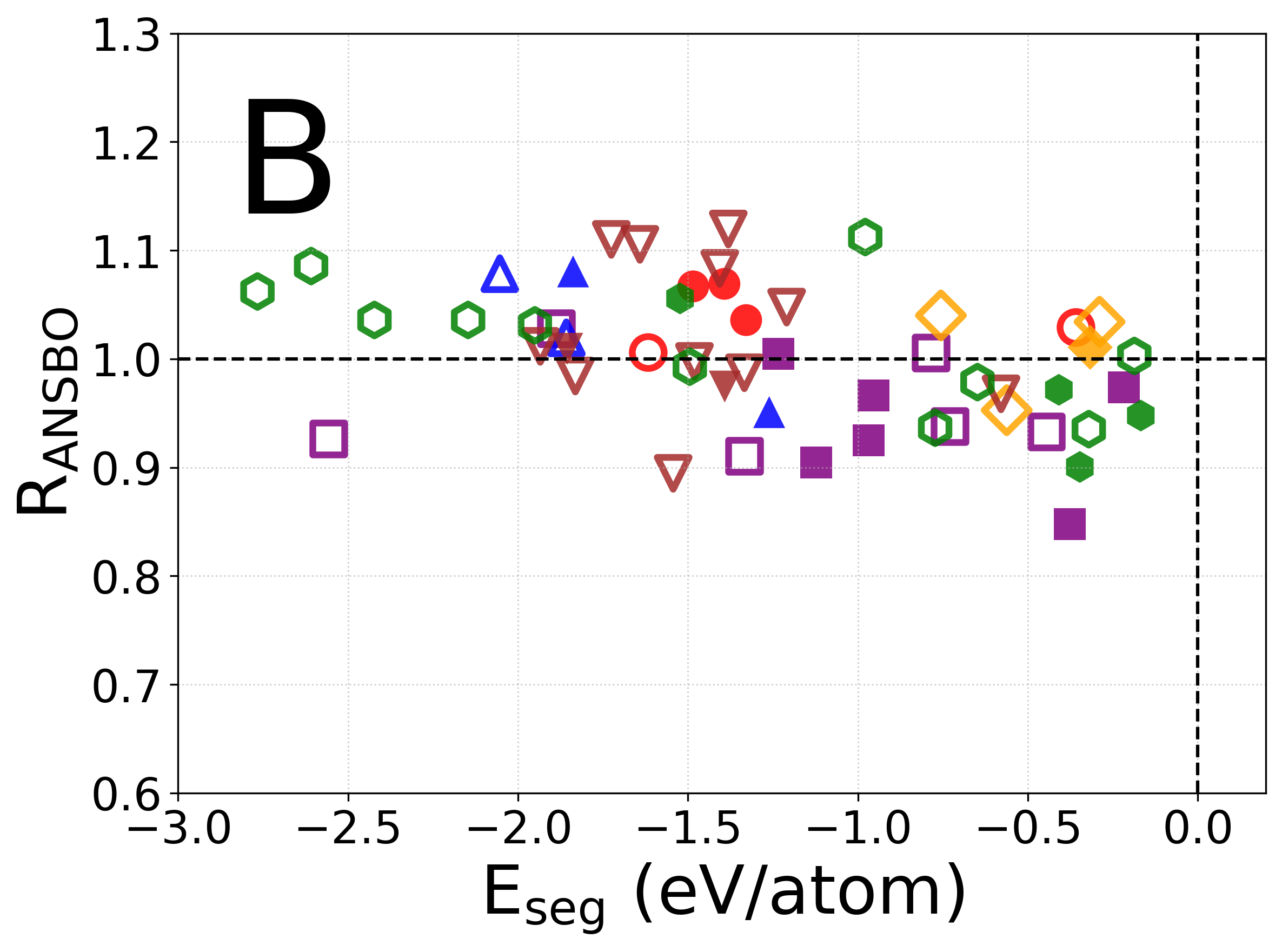}
		\caption{}
		\label{fig:SegEng_ANSBO_B}
	\end{subfigure}\hfill
	\begin{subfigure}{0.48\linewidth}
		\centering
		\includegraphics[width=\linewidth]{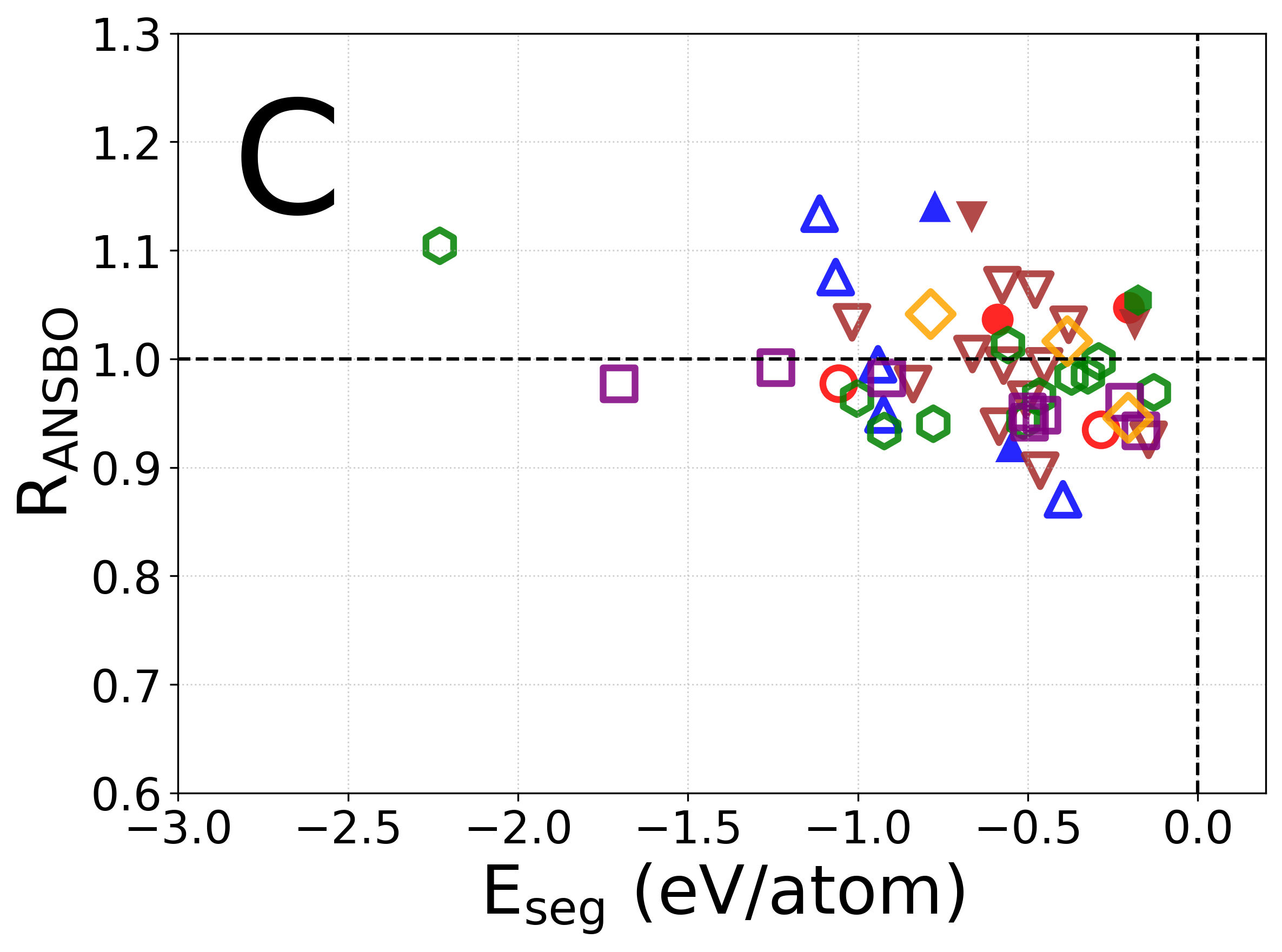}
		\caption{}
		\label{fig:SegEng_ANSBO_C}
	\end{subfigure}	
	\begin{subfigure}{0.48\linewidth}
		\centering
		\includegraphics[width=\linewidth]{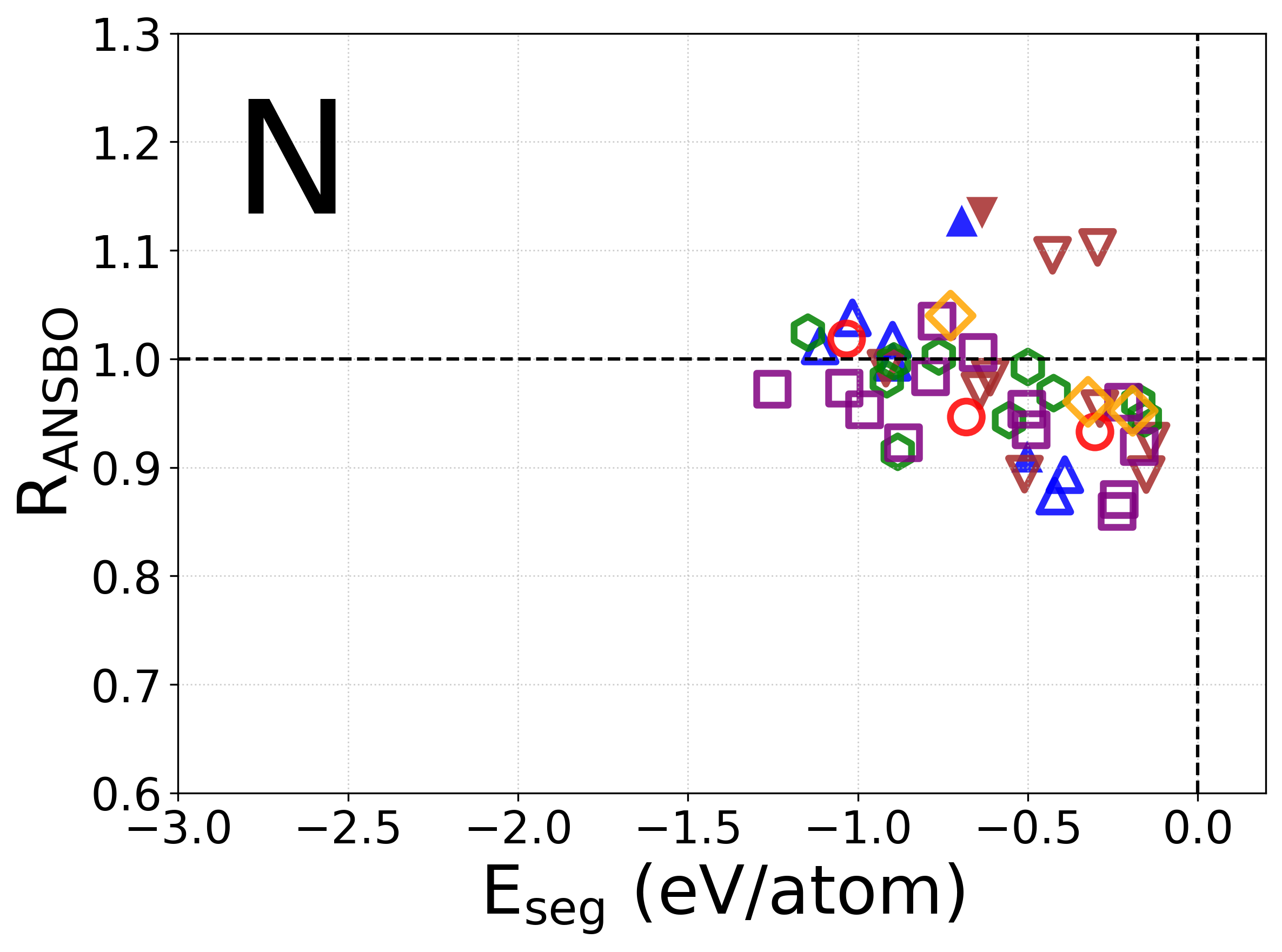}
		\caption{}
		\label{fig:SegEng_ANSBO_N}
	\end{subfigure}\hfill
	\begin{subfigure}{0.48\linewidth}
		\centering
		\includegraphics[width=\linewidth]{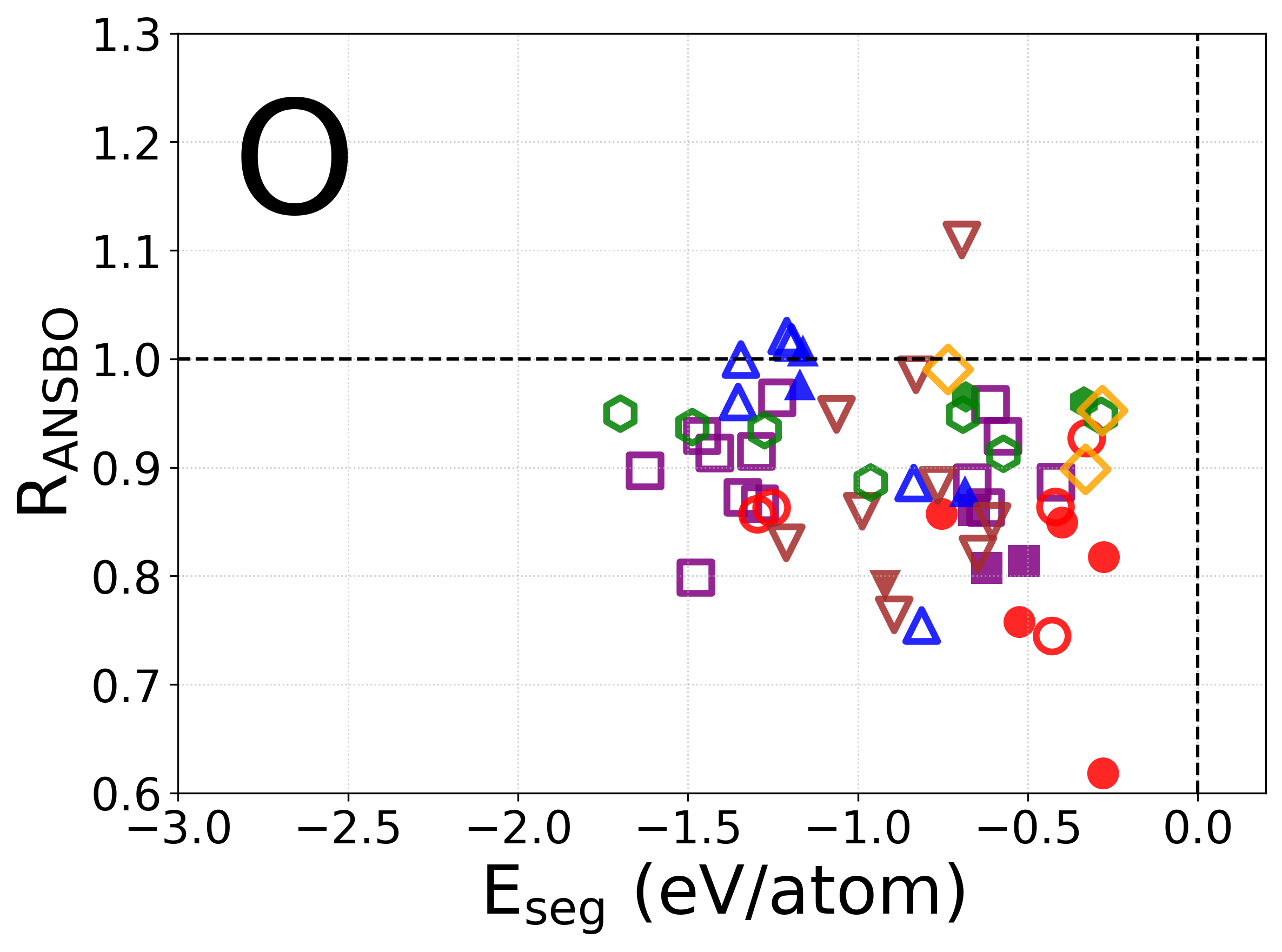}
		\caption{}
		\label{fig:SegEng_ANSBO_O}
	\end{subfigure}
	\begin{subfigure}{0.48\linewidth}
		\centering
		\includegraphics[width=\linewidth]{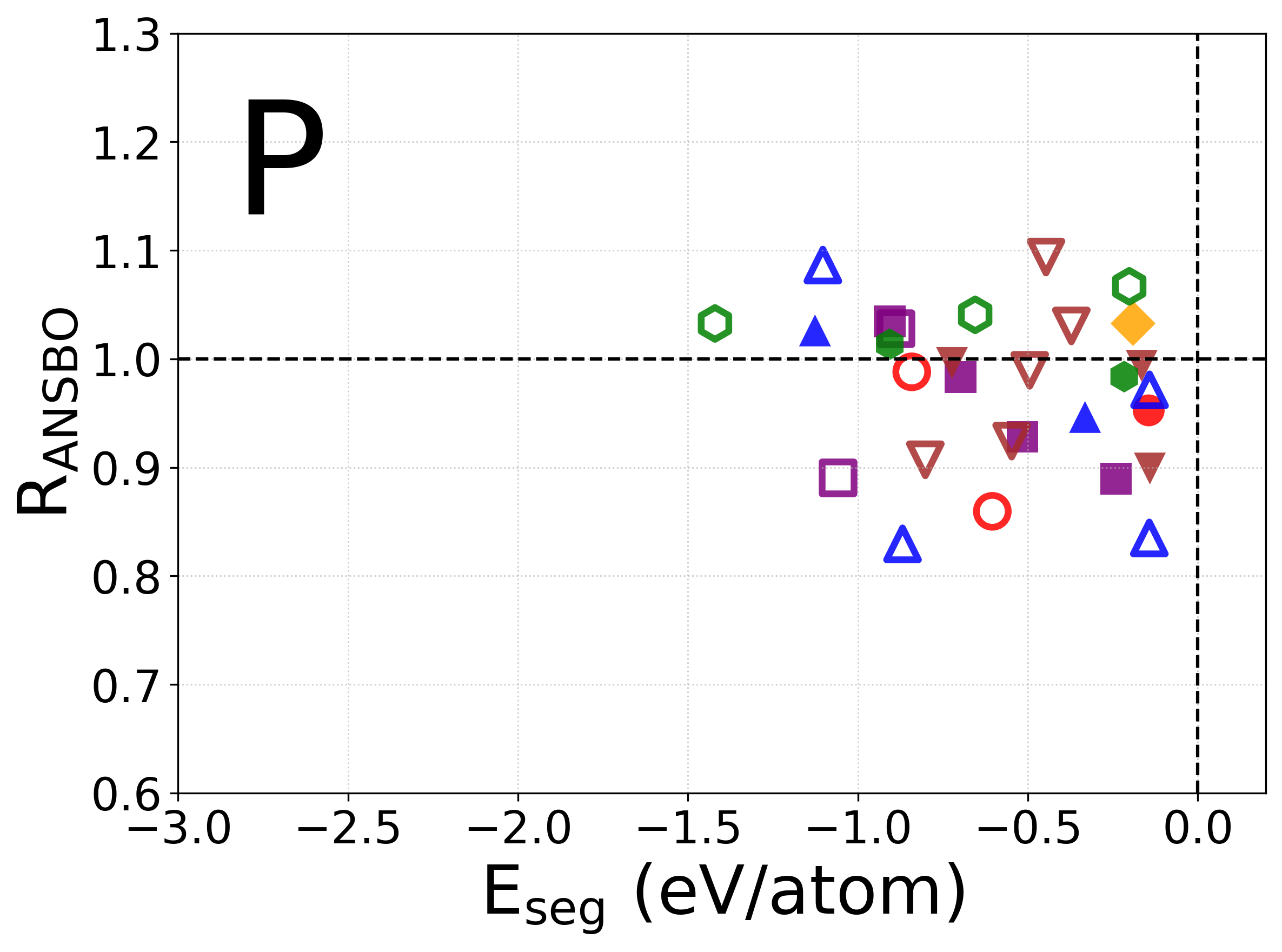}
		\caption{}
		\label{fig:SegEng_ANSBO_P}
	\end{subfigure}\hfill
	\begin{subfigure}{0.48\linewidth}
		\centering
		\includegraphics[width=\linewidth]{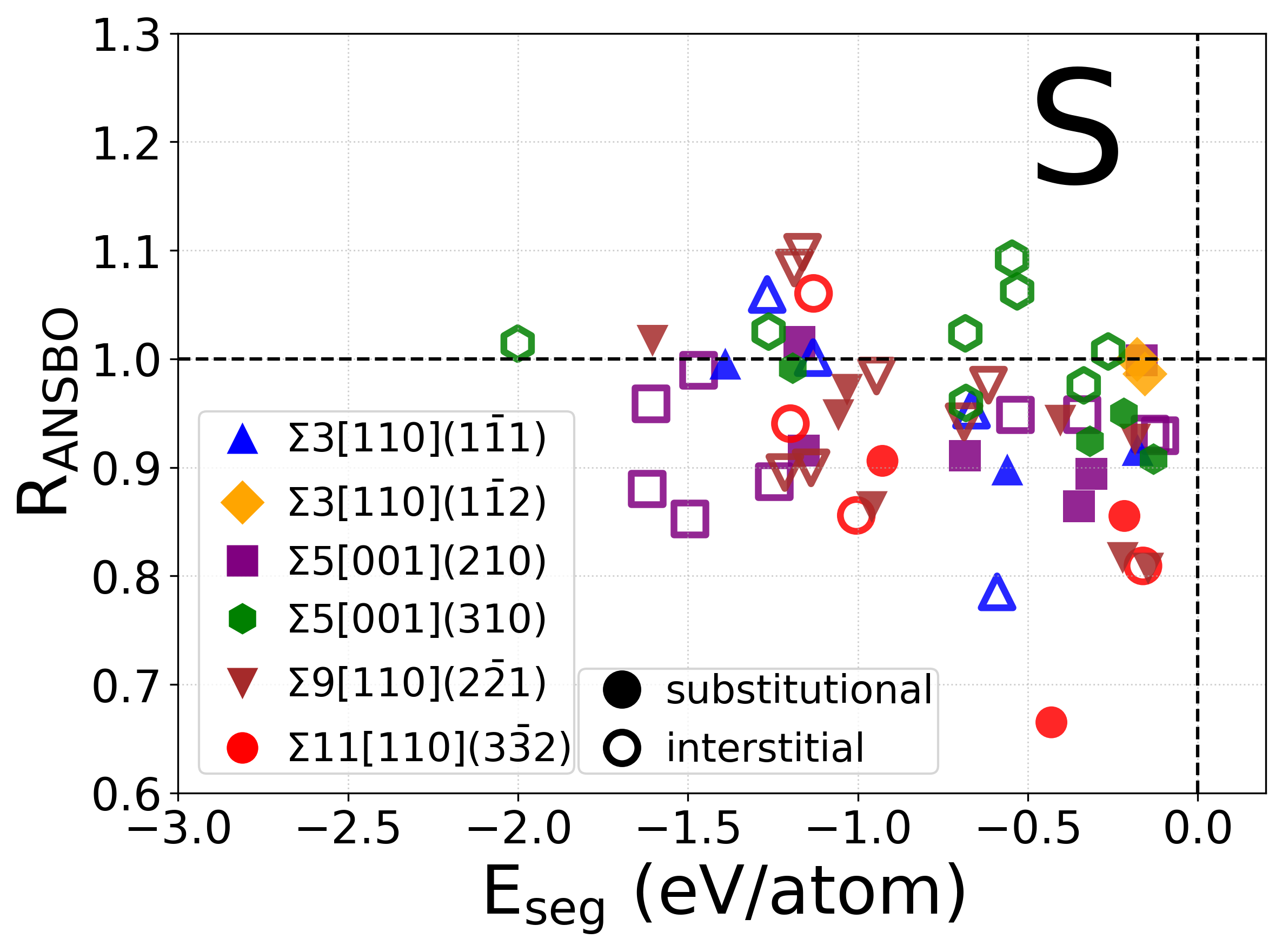}
		\caption{}
		\label{fig:SegEng_ANSBO_S}
	\end{subfigure}
	
	\caption{Segregation engineering maps (segregation binding strength vs.\ GB cohesion modifier) for solutes in the same GB: (\ref{fig:SegEng_ANSBO_H}) H; (\ref{fig:SegEng_ANSBO_He}) He; (\ref{fig:SegEng_ANSBO_B}) B; (\ref{fig:SegEng_ANSBO_C}) C; (\ref{fig:SegEng_ANSBO_N}) N; (\ref{fig:SegEng_ANSBO_O}) O; (\ref{fig:SegEng_ANSBO_P}) P; (\ref{fig:SegEng_ANSBO_S}) S. Each panel shows the minimum segregation energy \(\min(E_{\rm seg})\) plotted against the cohesive strength modifier \(\eta\) at that element’s preferred site.}
	\label{fig:SegregationEngineering_ANSBO_All}
\end{figure}

For understanding the behaviour of elements for use in GB engineering, it is useful to plot the cohesive effect of an element at the interface against the likelihood of the element segregating at the site. Such plots are presented in Figs. \ref{fig:SegregationEngineering_Wsep_All}, \ref{fig:SegregationEngineering_ANSBO_All} for the segregation energy Rice-Wang work of separation and ANSBO cohesion frameworks, respectively. Here, it is important to note that our computed rigid Rice-Wang work of separation quantities generally do not have comparable quantities to most of the literature which have used the relaxed surface formulation.
\\\\
In terms of elemental effects on interfacial cohesion, we find that O, S and He are the most potent decohesion agents at Fe GBs; P, N and H are milder decohesive agents in comparison. B is a strengthener of GB cohesion; C is comparatively a milder strengthening agent on GB cohesion. We find that S has a stronger segregation binding at Fe GBs compared to P, and it also enacts a more potent deleterious effect on GB cohesion. As such, we expect elemental S to be a much more potent embrittler of GBs compared to P. The anomalous strengthening effect predicted by Rice-Wang is discussed below - here we take the ANSBO value in our discussion of relative elemental cohesive effects. 
\\\\

Here we note the somewhat peculiar results that can be observed for He in the Rice-Wang framework. He is a potent decohesion agent since it is a noble gas that does not typically form bonds, and hence should enact decohesion at the GB. However, its effects are not accurately reflected, as observed in Fig. \ref{fig:SegEng_Wsep_He}. One should be cautious when interpreting results for this as Rice-Wang theory is generally used underneath the assumption that thermodynamical equilibrium conditions are present. When He is present in steels, this is likely never the case, as it tends to be kinetically trapped in irradiated environments, violating the assumption of thermodynamic equilibrium in the Rice-Wang theory. The strange unphysical strengthening results observed are due to the occurrence of high-energy states on cleaved surfaces for He, as has been discussed by others \cite{mcmahonTheoryEmbrittlementSteels1978, briantChemistryGrainBoundary1990}, as well as in our prior work \cite{maiHighthroughputInitioStudy2025a}. Its decohesive effect is clearly visible in Fig. \ref{fig:SegEng_ANSBO_He}, when we compute cohesion in the bonding-based framework.
\\\\
In Figs. \ref{fig:SegregationEngineering_Wsep_All}, \ref{fig:SegregationEngineering_ANSBO_All}, there are some features of interest for interface engineering. Note that the $\Sigma5[001](310)$ (green markers) is generally the most resistant to decohesive elements, with the smallest drop in cohesion observed relative to other GBs. This is consistent with the results of our prior study which studied substitutional sites for most elements in the periodic table \cite{maiHighthroughputInitioStudy2025a}, indicating that this resistance to decohesion by certain GBs is insensitive to the type of site sampling strategy. On the other hand, there are also certain GB archetypes which are comparatively much more sensitive to decohesive effects enacted by segregants, e.g. the twin $\Sigma3[1\bar{1}0](112)$ GB, also consistent with the results observed prior for substitutional segregants. 

\subsection{Site type distinction: substitutional vs.\ interstitial}
\label{sec:site_type_distinction}
\begin{figure}[h!]
	\centering
	\includegraphics[height=8cm]{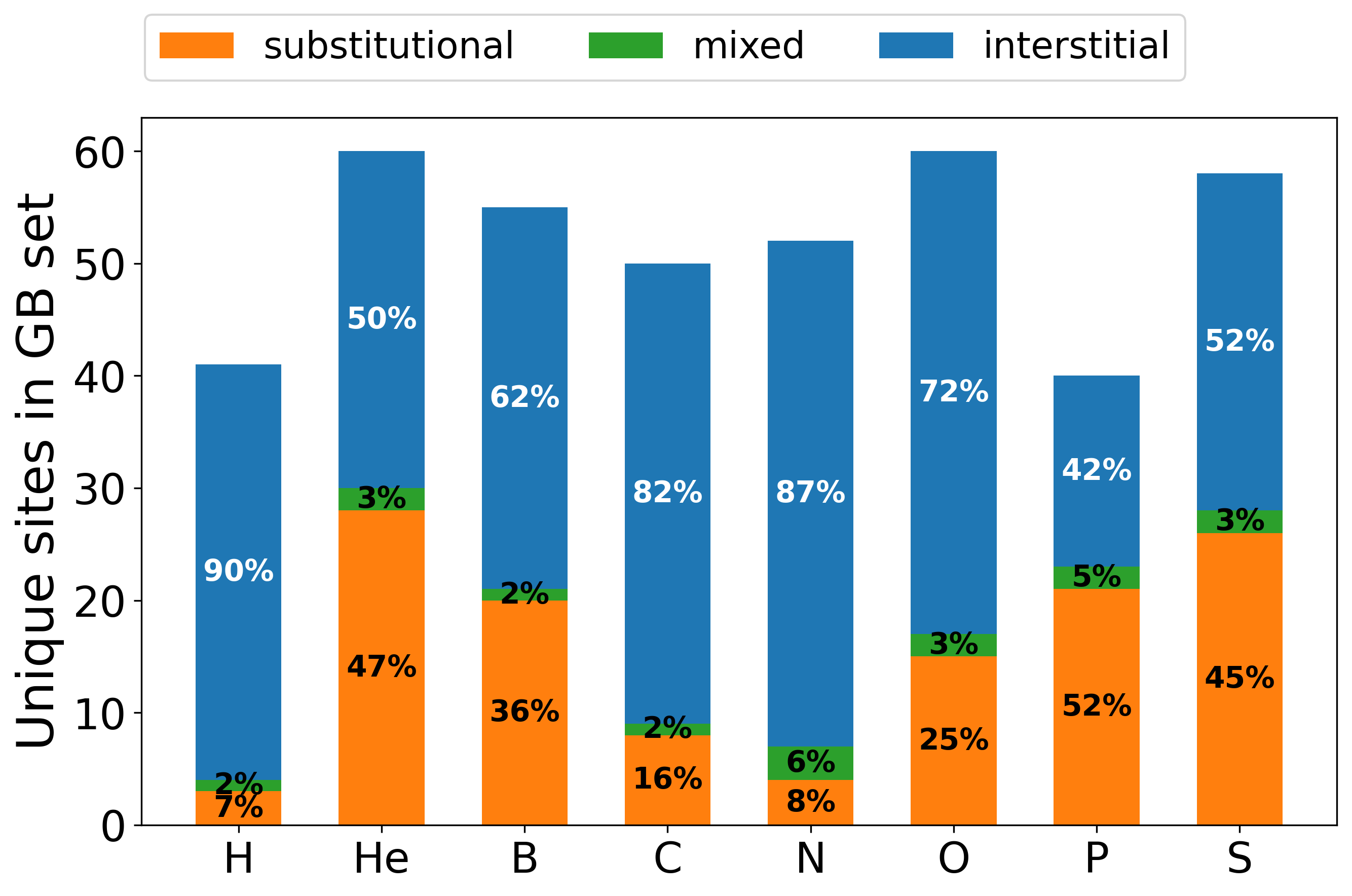}
	\caption{A numerical breakdown of the types of unique, energetically favourable sites (E$_\text{seg} < -0.1$ eV) for each element in the GB set considered in our study. A total of 416 unique site environments across all elements were retained after our duplicate removal process. The individual colours in each bar indicate the starting position of that site in our GB set (substitutional = orange, interstitial = blue, mixed = green). "Mixed" sites were termed as such when duplicates of the unique site were found to contain sites which started from the other type of starting site (i.e. both "substitutional" and "interstitial" sites relaxed into the same final structure).}
	\label{fig:site_classification}
\end{figure}

Much time has been devoted to efforts in classifying and discussing the so-called interstitial and substitutional site positions at GBs \cite{cernySegregationPhosphorusSilicon2024, wachowiczEffectImpuritiesStructural2011, wagihSpectrumInterstitialSolute2023, reiners-sakicInterstitialsKeyIngredient2025, mengHighlyTransferableEfficient2024a, mengNeuralNetworkInteratomic2025, lejcekInterstitialSubstitutionalSolute2016}. Generally, it is assumed that light elements occupy the so-called "voids" at GBs. In our study, we have surveyed these classes of sites over the 6 model GBs, comprehensively sampling the available site environments in these GBs. Our results indicate that such distinctions in starting site characterisation are not generally useful in predicting site characteristics, like their resultant final segregation energies or cohesive effects for the light elements. 
\\\\
Here, our data enables us to investigate the nature and distribution of site types for segregation of the light elements. We have plotted the histograms for the distribution of segregation energies across the unique sites in each GB in our study in Fig. \ref{fig:Eseg_histograms}. For easier visualisation of the site type makeup of energetically favourable unique sites, we have also plotted the percentages of each site type in the overall distribution of unique sites for each element in Fig. \ref{fig:site_classification}. 
\\\\
\\\\
We can observe that while starting "interstitial" sites form the majority of the favourable sites for the elements of H, B, C, N, O, "substitutional" sites form a significant part of the distribution for the elements of He, B, P, S. Up to half of the unique sites are substitutional for both P/S/He. Note the presence of "mixed" sites which are sites which possess the same structure after relaxation from either kind of starting site. This is significant as it implies that there are sites which are accessible to either starting configuration which are not easily classified into either "substitutional" or "interstitial", but rather exist in a more ambiguous space. As such, there is likely no true \textit{significant} binary distinction between the two classes of sites at more random environments, i.e. in random GBs. 
\\\\
Our results therefore demonstrate that any attempt to comprehensively survey the spectra of segregation energies for light elements, or any elements that ostensibly occupy interstitial sites, necessitates the employment of sampling strategies that survey both kinds of sites systematically. In this study, this is achieved via the substitutional + Voronoi site sampling strategy. We recommend that practitioners in future studies employ such strategies to ensure a complete picture of the segregation behaviour of these elements.
\\\\
Only in the elements of H, B, C, N, O are interstitial starting positions clearly energetically preferred for segregation [Figs. \ref{fig:Eseg_histograms} and \ref{fig:SegregationEngineering_Wsep_All}]. However, segregation at substitutional sites is energetically significant and makes up a substantial part of their segregation binding spectra, with the exception of H. He, P, S exhibit much more ambivalent behaviour with respect to preferred binding sites. Often, their induced cohesive effects across either kind of starting site can be quite similar.
\\\\
Beyond the importance of reporting site classifications in the reproduction of previously computed results, we argue that its relevance is effectively meaningless in the context of true polycrystalline materials, where most GB sites are expected to be much more amorphous/anisotropic than the highly symmetric GBs studied here. Here we have shown that ambiguity is already present in these highly symmetric GBs, with even "obvious" interstitials such as O, N, C yielding sites with substantial segregation binding at "substitutional" starting positions. Thus, distinguishing between "substitutional" and "interstitial" sites at these highly random site environments at GBs does not provide new insight and is often misleading. This is further supported by our similarity filtering approach which finds that many so-called substitutional sites relax into the same final states as interstitials and vice-versa. 

\section{Conclusion}
We have studied the segregation behaviours for the light elements: H, He, B, C, N, O, P and S, which are light elements that are technologically relevant solutes and/or impurities for steels. Our work demonstrates that \textit{complexity} is intrinsic to segregation, and that common assumptions in the design of previous computational studies can lead to incomplete descriptions of the phenomena. In terms of relative segregation trapping at GBs, we find that B$>$O, S$>$C, He$>$P, N$>$H. The results of our cohesion investigation indicate that B is a potent GB cohesion enhancer, C is a comparatively milder one, N, P, H are mild decohesion agents, and O, He and S are potent decohesive agents.
\\\\
In comprehensively surveying both the interstitial and substitutional sites in 6 carefully selected CSL GBs, we demonstrate that both kinds of starting sites can yield distinct but energetically favourable relaxations for these solutes. We find that relaxation of interstitial starting positions can yield substitutional-like structures with similar energies, and vice versa. As distinctions between these sites are shown to be already somewhat ambiguous in these highly symmetric GBs, we argue that they should not be meaningful in the more general case, where structural disorder is larger, e.g. in general GBs. This implies that thorough sampling strategies that make fewer assumptions on the nature of the starting sites are likely necessary to accurately capture the spectra of these elements in polycrystalline environments.
\\\\
We show that GB energy is not correlated with the strength of segregation trapping for these light elements in ferritic iron. Furthermore, we show that the strength of their segregation trapping is not correlated with the available volume at the host site, but rather its ability to maximise the nearest-neighbour bond lengths. It may be understood as the ability of the local site environment to absorb strain energy of the shortest bonds, its "softness". Rigid atomic environments that cannot deform to maximise nearest neighbour distances for the segregating atom, tend to be correspondingly weaker traps for the light elements. 
\\\\
It is envisioned that the present work, combined with our previous study \cite{maiHighthroughputInitioStudy2025a}, will serve as a foundational basis for understanding segregation and cohesion in ferritic steels. The studies effectively provide a complete and thorough treatment of binary segregation and cohesion in ferromagnetic bcc Fe GBs to the extent reasonably allowed at present by pure ab-initio techniques. The data from these studies are freely available for download, rendering them FAIR and a suitable starting point for more complex case studies of segregation, i.e. ternary segregation and beyond, or for benchmarking/fine-tuning of MLIPs. The insights in this study enabled by our extensive data have allowed a comprehensive reexamination of old assumptions on the nature of segregation and revealed new aspects and complexities for future investigation. 
\clearpage
\section{Supplementary information and data availability}
The complete DFT dataset, analysis code, and figure generation scripts are available in an open-access GitHub repository at \url{https://github.com/ligerzero-ai/FeGB_LightEleSegregation}. The repository contains:
\begin{itemize}
	\item \textbf{DFT results:} All segregation, cleavage, and bond-order data stored as pandas DataFrames (\texttt{data/}), including interstitial and substitutional GB calculations, bulk references, cleaved-surface energies, and DDEC6 Chargemol bond orders.
	\item \textbf{Structures:} Relaxed pure GB structures and starting interstitial site positions in VASP POSCAR format (\texttt{structures/}).
	\item \textbf{Analysis scripts:} Segregation energy computation (\texttt{segregation.py}), Voronoi and SOAP featurisation (\texttt{featurisers.py}), and publication plotting utilities (\texttt{plotters.py}). The full analysis pipeline, including the KP vs KS k-point comparison, is in \texttt{analysis/}.
	\item \textbf{Figure generation:} Partitioned scripts for reproducing all main-text figures (\texttt{scripts/MainFigures/}) and all Supplementary Information figures (\texttt{scripts/SupplementaryFigures/}).
\end{itemize}
The repository is available under the MIT license. 
\section{Pseudopotential details}
\begin{table}[htbp]
	\centering
	\caption{POTCAR pseudopotentials used for each element.}
	\label{tab:potcar}
	\begin{tabular}{lll}
		\hline
		Element & POTCAR & Valence Electrons \\
		\hline
		H  & PAW\_PBE H 15Jun2001  & 1 \\
		He & PAW\_PBE He 05Jan2001 & 2 \\
		Be & PAW\_PBE Be 06Sep2000 & 2 \\
		B  & PAW\_PBE B 06Sep2000  & 3 \\
		C  & PAW\_PBE C 08Apr2002  & 4 \\
		N  & PAW\_PBE N 08Apr2002  & 5 \\
		O  & PAW\_PBE O 08Apr2002  & 6 \\
		P  & PAW\_PBE P 06Sep2000  & 5 \\
		S  & PAW\_PBE S 06Sep2000  & 6 \\
		Fe & PAW\_PBE Fe 06Sep2000 & 8 \\
		\hline
	\end{tabular}
\end{table}

\section{Acknowledgements}
H.L. Mai, T. Hickel, J. Neugebauer acknowledge the support provided by the German Federal Ministry of Education and Research (BMBF) through the project Innovation-Platform MaterialDigital [Grant no. 13XP5094C]. T. Hickel, J. Neugebauer acknowledge financial support by the Deutsche Forschungsgemeinschaft (DFG) from CRC1394 “Structural and Chemical Atomic Complexity – From Defect Phase Diagrams to Material Properties”, project ID 409476157. This work was supported by computational resources provided by the Australian Government through the National Computational Infrastructure (Gadi) and the Pawsey Supercomputing Centre (Setonix) under the National Computational Merit Allocation Scheme. The Pawsey Supercomputing Centre is also supported by funding from the Government of Western Australia. Support and facilitation of our access to these compute resources from the Sydney Informatics Hub at the University of Sydney is gratefully acknowledged. S. Ringer acknowledges gratefully partial funding from the Australian Research Council.
\bibliography{references}

@article{azocarguzmanEffectsMechanicalStress2024,
  title = {Effects of Mechanical Stress, Chemical Potential, and Coverage on Hydrogen Solubility during Hydrogen-Enhanced Decohesion of Ferritic Steel Grain Boundaries: {{A}} First-Principles Study},
  shorttitle = {Effects of Mechanical Stress, Chemical Potential, and Coverage on Hydrogen Solubility during Hydrogen-Enhanced Decohesion of Ferritic Steel Grain Boundaries},
  author = {Az{\'o}car Guzm{\'a}n, Abril and Janisch, Rebecca},
  year = 2024,
  month = jul,
  journal = {Physical Review Materials},
  volume = {8},
  number = {7},
  pages = {073601},
  issn = {2475-9953},
  doi = {10.1103/PhysRevMaterials.8.073601},
  urldate = {2025-04-01},
  langid = {english}
}

@article{bechtleGrainboundaryEngineeringMarkedly2009,
  title = {Grain-Boundary Engineering Markedly Reduces Susceptibility to Intergranular Hydrogen Embrittlement in Metallic Materials},
  author = {Bechtle, S. and Kumar, M. and Somerday, B. P. and Launey, M. E. and Ritchie, R. O.},
  year = 2009,
  month = aug,
  journal = {Acta Materialia},
  volume = {57},
  number = {14},
  pages = {4148--4157},
  issn = {1359-6454},
  doi = {10.1016/j.actamat.2009.05.012},
  urldate = {2023-01-24},
  langid = {english}
}

@article{blochlProjectorAugmentedwaveMethod1994,
  title = {Projector Augmented-Wave Method},
  author = {Bl{\"o}chl, P. E.},
  year = 1994,
  month = dec,
  journal = {Physical Review B},
  volume = {50},
  number = {24},
  pages = {17953--17979},
  publisher = {American Physical Society},
  doi = {10.1103/PhysRevB.50.17953},
  urldate = {2023-01-24}
}

@article{briantChemistryGrainBoundary1990,
  title = {On the Chemistry of Grain Boundary Segregation and Grain Boundary Fracture},
  author = {Briant, C. L.},
  year = 1990,
  month = sep,
  journal = {Metallurgical Transactions A},
  volume = {21},
  number = {9},
  pages = {2339--2354},
  issn = {1543-1940},
  doi = {10.1007/BF02646981},
  urldate = {2021-10-27},
  langid = {english}
}

@article{cernySegregationPhosphorusSilicon2024,
  title = {Segregation of {{Phosphorus}} and {{Silicon}} at the {{Grain Boundary}} in {{Bcc Iron}} via {{Machine-Learned Force Fields}}},
  author = {{\v C}ern{\'y}, Miroslav and {\v S}est{\'a}k, Petr},
  year = 2024,
  month = jan,
  journal = {Crystals},
  volume = {14},
  number = {1},
  pages = {74},
  publisher = {Multidisciplinary Digital Publishing Institute},
  issn = {2073-4352},
  doi = {10.3390/cryst14010074},
  urldate = {2025-06-02},
  copyright = {http://creativecommons.org/licenses/by/3.0/},
  langid = {english}
}

@article{dengSystematicSofteningUniversal2025,
  title = {Systematic Softening in Universal Machine Learning Interatomic Potentials},
  author = {Deng, Bowen and Choi, Yunyeong and Zhong, Peichen and Riebesell, Janosh and Anand, Shashwat and Li, Zhuohan and Jun, KyuJung and Persson, Kristin A. and Ceder, Gerbrand},
  year = 2025,
  month = jan,
  journal = {npj Computational Materials},
  volume = {11},
  number = {1},
  pages = {9},
  publisher = {Nature Publishing Group},
  issn = {2057-3960},
  doi = {10.1038/s41524-024-01500-6},
  urldate = {2025-10-21},
  copyright = {2025 The Author(s)},
  langid = {english}
}

@article{duFirstprinciplesStudyInteraction2011a,
  title = {First-Principles Study on the Interaction of {{H}} Interstitials with Grain Boundaries in \$\textbackslash ensuremath\textbraceleft\textbackslash alpha\textbraceright\$- and \$\textbackslash ensuremath\textbraceleft\textbackslash gamma\textbraceright\$-{{Fe}}},
  author = {Du, Yaojun A. and Ismer, Lars and Rogal, Jutta and Hickel, Tilmann and Neugebauer, J{\"o}rg and Drautz, Ralf},
  year = 2011,
  month = oct,
  journal = {Physical Review B},
  volume = {84},
  number = {14},
  pages = {144121},
  publisher = {American Physical Society},
  doi = {10.1103/PhysRevB.84.144121},
  urldate = {2022-01-24}
}

@article{echeverrirestrepoApplicabilityUniversalMachine2025,
  title = {Applicability of Universal Machine Learning Interatomic Potentials to the Simulation of Steels},
  author = {Echeverri Restrepo, Sebasti{\'a}n and Mohandas, Naveen K and Sluiter, Marcel H F and Paxton, Anthony T},
  year = 2025,
  month = mar,
  journal = {Modelling and Simulation in Materials Science and Engineering},
  volume = {33},
  number = {3},
  pages = {035003},
  publisher = {IOP Publishing},
  issn = {0965-0393},
  doi = {10.1088/1361-651X/adb483},
  urldate = {2025-10-21},
  langid = {english}
}

@article{hatcherParameterizedElectronicDescription2014,
  title = {Parameterized Electronic Description of Carbon Cohesion in Iron Grain Boundaries},
  author = {Hatcher, Nicholas and Madsen, Georg K H and Drautz, Ralf},
  year = 2014,
  month = mar,
  journal = {Journal of Physics: Condensed Matter},
  volume = {26},
  number = {14},
  pages = {145502},
  publisher = {IOP Publishing},
  issn = {0953-8984},
  doi = {10.1088/0953-8984/26/14/145502},
  urldate = {2025-10-22},
  langid = {english}
}

@article{heFirstprinciplesInvestigationEffect2013,
  title = {First-Principles Investigation into the Effect of {{Cr}} on the Segregation of Multi-{{H}} at the {{Fe $\Sigma$3}} (111) Grain Boundary},
  author = {He, Bingling and Xiao, Wei and Hao, Wei and Tian, Zhixue},
  year = 2013,
  month = oct,
  journal = {Journal of Nuclear Materials},
  volume = {441},
  number = {1},
  pages = {301--305},
  issn = {0022-3115},
  doi = {10.1016/j.jnucmat.2013.06.015},
  urldate = {2021-07-31},
  langid = {english}
}

@article{himanenDScribeLibraryDescriptors2020,
  title = {{{DScribe}}: {{Library}} of Descriptors for Machine Learning in Materials Science},
  shorttitle = {{{DScribe}}},
  author = {Himanen, Lauri and J{\"a}ger, Marc O.J. and Morooka, Eiaki V. and Federici Canova, Filippo and Ranawat, Yashasvi S. and Gao, David Z. and Rinke, Patrick and Foster, Adam S.},
  year = 2020,
  month = feb,
  journal = {Computer Physics Communications},
  volume = {247},
  pages = {106949},
  issn = {00104655},
  doi = {10.1016/j.cpc.2019.106949},
  urldate = {2026-03-06},
  langid = {english}
}

@article{itoElectronicOriginGrain2020,
  title = {Electronic Origin of Grain Boundary Segregation of {{Al}}, {{Si}}, {{P}}, and {{S}} in Bcc-{{Fe}}: Combined Analysis of Ab Initio Local Energy and Crystal Orbital {{Hamilton}} Population},
  shorttitle = {Electronic Origin of Grain Boundary Segregation of {{Al}}, {{Si}}, {{P}}, and {{S}} in Bcc-{{Fe}}},
  author = {Ito, Kazuma and Sawada, Hideaki and Tanaka, Shingo and Ogata, Shigenobu and Kohyama, Masanori},
  year = 2020,
  month = nov,
  journal = {Modelling and Simulation in Materials Science and Engineering},
  volume = {29},
  number = {1},
  pages = {015001},
  publisher = {IOP Publishing},
  issn = {0965-0393},
  doi = {10.1088/1361-651X/abc04c},
  urldate = {2024-04-19},
  langid = {english}
}

@article{itoMachineLearningInteratomic2024,
  title = {Machine Learning Interatomic Potential with {{DFT}} Accuracy for General Grain Boundaries in {$\alpha$}-{{Fe}}},
  author = {Ito, Kazuma and Yokoi, Tatsuya and Hyodo, Katsutoshi and Mori, Hideki},
  year = 2024,
  month = nov,
  journal = {npj Computational Materials},
  volume = {10},
  number = {1},
  pages = {255},
  publisher = {Nature Publishing Group},
  issn = {2057-3960},
  doi = {10.1038/s41524-024-01451-y},
  urldate = {2025-10-21},
  copyright = {2024 The Author(s)},
  langid = {english}
}

@article{janssenPyironIntegratedDevelopment2019,
  title = {Pyiron: {{An}} Integrated Development Environment for Computational Materials Science},
  shorttitle = {Pyiron},
  author = {Janssen, Jan and Surendralal, Sudarsan and Lysogorskiy, Yury and Todorova, Mira and Hickel, Tilmann and Drautz, Ralf and Neugebauer, J{\"o}rg},
  year = 2019,
  month = jun,
  journal = {Computational Materials Science},
  volume = {163},
  pages = {24--36},
  issn = {0927-0256},
  doi = {10.1016/j.commatsci.2018.07.043},
  urldate = {2024-05-03}
}

@article{johnsonIIRemarkableChanges1875,
  title = {{{II}}. {{On}} Some Remarkable Changes Produced in Iron and Steel by the Action of Hydrogen and Acids},
  author = {Johnson, William H.},
  year = 1875,
  journal = {Proceedings of the Royal Society of London},
  volume = {23},
  number = {156-163},
  pages = {168--179},
  publisher = {The Royal Society London}
}

@article{johnsonRemarkableChangesProduced1875,
  title = {On Some Remarkable Changes Produced in Iron and Steel by the Action of Hydrogen and Acids},
  author = {Johnson, William H.},
  year = 1875,
  journal = {Nature},
  volume = {11},
  number = {281},
  pages = {393}
}

@article{kalderonSteamTurbineFailure1972,
  title = {Steam {{Turbine Failure}} at {{Hinkley Point}} `{{A}}'},
  author = {Kalderon, D.},
  year = 1972,
  month = jun,
  journal = {Proceedings of the Institution of Mechanical Engineers},
  volume = {186},
  number = {1},
  pages = {341--377},
  publisher = {IMECHE},
  issn = {0020-3483},
  doi = {10.1243/PIME_PROC_1972_186_037_02},
  urldate = {2022-06-10},
  langid = {english}
}

@article{kholtobinaEffectAlloyingElements2021,
  title = {Effect of Alloying Elements on Hydrogen Enhanced Decohesion in Bcc Iron},
  author = {Kholtobina, Anastasiia S. and Ecker, Werner and Pippan, Reinhard and Razumovskiy, Vsevolod I.},
  year = 2021,
  month = feb,
  journal = {Computational Materials Science},
  volume = {188},
  pages = {110215},
  issn = {0927-0256},
  doi = {10.1016/j.commatsci.2020.110215},
  urldate = {2021-07-30},
  langid = {english}
}

@article{kresseEfficiencyAbinitioTotal1996,
  title = {Efficiency of Ab-Initio Total Energy Calculations for Metals and Semiconductors Using a Plane-Wave Basis Set},
  author = {Kresse, G. and Furthm{\"u}ller, J.},
  year = 1996,
  month = jul,
  journal = {Computational Materials Science},
  volume = {6},
  number = {1},
  pages = {15--50},
  issn = {0927-0256},
  doi = {10.1016/0927-0256(96)00008-0},
  urldate = {2021-08-07},
  langid = {english}
}

@article{kresseEfficientIterativeSchemes1996,
  title = {Efficient Iterative Schemes for Ab Initio Total-Energy Calculations Using a Plane-Wave Basis Set},
  author = {Kresse, G. and Furthm{\"u}ller, J.},
  year = 1996,
  month = oct,
  journal = {Physical Review B},
  volume = {54},
  number = {16},
  pages = {11169--11186},
  doi = {10.1103/PhysRevB.54.11169},
  urldate = {2020-01-28}
}

@article{lejcekInterfacialSegregationGrain2017,
  title = {Interfacial Segregation and Grain Boundary Embrittlement: {{An}} Overview and Critical Assessment of Experimental Data and Calculated Results},
  shorttitle = {Interfacial Segregation and Grain Boundary Embrittlement},
  author = {Lej{\v c}ek, Pavel and {\v S}ob, Mojm{\'i}r and Paidar, V{\'a}clav},
  year = 2017,
  month = jun,
  journal = {Progress in Materials Science},
  volume = {87},
  pages = {83--139},
  issn = {0079-6425},
  doi = {10.1016/j.pmatsci.2016.11.001},
  urldate = {2021-07-13},
  langid = {english}
}

@article{lejcekInterstitialSubstitutionalSolute2016,
  title = {Interstitial and Substitutional Solute Segregation at Individual Grain Boundaries of {$\alpha$}-Iron: Data Revisited},
  shorttitle = {Interstitial and Substitutional Solute Segregation at Individual Grain Boundaries of {$\alpha$}-Iron},
  author = {Lej{\v c}ek, Pavel and Hofmann, Siegfried},
  year = 2016,
  month = jan,
  journal = {Journal of Physics: Condensed Matter},
  volume = {28},
  number = {6},
  pages = {064001},
  publisher = {IOP Publishing},
  issn = {0953-8984},
  doi = {10.1088/0953-8984/28/6/064001},
  urldate = {2024-04-17},
  langid = {english}
}

@article{liProbingOutofdistributionGeneralization2025,
  title = {Probing Out-of-Distribution Generalization in Machine Learning for Materials},
  author = {Li, Kangming and Rubungo, Andre Niyongabo and Lei, Xiangyun and Persaud, Daniel and Choudhary, Kamal and DeCost, Brian and Dieng, Adji Bousso and {Hattrick-Simpers}, Jason},
  year = 2025,
  month = jan,
  journal = {Communications Materials},
  volume = {6},
  number = {1},
  pages = {9},
  publisher = {Nature Publishing Group},
  issn = {2662-4443},
  doi = {10.1038/s43246-024-00731-w},
  urldate = {2025-09-29},
  copyright = {2025 The Author(s)},
  langid = {english}
}

@article{maiHighthroughputInitioStudy2025a,
  title = {A High-Throughput Ab Initio Study of Elemental Segregation and Cohesion at Ferritic-Iron Grain Boundaries},
  author = {Mai, Han Lin and Cui, Xiang-Yuan and Hickel, Tilmann and Neugebauer, J{\"o}rg and Ringer, Simon P.},
  year = 2025,
  month = sep,
  journal = {Acta Materialia},
  volume = {297},
  pages = {121288},
  issn = {1359-6454},
  doi = {10.1016/j.actamat.2025.121288},
  urldate = {2025-11-04}
}

@article{maiPhosphorusTransitionMetal2023,
  title = {Phosphorus and Transition Metal Co-Segregation in Ferritic Iron Grain Boundaries and Its Effects on Cohesion},
  author = {Mai, Han Lin and Cui, Xiang-Yuan and Scheiber, Daniel and Romaner, Lorenz and Ringer, Simon P.},
  year = 2023,
  month = may,
  journal = {Acta Materialia},
  volume = {250},
  pages = {118850},
  issn = {1359-6454},
  doi = {10.1016/j.actamat.2023.118850},
  urldate = {2023-09-13}
}

@article{maiSegregationTransitionMetals2022,
  title = {The Segregation of Transition Metals to Iron Grain Boundaries and Their Effects on Cohesion},
  author = {Mai, Han Lin and Cui, Xiang-Yuan and Scheiber, Daniel and Romaner, Lorenz and Ringer, Simon P.},
  year = 2022,
  month = jun,
  journal = {Acta Materialia},
  volume = {231},
  pages = {117902},
  issn = {1359-6454},
  doi = {10.1016/j.actamat.2022.117902},
  urldate = {2022-04-11},
  langid = {english}
}

@article{manzIntroducingDDEC6Atomic2017,
  title = {Introducing {{DDEC6}} Atomic Population Analysis: Part 3. {{Comprehensive}} Method to Compute Bond Orders},
  shorttitle = {Introducing {{DDEC6}} Atomic Population Analysis},
  author = {Manz, Thomas A.},
  year = 2017,
  journal = {RSC advances},
  volume = {7},
  number = {72},
  pages = {45552--45581},
  publisher = {Royal Society of Chemistry}
}

@article{mcleanGrainBoundariesMetals1958,
  title = {Grain {{Boundaries}} in {{Metals}}},
  author = {McLean, D. and Maradudin, A.},
  year = 1958,
  month = jan,
  journal = {Physics Today},
  volume = {11},
  pages = {35},
  issn = {0031-9228},
  doi = {10.1063/1.3062658},
  urldate = {2021-07-13}
}

@article{mcmahonTheoryEmbrittlementSteels1978,
  title = {On the Theory of Embrittlement of Steels by Segregated Impurities},
  author = {McMahon, C. J. and Vitek, V. and Belton, G. R.},
  year = 1978,
  month = sep,
  journal = {Scripta Metallurgica},
  volume = {12},
  number = {9},
  pages = {785--789},
  issn = {0036-9748},
  doi = {10.1016/0036-9748(78)90036-4},
  urldate = {2021-10-27},
  langid = {english}
}

@article{mengHighlyTransferableEfficient2024a,
  title = {A Highly Transferable and Efficient Machine Learning Interatomic Potentials Study of {$\alpha<$}math{$><$}mi Is="true"{$>\alpha<$}/Mi{$><$}/Math{$>$}-{{Fe}}--{{C}} Binary System},
  author = {Meng, Fan-Shun and Shinzato, Shuhei and Zhang, Shihao and Matsubara, Kazuki and Du, Jun-Ping and Yu, Peijun and Geng, Wen-Tong and Ogata, Shigenobu},
  year = 2024,
  month = dec,
  journal = {Acta Materialia},
  volume = {281},
  pages = {120408},
  issn = {1359-6454},
  doi = {10.1016/j.actamat.2024.120408},
  urldate = {2024-10-13}
}

@article{mengNeuralNetworkInteratomic2025,
  title = {A {{Neural Network Interatomic Potential}} for the {{Ternary}} {$\alpha$}-{{Fe-C-H System}}: {{Toward Million-Atom Simulations}} of {{Hydrogen Embrittlement}} in {{Steel}}},
  shorttitle = {A {{Neural Network Interatomic Potential}} for the {{Ternary}} {$\alpha$}-{{Fe-C-H System}}},
  author = {Meng, Fan-Shun and Shinzato, Shuhei and Matsubara, Kazuki and Du, Jun-Ping and Yu, Peijun and Ogata, Shigenobu},
  year = 2025,
  month = nov,
  journal = {JOM},
  volume = {77},
  number = {11},
  pages = {8101--8117},
  issn = {1543-1851},
  doi = {10.1007/s11837-025-07721-4},
  urldate = {2025-10-22},
  langid = {english}
}

@article{menonPyscalPythonModule2019,
  title = {Pyscal: {{A}} Python Module for Structural Analysis of Atomic Environments},
  shorttitle = {Pyscal},
  author = {Menon, Sarath and Leines, Grisell and Rogal, Jutta},
  year = 2019,
  month = nov,
  journal = {Journal of Open Source Software},
  volume = {4},
  number = {43},
  pages = {1824},
  issn = {2475-9066},
  doi = {10.21105/joss.01824},
  urldate = {2025-04-01},
  copyright = {http://creativecommons.org/licenses/by/4.0/}
}

@article{mirzaevInitioModellingInteraction2016,
  title = {Ab Initio Modelling of the Interaction of {{H}} Interstitials with Grain Boundaries in Bcc {{Fe}}},
  author = {Mirzaev, D.A. and Mirzoev, A.A. and Okishev, {\relax K.Yu}. and Verkhovykh, A.V.},
  year = 2016,
  month = may,
  journal = {Molecular Physics},
  volume = {114},
  number = {9},
  pages = {1502--1512},
  publisher = {Taylor \& Francis},
  issn = {0026-8976},
  doi = {10.1080/00268976.2015.1136439},
  urldate = {2026-04-04}
}

@article{mommaVESTAThreedimensionalVisualization2008,
  title = {{{VESTA}}: A Three-Dimensional Visualization System for Electronic and Structural Analysis},
  shorttitle = {{{VESTA}}},
  author = {Momma, K. and Izumi, F.},
  year = 2008,
  month = jun,
  journal = {Journal of Applied Crystallography},
  volume = {41},
  number = {3},
  pages = {653--658},
  publisher = {International Union of Crystallography},
  issn = {0021-8898},
  doi = {10.1107/S0021889808012016},
  urldate = {2024-04-24},
  langid = {english}
}

@article{ongPythonMaterialsGenomics2013,
  title = {Python {{Materials Genomics}} (Pymatgen): {{A}} Robust, Open-Source Python Library for Materials Analysis},
  shorttitle = {Python {{Materials Genomics}} (Pymatgen)},
  author = {Ong, Shyue Ping and Richards, William Davidson and Jain, Anubhav and Hautier, Geoffroy and Kocher, Michael and Cholia, Shreyas and Gunter, Dan and Chevrier, Vincent L. and Persson, Kristin A. and Ceder, Gerbrand},
  year = 2013,
  month = feb,
  journal = {Computational Materials Science},
  volume = {68},
  pages = {314--319},
  issn = {0927-0256},
  doi = {10.1016/j.commatsci.2012.10.028},
  urldate = {2022-10-02},
  langid = {english}
}

@article{perdewGeneralizedGradientApproximation1996,
  title = {Generalized {{Gradient Approximation Made Simple}}},
  author = {Perdew, John P. and Burke, Kieron and Ernzerhof, Matthias},
  year = 1996,
  month = oct,
  journal = {Physical Review Letters},
  volume = {77},
  number = {18},
  pages = {3865--3868},
  doi = {10.1103/PhysRevLett.77.3865},
  urldate = {2020-01-28}
}

@article{reiners-sakicInterstitialsKeyIngredient2025,
  title = {Interstitials as a Key Ingredient for {{P}} Segregation to Grain Boundaries in Polycrystalline {$\alpha$}-{{Fe}}},
  author = {{Reiners-Sakic}, Amin and Reichmann, Alexander and D{\"o}singer, Christoph and Romaner, Lorenz and Holec, David},
  year = 2025,
  month = nov,
  journal = {Scripta Materialia},
  volume = {268},
  pages = {116864},
  issn = {1359-6462},
  doi = {10.1016/j.scriptamat.2025.116864},
  urldate = {2025-10-22}
}

@article{rycroftVoroThreedimensionalVoronoi2009,
  title = {Voro++: A Three-Dimensional {{Voronoi}} Cell Library in {{C}}++},
  shorttitle = {Voro++},
  author = {Rycroft, Chris},
  year = 2009,
  month = feb,
  urldate = {2025-04-01},
  langid = {english}
}

@article{sakicInterplayAlloyingTramp2024,
  title = {Interplay between Alloying and Tramp Element Effects on Temper Embrittlement in Bcc Iron: {{DFT}} and Thermodynamic Insights},
  shorttitle = {Interplay between Alloying and Tramp Element Effects on Temper Embrittlement in Bcc Iron},
  author = {Sakic, Amin and Schnitzer, Ronald and Holec, David},
  year = 2024,
  month = aug,
  journal = {Acta Materialia},
  volume = {275},
  pages = {120044},
  issn = {1359-6454},
  doi = {10.1016/j.actamat.2024.120044},
  urldate = {2025-02-15}
}

@article{scheiberImpactSegregationEnergy2021,
  title = {Impact of the Segregation Energy Spectrum on the Enthalpy and Entropy of Segregation},
  author = {Scheiber, Daniel and Romaner, Lorenz},
  year = 2021,
  month = dec,
  journal = {Acta Materialia},
  volume = {221},
  pages = {117393},
  issn = {1359-6454},
  doi = {10.1016/j.actamat.2021.117393},
  urldate = {2021-11-11},
  langid = {english}
}

@article{shuangUniversalMachineLearning2025,
  title = {Universal Machine Learning Interatomic Potentials Poised to Supplant {{DFT}} in Modeling General Defects in Metals and Random Alloys},
  author = {Shuang, Fei and Wei, Zixiong and Liu, Kai and Gao, Wei and Dey, Poulumi},
  year = 2025,
  month = jul,
  journal = {Machine Learning: Science and Technology},
  volume = {6},
  number = {3},
  pages = {030501},
  publisher = {IOP Publishing},
  issn = {2632-2153},
  doi = {10.1088/2632-2153/adea2d},
  urldate = {2025-10-21},
  langid = {english}
}

@article{suzudoAtomisticModelingHe2013,
  title = {Atomistic Modeling of {{He}} Embrittlement at Grain Boundaries of {$\alpha$}-{{Fe}}: A Common Feature over Different Grain Boundaries},
  shorttitle = {Atomistic Modeling of {{He}} Embrittlement at Grain Boundaries of {$\alpha$}-{{Fe}}},
  author = {Suzudo, T. and Yamaguchi, M. and Tsuru, T.},
  year = 2013,
  month = nov,
  journal = {Modelling and Simulation in Materials Science and Engineering},
  volume = {21},
  number = {8},
  pages = {085013},
  publisher = {IOP Publishing},
  issn = {0965-0393},
  doi = {10.1088/0965-0393/21/8/085013},
  urldate = {2024-05-06},
  langid = {english}
}

@article{tahirHydrogenEmbrittlementCarbon2014,
  title = {Hydrogen Embrittlement of a Carbon Segregated \$\textbackslash{{Sigma}}\$5(310)[001] Symmetrical Tilt Grain Boundary in \$\textbackslash alpha\$-{{Fe}}},
  author = {Tahir, A. M. and Janisch, R. and Hartmaier, A.},
  year = 2014,
  month = aug,
  journal = {Materials Science and Engineering: A},
  volume = {612},
  pages = {462--467},
  issn = {0921-5093},
  doi = {10.1016/j.msea.2014.06.071},
  urldate = {2021-09-08},
  langid = {english}
}

@article{tehranchiAtomisticStudyHydrogen2017,
  title = {Atomistic Study of Hydrogen Embrittlement of Grain Boundaries in Nickel: {{I}}. {{Fracture}}},
  shorttitle = {Atomistic Study of Hydrogen Embrittlement of Grain Boundaries in Nickel},
  author = {Tehranchi, A. and Curtin, W. A.},
  year = 2017,
  month = jan,
  journal = {Journal of the Mechanics and Physics of Solids},
  volume = {101},
  pages = {150--165},
  issn = {0022-5096},
  doi = {10.1016/j.jmps.2017.01.020},
  urldate = {2021-08-07},
  langid = {english}
}

@article{tuchindaGrainBoundaryEmbrittlement2025,
  title = {A Grain Boundary Embrittlement Genome for Substitutional Cubic Alloys},
  author = {Tuchinda, Nutth and Olson, Gregory B. and Schuh, Christopher A.},
  year = 2025,
  month = apr,
  journal = {Applied Physics Letters},
  volume = {126},
  number = {17},
  publisher = {AIP Publishing},
  issn = {0003-6951},
  doi = {10.1063/5.0264543},
  urldate = {2025-10-21},
  langid = {english}
}

@article{wachowiczEffectImpuritiesStructural2011,
  title = {Effect of Impurities on Structural, Cohesive and Magnetic Properties of Grain Boundaries in \$\textbackslash upalpha\$-{{Fe}}},
  author = {Wachowicz, E. and Kiejna, A.},
  year = 2011,
  month = jan,
  volume = {19},
  number = {2},
  pages = {025001},
  publisher = {IOP Publishing},
  issn = {0965-0393},
  doi = {10.1088/0965-0393/19/2/025001},
  urldate = {2021-11-11},
  langid = {english}
}

@article{wagihGrainBoundarySegregation2024,
  title = {Grain {{Boundary Segregation Predicted}} by {{Quantum-Accurate Segregation Spectra}} but Not by {{Classical Models}}},
  author = {Wagih, Malik and Naunheim, Yannick and Lei, Tianjiao and Schuh, Christopher},
  year = 2024,
  month = jan,
  journal = {Acta Materialia},
  pages = {119674},
  doi = {10.1016/j.actamat.2024.119674}
}

@article{wagihLearningGrainBoundary2020,
  title = {Learning Grain Boundary Segregation Energy Spectra in Polycrystals},
  author = {Wagih, Malik and Larsen, Peter M. and Schuh, Christopher A.},
  year = 2020,
  month = dec,
  journal = {Nature Communications},
  volume = {11},
  number = {1},
  pages = {6376},
  publisher = {Nature Publishing Group},
  issn = {2041-1723},
  doi = {10.1038/s41467-020-20083-6},
  urldate = {2022-02-23},
  copyright = {2020 The Author(s)},
  langid = {english}
}

@article{wagihSpectrumInterstitialSolute2023,
  title = {The Spectrum of Interstitial Solute Energies in Polycrystals},
  author = {Wagih, Malik and Schuh, Christopher A.},
  year = 2023,
  journal = {Scripta Materialia},
  volume = {235},
  pages = {115631},
  publisher = {Elsevier},
  urldate = {2024-04-19}
}

@article{wagihViewpointCanSymmetric2023,
  title = {Viewpoint: {{Can}} Symmetric Tilt Grain Boundaries Represent Polycrystals?},
  shorttitle = {Viewpoint},
  author = {Wagih, Malik and Schuh, Christopher A.},
  year = 2023,
  month = dec,
  journal = {Scripta Materialia},
  volume = {237},
  pages = {115716},
  issn = {1359-6462},
  doi = {10.1016/j.scriptamat.2023.115716},
  urldate = {2024-04-18}
}

@article{wangFirstprinciplesStudyCarbon2016a,
  title = {First-Principles Study of Carbon Segregation in Bcc Iron Symmetrical Tilt Grain Boundaries},
  author = {Wang, Jingliang and Janisch, Rebecca and Madsen, Georg K. H. and Drautz, Ralf},
  year = 2016,
  month = aug,
  journal = {Acta Materialia},
  volume = {115},
  pages = {259--268},
  issn = {1359-6454},
  doi = {10.1016/j.actamat.2016.04.058},
  urldate = {2021-08-31},
  langid = {english}
}

@article{wilkinsonFAIRGuidingPrinciples2016,
  title = {The {{FAIR Guiding Principles}} for Scientific Data Management and Stewardship},
  author = {Wilkinson, Mark D. and Dumontier, Michel and Aalbersberg, IJsbrand Jan and Appleton, Gabrielle and Axton, Myles and Baak, Arie and Blomberg, Niklas and Boiten, Jan-Willem and {da Silva Santos}, Luiz Bonino and Bourne, Philip E. and Bouwman, Jildau and Brookes, Anthony J. and Clark, Tim and Crosas, Merc{\`e} and Dillo, Ingrid and Dumon, Olivier and Edmunds, Scott and Evelo, Chris T. and Finkers, Richard and {Gonzalez-Beltran}, Alejandra and Gray, Alasdair J. G. and Groth, Paul and Goble, Carole and Grethe, Jeffrey S. and Heringa, Jaap and {'t Hoen}, Peter A. C. and Hooft, Rob and Kuhn, Tobias and Kok, Ruben and Kok, Joost and Lusher, Scott J. and Martone, Maryann E. and Mons, Albert and Packer, Abel L. and Persson, Bengt and {Rocca-Serra}, Philippe and Roos, Marco and {van Schaik}, Rene and Sansone, Susanna-Assunta and Schultes, Erik and Sengstag, Thierry and Slater, Ted and Strawn, George and Swertz, Morris A. and Thompson, Mark and {van der Lei}, Johan and {van Mulligen}, Erik and Velterop, Jan and Waagmeester, Andra and Wittenburg, Peter and Wolstencroft, Katherine and Zhao, Jun and Mons, Barend},
  year = 2016,
  month = mar,
  journal = {Scientific Data},
  volume = {3},
  number = {1},
  pages = {160018},
  publisher = {Nature Publishing Group},
  issn = {2052-4463},
  doi = {10.1038/sdata.2016.18},
  urldate = {2026-04-04},
  copyright = {2016 The Author(s)},
  langid = {english}
}

@article{xuEffectAlloyingSolutes2024,
  title = {Effect of Alloying Solutes on Hydrogen Segregation at Pure Iron {{$\Sigma$3}}(111) Grain Boundary: {{First-principles}} Calculation},
  shorttitle = {Effect of Alloying Solutes on Hydrogen Segregation at Pure Iron {{$\Sigma$3}}(111) Grain Boundary},
  author = {Xu, Zemin and Cheng, Lin and Xia, Kai and Hu, Chengyang and Wu, Kaiming},
  year = 2024,
  month = sep,
  journal = {International Journal of Hydrogen Energy},
  volume = {84},
  pages = {321--333},
  issn = {0360-3199},
  doi = {10.1016/j.ijhydene.2024.08.232},
  urldate = {2025-10-21}
}

@article{yamaguchiDecohesionIronGrain2007,
  title = {Decohesion of Iron Grain Boundaries by Sulfur or Phosphorous Segregation: {{First-principles}} Calculations},
  shorttitle = {Decohesion of Iron Grain Boundaries by Sulfur or Phosphorous Segregation},
  author = {Yamaguchi, Masatake and Nishiyama, Yutaka and Kaburaki, Hideo},
  year = 2007,
  month = jul,
  journal = {Physical Review B},
  volume = {76},
  number = {3},
  pages = {035418},
  publisher = {American Physical Society},
  doi = {10.1103/PhysRevB.76.035418},
  urldate = {2021-10-28}
}

@article{yamaguchiFirstprinciplesStudyGrain2011,
  title = {First-Principles Study on the Grain Boundary Embrittlement of Metals by Solute Segregation: {{Part II}}. {{Metal}} ({{Fe}}, {{Al}}, {{Cu}})-Hydrogen ({{H}}) Systems},
  shorttitle = {First-Principles Study on the Grain Boundary Embrittlement of Metals by Solute Segregation},
  author = {Yamaguchi, Masatake and Ebihara, Ken-Ichi and Itakura, Mitsuhiro and Kadoyoshi, Tomoko and Suzudo, Tomoaki and Kaburaki, Hideo},
  year = 2011,
  journal = {Metallurgical and Materials Transactions A},
  volume = {42},
  number = {2},
  pages = {330--339},
  publisher = {Springer}
}

@article{yamaguchiFirstprinciplesStudyGrain2011a,
  title = {First-Principles Study on the Grain Boundary Embrittlement of Metals by Solute Segregation: {{Part I}}. Iron ({{Fe}})-Solute ({{B}}, {{C}}, {{P}}, and {{S}}) Systems},
  shorttitle = {First-Principles Study on the Grain Boundary Embrittlement of Metals by Solute Segregation},
  author = {Yamaguchi, Masatake},
  year = 2011,
  journal = {Metallurgical and Materials Transactions A},
  volume = {42},
  number = {2},
  pages = {319--329},
  publisher = {Springer}
}

@article{yamaguchiIntergranularDecohesionInduced2014,
  title = {Intergranular {{Decohesion Induced}} by {{Mobile Hydrogen}} in {{Iron}} with and without {{Segregated Carbon}}: {{First-Principles Calculations}}},
  shorttitle = {Intergranular {{Decohesion Induced}} by {{Mobile Hydrogen}} in {{Iron}} with and without {{Segregated Carbon}}},
  author = {Yamaguchi, Masatake and Kameda, Jun},
  year = 2014,
  month = jan,
  doi = {10.1115/1.860298_ch80},
  urldate = {2024-10-21},
  langid = {english}
}

@article{yamaguchiMobileEffectHydrogen2012,
  title = {Mobile Effect of Hydrogen on Intergranular Decohesion of Iron: First-Principles Calculations},
  shorttitle = {Mobile Effect of Hydrogen on Intergranular Decohesion of Iron},
  author = {Yamaguchi, Masatake and Kameda, Jun and Ebihara, Ken-Ichi and Itakura, Mitsuhiro and Kaburaki, Hideo},
  year = 2012,
  month = apr,
  journal = {Philosophical Magazine},
  volume = {92},
  number = {11},
  pages = {1349--1368},
  publisher = {Taylor \& Francis},
  issn = {1478-6435},
  doi = {10.1080/14786435.2011.645077},
  urldate = {2023-01-24}
}

@article{zhangFirstprinciplesStudyHe2010,
  title = {First-Principles Study of {{He}} Effects in a Bcc {{Fe}} Grain Boundary: Site Preference, Segregation and Theoretical Tensile Strength},
  shorttitle = {First-Principles Study of {{He}} Effects in a Bcc {{Fe}} Grain Boundary},
  author = {Zhang, Lei and Shu, Xiaolin and Jin, Shuo and Zhang, Ying and Lu, Guang-Hong},
  year = 2010,
  month = aug,
  journal = {Journal of Physics: Condensed Matter},
  volume = {22},
  number = {37},
  pages = {375401},
  issn = {0953-8984},
  doi = {10.1088/0953-8984/22/37/375401},
  urldate = {2024-05-06},
  langid = {english}
}

@article{zhangFirstprinciplesStudyHelium2009,
  title = {First-Principles Study of Helium Effect in a Ferromagnetic Iron Grain Boundary: {{Energetics}}, Site Preference and Segregation},
  shorttitle = {First-Principles Study of Helium Effect in a Ferromagnetic Iron Grain Boundary},
  author = {Zhang, Ying and Feng, Wen-Qiang and Liu, Yue-Lin and Lu, Guang-Hong and Wang, Tianmin},
  year = 2009,
  month = sep,
  journal = {Nuclear Instruments and Methods in Physics Research Section B: Beam Interactions with Materials and Atoms},
  series = {Proceedings of the {{Ninth International Conference}} on {{Computer Simulation}} of {{Radiation Effects}} in {{Solids}}},
  volume = {267},
  number = {18},
  pages = {3200--3203},
  issn = {0168-583X},
  doi = {10.1016/j.nimb.2009.06.064},
  urldate = {2024-05-06}
}
\end{document}